%% file: dpm_vs_9.tex
\documentclass[reqno,12pt]{article}

\usepackage{amsmath}
\usepackage[numbers,super,comma,compress]{natbib}
\makeatletter
\renewcommand\@biblabel[1]{#1.}
\makeatother

\usepackage{times,bm}
\usepackage{amssymb,graphicx,setspace,verbatim}
\usepackage[small]{caption}
\usepackage{subcaption}
\usepackage{wrapfig}
\usepackage{multirow}
\usepackage{rotating}
\usepackage{color}
\usepackage{adjustbox}
\usepackage{scalerel}
\usepackage{longtable}
\usepackage{changepage}
\usepackage{geometry}
\usepackage{enumitem}

\renewcommand{\baselinestretch}{1.75} %also used to set spacing
\renewcommand{\arraystretch}{.95}

\newcommand{\bzero}{\mbox{\boldmath $0$}}

\newcommand{\bb}{\mbox{\boldmath $b$}}

\newcommand{\bd}{\mbox{\boldmath $d$}}

\newcommand{\bx}{\mbox{\boldmath $x$}}

\newcommand{\by}{\mbox{\boldmath $y$}}
\newcommand{\bz}{\mbox{\boldmath $z$}}
\newcommand{\bA}{\mbox{\boldmath $A$}}
\newcommand{\bB}{\mbox{\boldmath $B$}}

\newcommand{\bI}{\mbox{\boldmath $I$}}

\newcommand{\bM}{\mbox{\boldmath $M$}}

\newcommand{\bP}{\mbox{\boldmath $P$}}
\newcommand{\bQ}{\mbox{\boldmath $Q$}}

\newcommand{\bU}{\mbox{\boldmath $U$}}
\newcommand{\bV}{\mbox{\boldmath $V$}}

\newcommand{\bY}{\mbox{\boldmath $Y$}}
\newcommand{\bZ}{\mbox{\boldmath $Z$}}

\newcommand{\cC}{{\cal C}}
\newcommand{\cD}{{\cal D}}

\newcommand{\cI}{{\cal I}}

\newcommand{\cM}{{\cal M}}
\newcommand{\cN}{{\cal N}}
\newcommand{\cO}{{\cal O}}

\newcommand{\cW}{{\cal W}}

\newcommand{\hP}{\hat{P}}

\newcommand{\hgamma}{\hat{\gamma}}
\newcommand{\hbgamma}{\mbox{\boldmath $\hgamma$}}

\newcommand{\hphi}{\hat{\phi}}
\newcommand{\hbphi}{\mbox{\boldmath $\hphi$}}

\newcommand{\beps}{\mbox{\boldmath $\varepsilon$}}

\newcommand{\bsV}{\mbox{\small \boldmath $V$}}
\newcommand{\bsPsi}{\mbox{\small \boldmath $\Psi$}}

\newcommand{\bphi}{\mbox{\boldmath $\phi$}}

\newcommand{\bgamma}{\mbox{\boldmath $\gamma$}}

\newcommand{\bmu}{\mbox{\boldmath $\mu$}}

\newcommand{\bSigma}{\mbox{\boldmath $\Sigma$}}

\newcommand{\bPsi}{\mbox{\boldmath $\Psi$}}

\newcommand{\launch}{{\mbox{\scriptsize launch}}}
\newcommand{\nmmi}{n_{m(\mbox{\scriptsize{-}}i)}}

\newcommand{\bdm}{\begin{displaymath}}
\newcommand{\edm}{\end{displaymath}}
\newcommand{\beq}{\begin{equation}}
\newcommand{\eeq}{\end{equation}}

\newcommand{\tth}{^{\mbox{\scriptsize th}}}

\long\def\symbolfootnote[#1]#2{\begingroup%
\def\thefootnote{\fnsymbol{footnote}}\footnote[#1]{#2}\endgroup}

\newcommand{\white}[1]{\textcolor{white}{#1}}

\newlength{\offsetpage}
\newenvironment{widepage}{\begin{adjustwidth}{-\offsetpage}{-\offsetpage}%
    \addtolength{\textwidth}{2\offsetpage}}%
{\end{adjustwidth}}

\begin{document}

\pagestyle{empty}

\pagenumbering{arabic}
\begin{center}
{\singlespacing
\begin{Large}{\bf
Clustering and Variable Selection in the Presence of Mixed Variable Types and Missing Data\\}
\vspace{.35in}
\end{Large}

C.~B.~Storlie$^\dag$, S.~M.~Myers$^\ddag$, S.~K.~Katusic$^\dag$, A.~L.~Weaver$^\dag$,  R.~Voigt$^\S$, \\[.1in] %R.~C.~Colligan$^\dag$,
P.~E.~Croarkin$^\dag$, R.~E.~Stoeckel$^\dag$, J.~D.~Port$^\dag$ \\[.25in]
$^\dag$Mayo Clinic $\;\;\;\;\;\;\;\;$ $^\S$Texas Children's Hospital\\[.1in]
$^\ddag$Geisinger Autism \& Developmental Medicine Institute\\[.3in]
}

\renewcommand{\baselinestretch}{1.47} %also used to set spacing  
\begin{abstract}
  \vspace{-.03in}We consider the problem of model-based clustering in the presence of many correlated, mixed continuous and discrete variables, some of which may have missing values.  Discrete variables are treated with a latent continuous variable approach and the Dirichlet process is used to construct a mixture model with an unknown number of components.  Variable selection is also performed to identify the variables that are most influential for determining cluster membership. The work is motivated by the need to cluster patients thought to potentially have autism spectrum disorder (ASD) on the basis of many cognitive and/or behavioral test scores.  There are a modest number of patients (486) in the data set along with many (55) test score variables (many of which are discrete valued and/or missing).  The goal of the work is to (i) cluster these patients into similar groups to help identify those with similar clinical presentation, and (ii) identify a sparse subset of tests that inform the clusters in order to eliminate unnecessary testing.   The proposed approach compares very favorably to other methods via simulation of problems of this type.  The results of the ASD analysis suggested three clusters to be most likely, while only four test scores had high ($>0.5$) posterior probability of being informative.  This will result in much more efficient and informative testing.  The need to cluster observations on the basis of many correlated, continuous/discrete variables with missing values, is a common problem in the health sciences as well as in many other disciplines.

\vspace{.065in}

\vspace{.065in}
\noindent
{\em Running title}: Clustering and Variable Selection with Mixed Variable Types

\vspace{.065in}
\noindent
{\em Corresponding Author}: Curtis Storlie, \verb1storlie.curt@mayo.edu1

\end{abstract}

\end{center}

\clearpage
\pagestyle{plain}
\setcounter{page}{1}

%%%%%%%%%%%%%%%%%%%%%%%%%%%%%%%%%%%%%%%%%%%%%%%%%%%%%%%%%%%%%%%%%%%%%%%%%%%%%%%%%%%%%%%%%%%

\vspace{-.35in}
\section{Introduction}
\vspace{-.1in}

Model-based clustering has become a very popular means for unsupervised learning \citep{Fraley02,Basu03,Quintana03,Tadesse2005}. This is due in part to the ability to use the model likelihood to inform, not only the cluster membership, but also the number of clusters $M$ which has been a heavily researched problem for many years.  %This can viewed as model selection problem via BIC or some other criterion, or by treating $M$ as a parameter to be estimated.
The most widely used model-based approach is the normal mixture model which is not suitable for mixed continuous/discrete variables.  For example, this work is motivated by the need to cluster patients thought to potentially have autism spectrum disorder (ASD) on the basis of many correlated test scores.  There are a modest number of patients (486) in the data set along with many (55) test score/self-report variables, many of which are discrete valued or have left or right boundaries.  Figure~\ref{fig:scatter_intro} provides a look at the data across three of the variables; Beery\_standard is discrete valued and ABC\_irritability is continuous, but with significant mass at the left boundary of zero.
The goals of this problem are to (i) cluster these patients into similar groups to help identify those with similar clinical presentation, and (ii) identify a sparse subset of tests that inform the clusters in an effort to eliminate redundant testing.  This problem is also complicated by the fact that many patients in the data have missing test scores.  The need to cluster incomplete observations on the basis of many correlated continuous/discrete variables is a common problem in the health sciences as well as in many other disciplines.

When clustering in high dimensions, it becomes critically important to use some form of dimension reduction or variable selection to achieve accurate cluster formation.  A common approach to deal with this is a principal components or factor approach \citep{Liu2003}.  However, such a solution does not address goal (ii) above for the ASD clustering problem.  The problem of variable selection in regression or conditional density estimation has been well studied from both the $L_1$ penalization \citep{Tibs96,Zou05,Lin06} and Bayesian perspectives \citep{George93,Reich09,Chung2012}.  However, variable selection in clustering is more challenging than that in regression as there is no response to guide (supervise) the selection.  Still, there have been several articles considering this topic; see \citet{Fop2017} for a review. For example, \citet{Raftery2006} propose a partition of the variables into {\em informative} (dependent on cluster membership even after conditioning on all of the other variables) and {\em non-informative} (conditionally independent of cluster membership given the values of the other variables).  They use BIC to accomplish variable selection with a greedy search which is implemented in the R package \verb1clustvarsel1.  Similar approaches are used by \citet{Maugis2009} and \citet{Fop2015}.  An efficient algorithm for identifying the {\em optimal} set of informative variables is provided by \citet{Marbac2017} and implemented in the R package \verb1VarSelLCM1.  Their approach also allows for mixed data types and missing data, however, it assumes both {\em local} and {\em global} independence (i.e., independence of variables within a cluster and unconditional independence of informative and non-informative variables, respectively).  The popular LASSO or L1 type penalization has also been applied to shrink cluster means together for variable selection \citep{Pan2007,Wang2008,Xie2008}.  There have also been several approaches developed for sparse K-means and distance based clustering \citep{Friedman2004,Hoff2006,Witten2012}.
%Some approaches such as \citet{Friedman2004} and \citet{Hoff2006} allow the sparsity of the features to be determined on a component to componenet basis (e.g., encourage variable $x_j$ to have the same mean in some components but not others).  However, this is not necessary to identify the subset of ``required'' tests in goal (ii) above.  Further,
%However, in order to effectively account for uncertainty due to missing values and variable selection, we seek to cast the proposed model into a fully Bayesian framework.

\begin{figure}
\vspace{-.05in}
\begin{center}
  \caption{About Here.}
\vspace{-.3in}
\end{center}
\end{figure}

In the Bayesian literature \citet{Tadesse2005} consider variable selection in the finite normal mixture model using reversible jump (RJ) Markov chain Monte Carlo (MCMC) \citep{Richardson1997}. \citet{Kim2006} extend that work to the nonparametric Bayesian mixture model via the Dirichlet process model (DPM) \citep{Ferguson73,Neal2000,Teh06,LidHjort10}.  The DPM has the advantage of allowing for a countably infinite number of possible components (thus making it nonparametric), while providing a posterior distribution for how many components have been {\em observed} in the data set at hand.  Both \citet{Tadesse2005} and \citet{Kim2006} use a point mass prior to achieve sparse representation of the informative variables.  However, for simplicity they assume all non-informative variables are (unconditionally) independent of the informative variables.  This assumption is frequently violated in practice and it is particularly problematic in the case of the ASD analysis as it would force far too many variables to be included into the informative set as is demonstrated later in this paper.

There is not a generally accepted best practice to %model-based
clustering with
mixed discrete and continuous variables.    \citet{Hunt2003}, \citet{Biernacki2015}, and \citet{Murray2015} meld mixtures of {\em independent} multinomials for the categorical variables and mixtures of Gaussian
%conditional on the categorical values
for the continuous variables.  However, it may not be desirable for the dependency between the discrete variables to be entirely represented by mixture components when clustering is the primary objective.  As pointed out in \citet{Hennig2013}, mixture models can approximate any distribution arbitrarily well so care must be taken to ensure the mixtures fall in line with the goals of clustering.
%In this framework, for example, if there are two correlated Gaussian variables, then only one component is needed.  If there are two dichotomous variables that are correlated (tend to be equal to zero and one together) then multiple components (i.e., clusters) will be needed to represent the dependency; i.e., discrete variables will necessarily result in more clusters.
When using mixtures of Gaussian combined with independent multinomials, a data set with many correlated discrete variables will tend to result in more clusters than a comparable dataset with mostly continuous variables.  A discrete variable measure of some quantity instead of the continuous version could therefore result in very different clusters.
Thus, a Gaussian latent variable approach \citep{Muthen83,Everitt88,Dunson2000,Ranalli2017,McParland2016} would seem more appropriate for treating discrete variables when clustering is the goal.  An observed ordinal variable $x_j$, for example, is assumed to be the result of thresholding a latent Gaussian variable $z_j$.  For binary variables, this reduces to the multivariate probit model \citep{Lesaffre1991,Chib1998}.  There are also extensions of this approach to allow for unordered categorical variables.
% \citep{Zhang08,Bhattacharya12,Storlie17a}, however, such variables are not present in the ASD data.  Thus, for ease of presentation, attention is restricted in this paper to discrete ordinal variables (which includes binary variables).

In this paper, we propose a Bayesian nonparametric approach to perform simultaneous estimation of the number of clusters, cluster membership, and variable selection while explicitly accounting for discrete variables and partially observed data.  %To the best of our knowledge, this is the first model-based clustering approach to allow for variable selection in a manner such that the informative and non-informative variables  be correlated.
The discrete variables as well as continuous variables with boundaries are treated with a Gaussian latent variable approach.  The informative variable construct of \citet{Raftery2006} for normal mixtures is then adopted.  However, in order to effectively handle the missing values and account for uncertainty in the variable selection and number of clusters, the proposed model is cast in a fully Bayesian framework via the Dirichlet process.  This is then similar to the work of \citet{Kim2006}, however, they did not consider discrete variables or missing data.  Further, a key result of this paper is a solution to allow for dependence between informative and non-informative variables in the nonparametric Bayesian mixture model.  Thus, this work overcomes the assumption of (global) independence between informative and non-informative variables. Furthermore, by using the latent variable approach it also overcomes the (local) independence assumption among the informative/clustering variables often assumed when clustering data of mixed type \citep{Fop2017}.

The solution takes a particularly simple form and also provides an intuitive means with which to define the prior distribution in a manner that decreases prior sensitivity.  The component parameters are marginalized out to facilitate more efficient MCMC sampling via a modified version of the split-merge algorithm of \citet{Jain04}.  Finally, missing data is then handled in a principled manner by treating missing values as unknown parameters in the Bayesian framework \citep{Storlie2015calibration,Storlie17a}.  This approach implicitly assumes a missing at random (MAR) mechanism \citep{Rubin1976}, which implies that the likelihood of a missing value {\em can} depend on the value of the unobserved variable(s), marginally,
just not after conditioning on the observed variables.
%; see \citep{Schafer2002} for a good description of MAR.

%This approach implicitly assumes the commonly misunderstood missing at random (MAR) mechanism \citep{Rubin1976}.  MAR implies that the likelihood of a missing value {\em can} depend on the value of the unobserved variable(s), marginally, just not after conditioning on the observed variables; see \citep{Schafer2002} for a good description of MAR.

%If there were a reason to suspect the data were missing not at random, a binary indicator for the variable being observed or not could easily be included in this framework (since it allows for discrete variables) to model the missing mechanism.

The rest of the paper is laid out as follows. Section~\ref{sec:model} describes the proposed nonparametric Bayesian approach to clustering observations of mixed discrete and continuous variables with variable selection.  Section~\ref{sec:sims} evaluates the performance of this approach when compared to other methods on several simulation cases.  The approach is then applied to the problem for which it was designed in Section~\ref{sec:ASD} where a comprehensive analysis of the ASD problem is presented.  Section~\ref{sec:conclusions} concludes the paper.  This paper also has supplementary material which contains derivations, full exposition of the proposed MCMC algorithm, and MCMC trace plots.

%This paper also has online supplementary material containing Markov chain Monte Carlo (MCMC) details.

%%%%%%%%%%%%%%%%%%%%%%%%%%%%%%%%%%%%%%%%%%%%%%%%%%%%%%%%%%%%%%%%%%%%%%%%%%%%%%%%%%%%%

\vspace{-.2in}
\section{Methodology}
\label{sec:model}
\vspace{-.1in}

\vspace{-.12in}
\subsection{Dirichlet Process Mixture Models}
\vspace{-.1in}
\label{sec:model_descr}

As discussed above, the proposed model for clustering uses mixture distributions with a countably infinite number of components via the Dirichlet process prior \citep{Ferguson73,Escobar1995,Maceachern1998}.
Let $\by=(y_{1},\dots,y_{p})$ be a $p$-variate random vector and let $\by_i$, $i=1,\dots,n$, denote the $i\tth$ observation of $\by$.  It is assumed that $\by_i$ are independent random vectors coming from distribution $F(\theta_i)$.
%Let $\by_i=(y_{i,1},\dots,\by_{i,p})$, $i=1,\dots,n$, be independent random vectors coming from distribution $F(\theta_i)$.
The model parameters $\theta_i$ are assumed to come from a mixing distribution $G$ which has a Dirichlet process prior, i.e., the familiar model,
\vspace{-.13in}\beq
\by_i \mid \theta_i  \sim  F(\theta_i), \;\;\;\;\;
\theta_i  \sim  G, \;\;\;\;\;
G  \sim  \mbox{DP}(G_0, \alpha), 
\label{eq:DPM1}
\vspace{-.15in}\eeq
where DP represents a Dirichlet Process distribution, $G_0$ is the base distribution and $\alpha$ is a precision parameter, determining the concentration of the prior for $G$ about $G_0$ \citep{Escobar1995}.  The prior distribution for $\theta_i$ in terms of successive conditional distributions is obtained by integrating over $G$, i.e.,
\vspace{-.15in}\beq
\theta_i  \mid  \theta_1, \dots, \theta_{i-1} \sim \frac{1}{i-1+\alpha} \sum_{i'=1}^{i-1}\delta(\theta_{i'}) + \frac{\alpha}{i-1+\alpha} G_0,\label{eq:DPM2}
\vspace{-.05in}\eeq
where $\delta(\theta)$ is a point mass distribution at $\theta$.  The representation in (\ref{eq:DPM2}) makes it clear that (\ref{eq:DPM1}) can be viewed as a countably infinite mixture model.  Alternatively, let $\Omega=[\omega_1,\omega_2,\dots]$ denote the unique values of the $\theta_i$ and let $\phi_i$ be the index for the component to which observation $i$ belongs, i.e., so that $\omega_{\phi_i} = \theta_i$.  The following model \citep{Neal2000} is equivalent to (\ref{eq:DPM2})
\vspace{-.12in}\begin{eqnarray}
P(\phi_i = m \mid \phi_1, \dots, \phi_{i-1} ) =  \left\{
\begin{array}{ll}
  1 & \mbox{if $i=1$ and $m=1$.} \\
  \frac{n_{i,m}}{i-1+\alpha} & \mbox{if $\phi_{i'} = m$ for any $i'<i$.} \\
  \frac{\alpha}{i-1+\alpha} & \mbox{if $m=$max$(\phi_1, \dots, \phi_{i-1})+1$.} \\
    0 & \mbox{otherwise,}
\end{array}
\right.\label{eq:DPM3}
\\[-.3in] \nonumber
\vspace{-.0in}
\end{eqnarray}
with $\by_i \mid \phi_i, \Omega  \sim  F(\omega_{\phi_i})$, $\omega_m  \sim  G_0$ and $n_{i,m}$ is the number of $\phi_{i'}=m$ for $i' < i$.  Thus, a new observation $i$ is allocated to an existing cluster with probability proportional to the cluster size or it is assigned to a new cluster with probability proportional to $\alpha$.  This is often called the Chinese restaurant representation of the Dirichlet process.  
It is common to assume that $F$ is a normal distribution in which case $\omega_m \!=\! (\bmu_m, \bSigma_m)$ describes the mean and covariance of the $m\tth$ component.
This results in a normal mixture model with a countably infinite number of components.
%If $G_o$ is a conjugate prior for $F$, i.e., the normal-inverse Wishart, then the component parameters $\omega_m$ can be integrated out to facilitate Gibbs sampling \citep{Neal2000,Jain04}.  More detail on computation is provided in Section~\ref{sec:MCMC}.

\vspace{-.2in}
\subsection{Discrete Variables and Boundaries/Censoring}
\vspace{-.1in}
\label{sec:discrete_vars}

Normal mixture models are not effective for clustering when some of the variables are too discretized as demonstrated in Section~\ref{sec:sims}.  This is also a problem when the data have left or right boundaries that can be achieved (e.g., several people score the minimum or maximum on a test). However, a Gaussian latent variable approach can be used to circumvent these issues.  Suppose that variables $y_j$ for $j \in \cD$ are discrete, ordinal variables taking on possible values $\bd_j = \{d_{j,1}, \dots, d_{j,L_j}\}$ and that $y_j$ for $j \in \cC = \cD^c$ are continuous variables with lower and upper limits of $b_j$ and $c_j$, which could be infinite.  Assume for some latent, $p$-variate, continuous random vector $\bz$ that
\vspace{-.15in}\beq
y_j = \left\{
\begin{array}{ll}
  \sum_{l=1}^{L_j} d_{j,l} I_{\{a_{j,l-1} < z_j \leq a_{j,l}\}} & \mbox{for $j \in \cD$} \\
    z_jI_{\{b_j \leq z_j \leq c_j\}} + b_jI_{\{z_j < b_j\}} + c_jI_{\{z_j > c_j\}}& \mbox{for $j \in \cC$}
\end{array}
\right.
\label{eq:z2y}
\vspace{-.07in}\eeq
where $I_A$ is the indicator function equal to 1 if $A$ and 0 otherwise, $a_{j,0}= -\infty$, $a_{j,L_j}= \infty$, and $a_{j,l}=d_{j,l}$ for $l=1,\dots,L_j-1$.  That is, the discrete $y_j$ are the result of thresholding the latent variable $z_j$ on the respective cut-points.  The continuous $y_j$ variables are simply equal to the $z_j$ unless the $z_j$ cross the left or right boundary of what can be observed for $y_j$.  That is, if there are finite limits for $y_j$, then $y_j$ is assumed to be a left and/or right censored version of $z_j$, thus producing a positive mass at the boundary values of $y_j$.
%For example, in Figure~\ref{fig:scatter_intro} ABC\_Irritability has a left boundary of zero.  In order to represent this behavior it is imaginied that there is a continuous variable $z_j$ that could be negative to potentially differentiate these zeros, i.e., not all zeros are identical and a more fine grained test might differentiate the zeros.
%The $y_j$ could be truly censored (e.g., a test is stopped after a subject gets 100 out of the 150 questions correct produceing a mass at 100) or it could be an artifact of a hard boundary being achieved (e.g., many subject achieve the maximum score of 100 on a test).  This distinction is purely philosphical, as the probabilistic representation of $y_j$ will be treated the same.

A joint mixture model for mixed discrete and continuous variables is then, %can then be represented as
\vspace{-.18in}\beq
\bz_i \mid \phi_i, \Omega \sim  N(\bmu_{\phi_i}, \bSigma_{\phi_i}), \label{eq:mixed_dpm}
\vspace{-.18in}\eeq
with prior distributions for $\omega_m$ and $\bphi=[\phi_1,\dots,\phi_n]'$ as in (\ref{eq:DPM3}).

Binary $y_j$ such as gender can be accommodated by setting $\bd_j=\{0,1\}$.  However, if there is only one cut-point then the model must be restricted for identifiability \citep{Chib1998}; namely, if $y_j$ is binary, then we must set $\bSigma_m(j,j)=1$.  The restriction that $\bSigma_m(j,j)=1$ for binary $y_j$ complicates posterior inference, however, this problem has been relatively well studied in the multinomial probit setting and various proposed solutions exist \citep{Imai2005}.
%These approaches rely on the more general parameter expansion for data augmentation strategy \citep{Liu1999,vanDyk2001} which can be adopted here as well.
It is also straight-forward to use the latent Gaussian variable approach to allow for unordered categorical variables \citep{McCulloch2000,Imai2005,Zhang08,Bhattacharya12}, however, inclusion of categorical variables also complicates notation and there are no such variables in the ASD data.  For brevity, attention is restricted here to continuous and ordinal discrete variables.% (including binary variables).

\vspace{-.2in}
\subsection{Variable Selection}
\vspace{-.1in}
\label{sec:variable_selection}

Variable selection in clustering problems is more challenging than in regression problems due to the lack of targeted information with which to guide the selection.  Using model-based clustering allows a likelihood based approach to model selection, but exactly how the parameter space should be restricted when a variable is ``out of the model'' requires some care.  \citet{Raftery2006} defined a variable $y_{j}$ to be {\em non-informative} if conditional on the values of the other variables, it is independent of cluster membership.  This implies that a non-informative $y_j$ may still be quite dependent on cluster membership through its dependency with other variables.  They assumed a Gaussian mixture distribution for the informative variables, with a conditional Gaussian distribution for the non-informative variables and used maximum likelihood to obtain the change in BIC between candidate models.  Thus, they accomplished variable selection with a greedy search to minimize BIC.  They further considered restricted covariance parameterizations to reduce the parameter dimensionality (e.g., diagonal, common volume, common shape, common orientation, etc.).  We instead take a Bayesian approach to this problem via Stochastic Search Variable Selection (SSVS) \citep{George93,George97} as this allows for straight-forward treatment of uncertainty in the selected variables and that due to missing values.  \citet{Kim2006} used such an approach with a DPM for infinite normal mixtures, however, due to the difficulty imposed they did not use the same definition as \citet{Raftery2006} for a non-informative variable.  They defined a non-informative variable to be one that is (unconditionally) independent of cluster membership and {\em all} other variables.  This is not reasonable in many cases, particularly in the ASD problem, and can result in negative consequences as seen in Section~\ref{sec:sims}.  Below, we layout a more flexible model specification akin to that taken in \citet{Raftery2006} to allow for (global) dependence between informative and non-informative variables in a DPM.
%While acknowledging that this may not be the case in practice, they say it is difficult to accommodate such structure.
%For example, in the ASD data, 70 of the 87 tests have a correlation above 0.7 with at least one other test.  Over half of the tests have correlation above 0.7 with at least 5 other tests; it is unreasonable to force the dependence between among all of these variables to 

Let the informative variables be represented by the {\em model} $\bgamma$, a vector of binary values such that $\{ y_j : \gamma_j=1\}$ is the set of informative variables.  A priori it is assumed that $\Pr(\gamma_j=1)=\rho_j$.  Without loss of generality assume that $\by$ has elements ordered such that
$\by = [\by^{(1)}, \by^{(2)} ]$, 
with $\by^{(1)} = \{y_j: \gamma_j=1\}$ and $\by^{(2)} = \{y_j: \gamma_j = 0\}$, and similarly for $\bz^{(1)}$ and $\bz^{(2)}$.  The model in (\ref{eq:mixed_dpm}) becomes,
\vspace{-.15in}\beq
\bz_i \mid \bgamma, \phi_i, \Omega \sim  N(\bmu_{\phi_i}, \bSigma_{\phi_i}),
\label{eq:vs_dpm_2}
\vspace{-.18in}\eeq
with
\vspace{-.23in}\begin{eqnarray}
\bmu_m= \left(
\begin{array}{c}
  \bmu_{m1}\\
  \bmu_{m2}
\end{array}
\right), \;\;
\bSigma_m = \left(
\begin{array}{cc}
  \bSigma_{m11} & \bSigma_{m12}\\
  \bSigma_{m21} & \bSigma_{m22}
\end{array}
\right).\\[-.4in] \nonumber
\end{eqnarray}
From standard multivariate normal theory, $[\bz^{(2)} \mid \bz^{(1)},\phi=m] \sim N(\bmu_{2\mid 1}, \bSigma_{2 \mid 1})$ with $\bmu_{2\mid 1} = \bmu_{m2} + \bSigma_{m21}\bSigma_{m11}^{-1}(\bz^{(1)} - \bmu_{m1})$ and $\bSigma_{2 \mid 1} = \bSigma_{m22} - \bSigma_{m21}\bSigma_{m11}^{-1}\bSigma_{m12}$.
Now in order for the non-informative variables to follow the definition of \citet{Raftery2006}, the $\bmu_m$ and $\bSigma_m$ must be parameterized so that $\bmu_{2\mid 1}, \bSigma_{2 \mid 1}$ do not depend on $m$.  In order to accomplish this, it is helpful to make use of the canonical parameterization of the Gaussian \citep{Rue05},
\vspace{-.175in}\bdm
\bz \mid \bgamma, \Omega, \phi=m \sim \cN_C(\bb_m, \bQ_m),
\vspace{-.175in}\edm
with precision $\bQ_m=\bSigma_m^{-1}$ and $\bb_m = \bQ_m \bmu_m$.  Partition the canonical parameters as,
\vspace{-.14in}\begin{eqnarray}
\bb_m= \left(
\begin{array}{c}
  \bb_{m1}\\
  \bb_{2}
\end{array}
\right), \;\;
\bQ_m = \left(
\begin{array}{cc}
  \bQ_{m11} & \bQ_{12}\\
  \bQ_{21} & \bQ_{22}
\end{array}
\right).
\label{eq:can_MVN}
\\[-.28in] \nonumber
\end{eqnarray}

\noindent
{\em Result 1.}  {\em The parameterization in $($\ref{eq:can_MVN}$)$ results in $(\bmu_{2\mid 1}, \bSigma_{2 \mid 1})$ that does not depend on $m$.}\\[-.19in]

\noindent
{\em Proof.} The inverse of a partitioned matrix directly implies that $\bSigma_{2 \mid 1} = \bQ_{22}^{-1}$, which does not depend on $m$.  It also implies that $-\bQ_{22}^{-1}\bQ_{21} = \bSigma_{m21}\bSigma_{m11}^{-1}$, and substituting $\bSigma_m\bb_m$ for $\bmu_m$ in $\bmu_{2 \mid 1}$ gives
$\bmu_{2 \mid 1} =\bQ_{22}^{-1} \left(\bb_2 - \bQ_{21} \bz^{(1)} \right)$,
which also does not depend on $m$. $\:\Box$\\[-.17in]
%\begin{flushright}\vspace{-.28in} $\Box$ \end{flushright}

\noindent
The $\bQ_{21}$ does not depend on $m$ which implies the same dependency structure across the mixture components.  This is a necessary assumption in order for $\bz^{(2)}$ to be non-informative variables, i.e., so that cluster membership conditional on $\bz^{(1)}$ is independent of $\bz^{(2)}$.

Now the problem reduces to defining a prior distribution for $\Omega$, i.e., $\omega_m = \{\bb_m, \bQ_m\}$, $m=1,2,\dots$, conditional on the model $\bgamma$, that maintains the form of (\ref{eq:can_MVN}).  Let $\omega^{(1)}_m = \{\bb_{m1}, \bQ_{m11}\}$ and $\omega^{(2)}_m = \omega^{(2)} = \{\bb_2, \bQ_{21},\bQ_{22}\}$.  The prior distribution for $\Omega$ will be defined first unconditionally for $\omega^{(2)}$ and then for $\omega^{(1)}_m$, $m=1,2,\dots,$ conditional on $\omega^{(2)}$.  There are several considerations in defining these distributions: (i) the resulting $\bQ_m$ must be positive definite, (ii) it is desirable for the marginal distribution of $(\mu_m,\bSigma_m)$ to remain unchanged for any model $\bgamma$ to limit the influence of the prior for $\omega_m$ on variable selection, and (iii) it is desirable for them to be conjugate to facilitate MCMC sampling \citep{Neal2000,Jain04}.

Let $\bPsi$ be a $p \times p$ positive definite matrix, partitioned just as $\bQ_m$, and for a given $\bgamma$ assume the following distribution for $\omega^{(2)}$,
\vspace{-.25in}\begin{eqnarray}
  \begin{array}{c}
\bQ_{22} \sim  \cW(\!\bPsi_{22 \mid 1}^{-1}, \eta), \;\;\;\;
  \bb_2 \! \mid \! \bQ_{22}   \sim  \cN(\bzero, \mbox{$\frac{1}{\lambda}$}\bQ_{22}),\\[.1in]
  \bQ_{21} \!\!\mid \! \bQ_{22}  \sim  \cM\cN\!\left(-\bQ_{22\:}\!\bPsi_{21}\!\bPsi_{11}^{-1}  , \bQ_{22}  \:,  \bPsi_{11}^{-1} \right),
  \end{array}
  \label{eq:comp_prior_1}
\\[-.38in] \nonumber
\end{eqnarray}
where $\cW$ denotes the Wishart distribution, and $\cM\cN$ denotes the matrix normal distribution.

The distribution of $\omega^{(1)}_m$, conditional on $\omega^{(2)}$ is defined implicitly below.  A prior distribution is {\em not} placed on $(\bb_{m1}, \bQ_{m11})$, directly.  It is helpful to reparameterize from $(\bb_2, \bQ_{22}, \bQ_{21}, \bb_{m1}, \bQ_{m11})$ to $(\bb_2, \bQ_{22}, \bQ_{21}, \bmu_{m1}, \bSigma_{m11})$.  By doing this, independent priors can be placed on $(\bb_2, \bQ_{22}, \bQ_{21})$ and $(\bmu_{m1}, \bSigma_{m11})$ and still maintain all of the desired properties as will be seen in Results~2~and~3.  

The prior distribution of $(\bmu_{m1}, \bSigma_{m11})$ is
\vspace{-.2in}\beq
  \bSigma_{m11}  \stackrel{iid}{\sim}  \cW^{-1}\left(\bPsi_{11}, \eta - p_2\right), \;\;\;\;
  \bmu_{m1} \mid \bSigma_{m11}  \stackrel{ind}{\sim}  \cN\left(\bzero, \mbox{$\frac{1}{\lambda}$}\bSigma_{m11}\right),
\label{eq:comp_prior_2}
\vspace{-.18in}\eeq
where $\cW^{-1}$ denotes the inverse-Wishart distribution and $(\bmu_{m1}, \bSigma_{m11})$ are independent of $\omega^{(2)}$.
The resulting distribution of $(\bb_{m1}, \bQ_{m11})$ conditional on $(\bb_2, \bQ_{22}, \bQ_{21})$ is not a common or named distribution, but it is well defined via the relations,
%This is {\em not} to say that $\omega^{(1)}_m=(\bb_{m1}, \bQ_{m11})$ is independent of $\omega^{(2)}$, rather the distribution imposed on $(\bb_{m1}, \bQ_{m11})$ via (\ref{eq:comp_prior_1}) and (\ref{eq:comp_prior_2}) is quite dependent on $\omega^{(2)}$ via the relations,
$\bb_{m1}  =  \bSigma_{m11}^{-1}\bmu_{m1} + \bQ_{12} \bQ_{22}^{-1} \bb_2$, and
$\bQ_{m11} =  \bSigma_{m11}^{-1} + \bQ_{12} \bQ_{22}^{-1} \bQ_{21}$.\\[-.15in]

\noindent
{\em Result 2.}
{\em The prior distribution defined in $($\ref{eq:comp_prior_1}$)$ and $($\ref{eq:comp_prior_2}$)$ results in a marginal distribution for $(\bmu_m, \bSigma_m)$ of $\cN\cI\cW ( \bzero, \lambda, \bPsi, \eta)$, $\!$i.e.,$\!$ the same normal-inverse-Wishart regardless of $\bgamma$.$\!\!\!\!$}\\[-.15in]

\noindent
{\em Proof.} It follows from Theorem 3 of \citet{Bodnar2008} that $\bSigma_m \sim \cI\cW (\eta, \bPsi)$.  It remains to show $\bmu_m \mid \bSigma_m \sim \cN(\bzero, (1/\lambda)\bSigma_{m})$. However, according to  (\ref{eq:comp_prior_1}) and (\ref{eq:comp_prior_2}) and the independence assumption,
\vspace{-.15in}\begin{eqnarray}
\left.\left(
\begin{array}{c}
  \bmu_{m1}\\
  \bb_{2}
\end{array}
\right) \right| \bSigma_m \sim \cN \left(
\left(
\begin{array}{c}
  \bzero\\
  \bzero
\end{array}
\right) ,
\frac{1}{\lambda}\left(
\begin{array}{cc}
  \bSigma_{m11} & \bzero\\
  \bzero & \bQ_{22}
\end{array}
\right)
\right).\nonumber \\[-.375in] \nonumber
\end{eqnarray}
Also, $\bb_m = \bQ_m \bmu_m$ implies,
%$\bmu_{m2} = \bQ_{22}^{-1}\left(\bb_2-\bQ_{21} \bmu_{m1} \right)$, which means
\vspace{-.15in}\bdm
\left(
\begin{array}{c}
  \bmu_{m1}\\
  \bmu_{m2}
\end{array}
\right) =
\left(
\begin{array}{cc}
  \bI & \bzero\\
  -\bQ_{22}^{-1} \bQ_{21} & \bQ_{22}^{-1}
\end{array}
\right)
\left(
\begin{array}{c}
  \bmu_{m1}\\
  \bb_{2}
\end{array}
\right).
\vspace{-.12in}\edm
Using the relation $\bA\bx \sim \cN(\bA\bmu, \bA\bSigma\bA')$ for $\bx \sim \cN(\bmu, \bSigma)$ gives the desired result. $\:\Box$\\[-.15in]
%\begin{flushright}\vspace{-.33in} $\Box$ \end{flushright}

As mentioned above, the normal-inverse-Wishart distribution is conjugate for $\omega_m$ in the unrestricted (no variable selection) setting.  It turns out that the distribution defined in (\ref{eq:comp_prior_1}) and (\ref{eq:comp_prior_2}) is conjugate for the parameterization in (\ref{eq:can_MVN}) as well, so that the component parameters can be integrated out of the likelihood.  Let the (latent) observations be denoted as $\bZ=[\bz_1',\dots,\bz_n']'$, and the data likelihood as $f(\bZ \mid \bgamma, \bphi, \Omega)$.\\[-.15in]

\noindent
{\em Result 3.}
{\em The marginal likelihood of $\bZ$ is given by
\vspace{-.15in}  \bdm
f(\bZ \!\mid \!\bgamma, \bphi) = \int f(\bZ \mid \bgamma, \bphi, \Omega) f(\Omega \mid \bgamma) d\Omega \;\;\;\;\;\;\;\;\;\;\;\;\;\;\;\;\;\;\;\;\;\;\;\;\;\;\;\;\;\;\;\;\;\;\;\;\;\;\;\;\;\;\;\;\;\;\;\;\;\;\;\;\;\;\;\;\;\;\;\;\;\;\;\;\;\;\;\;\;\;\;\;\;\;\;\;\;\;\;\;\;\;\;\;\;\;\;\;\;\;\;
\vspace{-.4in} \edm
\begin{eqnarray}
\;\;\;\;\;\;\;\;\;\;  = \pi^{-\frac{np}{2}}\!\prod_{m=1}^M \!\left[\!
    \left(\frac{\lambda}{n_m\!+\!\lambda} \right)^{\!\!\frac{p_1}{2}}\!
    \frac{|\bPsi_{\!11}|^{\frac{\eta-p_2}{2}}\Gamma_{\!p_1}\!\!\left(\frac{n_m+\eta-p_2}{2}\right)}
         {|\bV_{\!m11}|^{\frac{n_m+\eta-p_2}{2}}\Gamma_{\!p_1}\!\!\left(\frac{\eta-p_2}{2}\right)}
         \!\right]\!
  \left[\!
    \left(\frac{\lambda}{n\!+\!\lambda} \right)^{\!\!\frac{p_2}{2}}\!
    \frac{|\bPsi_{\!11}|^{\frac{p_2}{2}}|\bPsi_{\!2\mid 1}|^{\frac{\eta}{2}}\Gamma_{\!p_2}\!\!\left(\frac{n+\eta}{2}\right)}
         {|\bV_{\!11}|^{\frac{p_2}{2}}|\bV_{\!2\mid 1}|^{\frac{n+\eta}{2}}\Gamma_{\!p_2}\!\!\left(\frac{\eta}{2}\right)}
         \!\right], \nonumber \\[-.35in] \nonumber
\end{eqnarray}
where (i) $M=\max(\bphi)$, i.e., the number of observed components, (ii) $p_1 = \sum \gamma_j$ is the number of informative variables, (iii) $p_2 = p - p_1$, (iv) $n_m$ is the number of $\phi_i = m$, (v) $\Gamma_p(\cdot)$ is the multivariate gamma function, and (vi) $\bV_{\!m11}$, $\bV_{\!11}$, $\bV_{\!2 \mid 1}$ are defined as,
\vspace{-.15in}\begin{eqnarray}
\bV_{\!m11} & = & \sum_{\phi_i =m} (\bz^{(1)}_{i} \!- \!\bar{\bz}_{m1})(\bz^{(1)}_{i} \!-\! \bar{\bz}_{m1})' \!+ \!\frac{n_m\lambda}{n_m\!+\!\lambda}\bar{\bz}_{m1}\bar{\bz}_{m1}' \!+\! \bPsi_{11}, \nonumber \\
\bV_{\!11} & = & \!\sum_{i=1}^n (\bz^{(1)}_{i} \!\!- \!\bar{\bz}_{1})(\bz^{(1)}_{i} \!\!- \!\bar{\bz}_{1})' \!+ \!\frac{n\lambda}{n\!+\!\lambda}\bar{\bz}_{1}\bar{\bz}_{1}' \!+ \!\bPsi_{\!11}, \nonumber \\
    \bV_{\!2 \mid 1} & = & \bV_{\!22} - \bV_{\!21}\bV_{\!11}^{-1}\bV_{\!21}',
   \nonumber\\[-.44in] \nonumber
\end{eqnarray}
with $\bar{\bz}_{m1} = \frac{1}{n_m}\sum_{\phi_i=m} \bz^{(1)}_{i}$, $\;\bar{\bz}_{1} = \frac{1}{n}\sum_{i=1}^n \bz^{(1)}_{i}$,  $\;\bar{\bz}_{2} = \frac{1}{n}\sum_{i=1}^n \bz^{(2)}_{i}$,
\vspace{-.1in}\begin{eqnarray}
\!\bV_{\!\!22}\!  &= & \!\!\sum_{i=1}^n (\bz^{(2)}_{\!i} \!\!\!-\! \bar{\bz}_{2})(\bz^{(2)}_{\!i} \!\!\!-\! \bar{\bz}_{2})' \!\!+\! \frac{n\lambda}{n\!+\!\lambda}\bar{\bz}_{2}\bar{\bz}_{2}' \!+\! \bPsi_{\!22},\;\mbox{and}\:\\[-.07in]
  \bV_{\!\!21} \! &=& \!\! \sum_{i=1}^n (\bz^{(2)}_{\!i} \!\!\!-\! \bar{\bz}_{2})(\bz^{(1)}_{\!i} \!\!\!-\! \bar{\bz}_{1})'\!\! +\! \frac{n\lambda}{n\!+\!\lambda}\bar{\bz}_{2}\bar{\bz}_{1}' \!+\! \bPsi_{\!21}.
  \nonumber\\[-.375in] \nonumber
\end{eqnarray}
}
\noindent
The derivation of Result~3 is provided in Web Appendix~\ref{sec:like_deriv}.

\vspace{-.2in}
\subsection{Hyper-Prior Distributions}
\vspace{-.1in}
\label{sec:priors}

%\citet{Kim2006} found there to be a lot of prior sensitivity due to their choice of prior for $\bSigma_{m11} \sim \cI\cW(a_1, \bPsi_{11})$ and $\bSigma_{22} = \sigma^2\bI$ with $\sigma^2 \sim \cI\cG(a_2,b)$.  This is in part due to the separate specification of $\bPsi_{11}$ and $b$.  The specification above treats $\bSigma_{m} \sim \cI\cW(\eta, \bPsi)$ all as one prior, so that the choice will not be sensitive to the interplay between $b$ and $\bPsi$.
\citet{Kim2006} found there to be a lot of prior sensitivity due to the choice of prior for the component parameters.  This is in part due to the separate prior specification for the parameters corresponding to informative and non-informative variables, respectively.  The specification above treats all component parameters collectively, in a single prior, so that the choice will not be sensitive to the interplay between the priors chosen for informative and non-informative variables.
A further stabilization can be obtained by rationale similar to that used in \citet{Raftery2006} for restricted forms of the covariance (such as equal shape, orientation, etc.).  We do not enforce such restrictions exactly, but one might expect the components to have similar covariances or similar means for some of the components.  Thus it makes sense to put hierarchical priors on $\lambda$, $\bPsi$, and $\eta$, to encourage such similarity if warranted by the data.  A Gamma prior is also placed on the concentration parameter $\alpha$, i.e.,
\vspace{-.12in}\beq
\begin{array}{ll}
\;\;\;\;\;\;\;\lambda  \sim  \mbox{Gamma}(A_\lambda, B_\lambda), \;\;\;\;\;\;&
\bPsi  \sim  \cW(\bP,N), \\
\!\!\!\!\!\!\!\!\!\eta \!-\!(p \!+ \!1) \sim  \mbox{Gamma}(A_\eta, B_\eta), &
\:\alpha  \sim \mbox{Gamma}(A_\alpha, B_\alpha).
\end{array}\label{eq:hyper_priors}
\vspace{-.1in}\eeq
In the analyses below, relatively vague priors were used with $A_\lambda\!=\!B_\lambda\!=\!A_\eta\!=\!B_\eta\!=\!2$.  The prior for $\alpha$ was set to $A_\alpha\!=\!2$, $B_\alpha\!=\!2$, to encourage anywhere from 1 to 15 clusters from 100 observations.  The results still have some sensitivity to the choice of $\bP$.  In addition, there are some drawbacks to Wishart priors which can be exaggerated when applied to variables of differing scale \citep{Gelman2006,Huang2013}.  In order to alleviate these issues, we recommend first standardizing the columns of the data to mean zero and unit variance, then using $N\!=\!p+2$, $\bP\!=\!(1/N)\bI$.  Finally, the prior probability for variable inclusion was set to $\rho_j\!=\!0.5$ for all $j$.
The data model in (\ref{eq:z2y}) and (\ref{eq:vs_dpm_2}), the component prior distribution in (\ref{eq:comp_prior_1}) and (\ref{eq:comp_prior_2}), along with the hyper-priors in (\ref{eq:hyper_priors}), completes the model specification.

\vspace{-.2in}
\subsection{MCMC Algorithm}
\vspace{-.1in}
\label{sec:MCMC_summary}

Complete MCMC details are provided in the Web Appendix~\ref{sec:MCMC}.  However, an overview is provided here to illustrate the main idea.  The complete list of parameters to be sampled in the MCMC are
$\Theta = \{
\bgamma, \bphi, \lambda, \eta, \bPsi, \alpha, \tilde{\bZ} 
\}$,
where $\tilde{\bZ}$ contains any latent element of $\bZ$ (i.e., either corresponding to missing data, discrete variable, or boundary/censored observation).  The only update that depends on the raw observed data $\bY=[\by_1',\dots,\by_n']'$ is the update of $\tilde{\bZ}$.  All other parameters, when conditioned on $\bY$ and $\bZ$, only depend on $\bZ$.  The $\tilde{\bZ}$ are block updated, each with a MH step, but with a proposal that looks almost conjugate, and is therefore accepted with high probability; the block size can be adjusted to trade-off between acceptance and speed (e.g., acceptance $\sim 40$\%).  A similar strategy is taken with the $\bPsi$ update, i.e., a nearly conjugate update is proposed and accepted/rejected via an MH step.  Because the component parameters are integrated out, the $\phi_i$ can be updated with simple Gibbs sampling \citep{Neal2000}, however, this approach has known mixing issues \citep{Jain04,Ishwaran2001}.  Thus, a modified split-merge algorithm \citep{Jain04} similar to that used in \citep{Kim2006} was developed to sample from the posterior distribution of $\bphi$. The remaining parameters are updated in a hybrid Gibbs, Metropolis Hastings (MH) fashion.  The $\bgamma$ vector is updated with MH by proposing an add, delete, or swap move \citep{George97}.  
The $\lambda, \eta, \alpha$ parameters have standard MH random walk updates on log-scale.  The MCMC routine then consists of applying each of the above updates in turn to complete a single MCMC iteration, with the exception that the $\bgamma$ update be applied $L_g$ times each iteration.

Two modifications were also made to the above strategy to improve mixing.  The algorithm above would at times have trouble breaking away from local modes when proposing $\bphi$ and $\bgamma$ updates separately.  Thus, an additional joint update is proposed for $\bphi$ and $\bgamma$ each iteration which substantially improved the chance of a move each iteration.  Also, as described in more detail in Web Appendix~\ref{sec:MCMC}, the traditional split merge algorithm proposes an update by first selecting two points, $i$ and $i'$, at random. If they are from the same cluster (according to the current $\bphi$) it then assigns them to separate clusters and assigns the remaining points from that cluster to each of the two new clusters at random.  It then conducts several ($L$) restricted (to one of the two clusters) Gibbs sampling updates to the remaining $\phi_h$ from the original cluster.  The resulting $\bphi^*$ then becomes the proposal for a split move.  We found that the following adjustment resulted in better acceptance of split/merge moves.  Instead of assigning the remaining points to the two clusters at random, simply assign them to the closest of the two observations $i$ or $i'$.  Then conduct $L$ restricted Gibbs sample updates to produce the proposal.  We found little performance gain beyond $L=3$.  
Lastly, it would be possible to instead use a finite mixture approximation via the kernel stick breaking representation of a DPM \citep{Sethuraman94,Ishwaran2001}.  However, this approach would be complicated by the dependency between $\bgamma$ and the structure and dimensionality of the component parameters.  This issue is entirely avoided with the proposed approach as the component parameters are integrated out.  The code to perform the MCMC for this model has been made available in a GitHub repository at \verb1https://github.com/cbstorlie/DPM-vs.git1.

\vspace{-.2in}
\subsection{Inference for $\bphi$ and $\bgamma$}
\vspace{-.1in}
\label{sec:inference}
The estimated cluster membership $\hphi$ for all of the methods was taken to be the respective mode of the estimated cluster membership probabilities.  For the DPM methods, the cluster membership probability matrix $P$ (which is an $n \times \infty$ matrix in principle) is not sampled in the MCMC, and is not identified due to many symmetric modes (thus their can be label switching in the posterior samples).  However, the information theoretic approach of \citet{Stephens2000} (applied to the DPM in \citet{Fu2013}) can be used to address this issue and relabel the posterior samples of $\bphi$ to provide an estimate of $P$.  The resulting estimate $\hP$ has $i\tth$ row, $m\tth$ column that can be thought of as the proportion of the relabeled posterior samples of $\phi_i$ that have the value $m$.  While technically $P$ is an $n \times \infty$ matrix, all columns after $M^*$ have zero entries in $\hP$, where $M^*$ is the maximum number of clusters observed in the posterior.
%They define $\hbphi$ to be the cluster orientation that minimizes the sum of the absolute deviations between the matrix of pairwise posterior probabilities $\Pr(\phi_i=\phi_{i'} | \bY)$ and the 0/1 association matrix corresponding to $\bphi$.  This search is NP-hard and $\hbphi$ therefore is taken to be the minimum over the $\bphi$ sampled during the MCMC.  This approximation can also be refined by switching the labels of the $\hphi_i$ in a greedy fashion until no improvements can be made.
For the results below, the point estimate of $\hbgamma$ is determined by $\hgamma_j = 1$ if $\Pr(\gamma_j = 1) > \rho_j = 0.5$, and $\hgamma_j = 0$ otherwise.

\vspace{-.2in}
\section{Simulation Results}
\vspace{-.1in}
\label{sec:sims}

In this section the performance of the proposed approach for clustering is evaluated on two simulation cases similar in nature to the ASD clustering problem.  Each of the cases is examined (i) without missing data or discrete variables/censoring, (ii) with missing data, but no discrete variables/censoring, (iii) with missing data and several discrete and/or censored variables.

The approaches to be compared are listed below.
\begin{itemize}[leftmargin=1.0in]
  \singlespacing
\item[{\bf DPM-vs} $-$] the proposed method.
\item[{\bf DPM-cont} $-$] the proposed method without accounting for discrete variables/censoring (i.e., assuming all continuous variables).
\item[{\bf DPM} $-$] the proposed method with variable selection turned off (i.e., a prior probability $\rho_j=1$).
\item[{\bf DPM-ind} $-$] the approach of \citet{Kim2006} when all variables are continuous (i.e., assuming non-informative variables are independent of the rest), but modified to treat discrete variables/censoring and missing data when applicable just as the proposed approach.
\item[{\bf Mclust-vs} $-$] the approach of \citet{Raftery2006} implemented with the \verb1clustvarsel1 package in R.  When there are missing data, Random Forest Imputation \citep{Stekhoven2012} implemented with the \verb1missForest1 package in R is used prior to application of \verb1clustvarsel1. However, the Mclust-vs approach does not treat discrete variables differently and thus treats all variables as continuous and uncensored.
\item[{\bf VarSelLCM} $-$] the approach of \citet{Marbac2017} implemented in the R package \verb1VarSelLCM1.  It allows for mixed data types and missing data, however, it assumes both {\em local} independence of variables within cluster and {\em global} independence between informative and non-informative variables.\\[-.29in]
\end{itemize}

Each simulation case is described below.  Figure~\ref{fig:sim_2c} provides a graphical depiction of the problem for the first eight variables from the first of the 100 realizations of Case 2(c). Case 1 simulations resulted in very similar data patterns as well.

\begin{itemize}[leftmargin=.9in]
  \singlespacing
\item[{\bf Case 1(a)} $-$] $n=150$, $p=10$. The true model has $M=3$ components with mixing proportions 0.5, 0.25, 0.25, respectively, and $\by \!\mid\! \phi$ is a multivariate normal with no censoring nor missing data.  Only two variables $\by^{(1)}=[y_1, y_2]'$ are informative, with means of $(2,0)$, $(0,2)$, $(-1.5,-1.5)$, unit variances, and correlations of 0.5, 0.5, -0.5 in each component, respectively. The non-informative variables $\by^{(2)}=[y_3,\dots,y_{10}]'$ are generated as {\em iid} $\cN(0,1)$.
\item[{\bf Case 1(b)} $-$] Same as the setup in 1(a) only the non-informative variables $\by^{(2)}$ are correlated with $\by^{(1)}$ through the relation $\by^{(2)} = \bB \by^{(1)} + \beps$, where $\bB$ is a $8 \times 2$ matrix whose elements are distributed as {\em iid} $\cN(0,0.3)$, and $\beps \sim \cN(0,Q^{-1}_{22})$, with $\bQ_{22} \sim \cW(\bI,10)$.    
%$y_3=y_1 - y_2 + \eps_3$, $y_4=0.5 y_1 + 1.5 y_2 + \eps_4$, with $\eps_3, \eps_4 \stackrel{iid}{\sim} \cN(0,1)$.
\item[{\bf Case 1(c)} $-$] Same as in 1(b), but variables $y_1, y_6$ are discretized to the closest integer, variables $y_2,y_9$ are left censored at -1.4 ($\sim$8\% of the observations), and $y_3, y_{10}$ are right censored at 1.4.  
 
\item[{\bf Case 1(d)} $-$] Same as 1(c), but the even numbered $y_j$ have $\sim 30$\% of the observations MAR.

\item[{\bf Case 2(a)} $-$] $n=300$, $p=30$. The true model has $M\!=\!3$ components with mixing proportions 0.5, 0.25, 0.25, respectively, $\by \!\mid\! \phi$ is a multivariate normal with no censoring nor missing data.  Only four variables $(y_1, y_2, y_3, y_4)$ are informative, with means of $(0.6,0,1.2,0)$, $(0,1.5,-0.6,1.9)$, $(-2,-2,0,0.6)$ and all variables with unit variance for each of the three components, respectively.  All correlations among informative variables are equal to 0.5 in components 1 and 2, while component 3 has correlation matrix, $\bSigma_{311}(i,j) \!=\! 0.5(-1)^{\|i+j\|}I_{\{i\neq j\}} \!+\! I_{\{i=j\}}$.
The non-informative variables $\by^{(2)}=[y_5,\dots,y_{30}]'$ are generated as {\em iid} $\cN(0,1)$.

\item[{\bf Case 2(b)} $-$] Same as the setup in Case 2(a) only the non-informative variables $\by^{(2)}$ are correlated with $\by^{(1)}$ through the relation $\by^{(2)} = \bB \by^{(1)} + \beps$, where $\bB$ is a $26 \times 4$ matrix whose elements are distributed as {\em iid} $\cN(0,0.3)$, and $\beps \sim \cN(0,Q^{-1}_{22})$, with $\bQ_{22} \sim \cW(\bI,30)$.    
%$y_3=y_1 - y_2 + \eps_3$, $y_4=0.5 y_1 + 1.5 y_2 + \eps_4$, with $\eps_3, \eps_4 \stackrel{iid}{\sim} \cN(0,1)$.

\item[{\bf Case 2(c)} $-$] Same setup as in Case 2(b), but now variables $y_1,y_6,y_{11}$ are discretized to the closest integer, variables $y_2, y_9, y_{10},y_{11}$ are left censored at -1.4 ($\sim$8\% of the observations), and variables $y_3, y_{12}, y_{13},y_{14}$ are right censored at 1.4.  

\item[{\bf Case 2(d)} $-$] Same as Case 2(c), but the even numbered $y_j$ have $\sim 30$\% MAR.\\[-.32in]
\end{itemize}

\begin{figure}[t!]
\vspace{-.1in}
\centering
\caption{About Here.}
\vspace{-.2in}
\end{figure}

For each of the eight simulation cases, 100 data sets were randomly generated and each of the five methods above was fit to each data set.  The methods are compared on the basis of the following statistics.

\begin{itemize}[leftmargin=.75in]
  \singlespacing
\item[{\em Acc $-$}] Accuracy calculated as the proportion of observations in the estimated clusters that are in the same group as they are in the true clusters, when put in the arrangement (relabeling) that best matches the true clusters.
\item[{\em FI} $-$] Fowlkes-Mallows index of $\hbphi$ relative to the true clusters.
%The Rand index ({\em RI}$\:$) of $\hbphi$ relative to the true clusters.
\item[{\em ARI} $-$] Adjusted Rand index.
\item[{\em M} $-$] The number of estimated clusters.  The estimated number of clusters for the Mclust-vs and VarSelLCM methods was chosen as the best of the possible $M=1,\dots8$ cluster models via BIC.  The number of clusters for the Bayesian methods is chosen as the posterior mode and is inherently allowed to be as large as $n$.
\item[$p_1$ $-$] The model size, $p_1 = \sum_j \hgamma_j$.
\item[{\em PVC} $-$] The proportion of variables correctly included/excluded from the model, \\{\em PVC}$\:$$=(1/p)\sum_j I_{\{\hgamma_j=\gamma_j\}}$.
\item[{\em CompT} $-$] The computation time in minutes (using 20,000 MCMC iterations for the Bayesian methods).\\[-.32in]
\end{itemize}
These measures are summarized in the columns of the tables below by their mean (and standard deviation) over the 100 data sets.  It appeared that 20,000 iterations (10,000 burn in and 10,000 posterior samples) was sufficient for the Bayesian methods to summarize the posterior in the simulation cases via several trial runs, however not every simulation result was inspected for convergence.

The simulation results from Cases 1(a)-(d) are summarized in Table~\ref{tab:case1}.  The summary score for the {\em best} method for each summary is in bold along with that for any other method that was not statistically different from the {\em best} method on the basis of the 100 trials (via an uncorrected paired $t$-test with $\alpha=0.05$).
%DPM-cont is identical to DPM-vs for cases 1(a), 1(b), 2(a), 2(b) and is therefore not listed separately.
As would be expected, DPM-ind is one of the best methods on Case 1(a), however, it is not significantly better than DPM-vs or Mclust-vs on any of the metrics.  VarSelLCM performs slightly worse than the top three methods in this case since the local independence assumption is being violated.  All of the other methods solidly outperform DPM though, which had a difficult time finding more than a single cluster since it had to include all 10 variables.  In Case 1(b) the assumptions of DPM-ind are now being violated and it is unable to perform adequate variable selection.  It must include far too many non-informative variables due to the correlation within $\by^{(2)}$ and between $\by^{(1)}$ and $\by^{(2)}$.  The clustering performance suffers as a result and like DPM, it has difficulty finding more than a single cluster.  VarSelLCM also struggles in this case for the same reason; the global independence assumption is being violated.  Mclust-vs still performs well in this case, but DPM-vs (and DPM-cont) is significantly better on two of the metrics.  In case 1(c) DPM-vs is now explicitly accounting for the discrete and left/right censored variables, while DPM-cont does not.   When the discrete variables are incorrectly assumed to be continuous it tends to create separate clusters at some of the unique values of the discrete variables.  This is because a very high likelihood can be obtained by normal distributions that are almost singular along the direction of the discrete variables.  %In fact, if not for restrictions on the covariance (either via the prior or covariance constraints in Mclust) the maximum likelihood estimate would be singular.  The latent variable approach is relatively immune to this concern.
Thus, DPM-vs substantially outperforms DPM-cont and Mclust-vs, demonstrating the importance of explicitly treating the discrete nature of the data when clustering.  Finally, Case 1(d) shows that the loss of 30\% of the data for half of the variables (including an informative variable) does not degrade the performance of DPM-vs by much.  In this case Mclust-vs uses Random Forest Imputation to first impute the data, then cluster.  The imputation procedure does not explicitly take into account of the cluster structure of the data, rather it could mask this structure.  This is another reason that the performance is worse than the proposed approach which incorporates the missingness directly into the clustering model.  Mclust-vs and VarSelLCM both have {\em much} faster run-times than the Bayesian methods, however, when there are local or global correlations and discrete variables and/or missing data, they did not perform nearly as well as DPM-vs.

\begin{table}[t!]
  \vspace{-.1in}
\centering
  \caption{About Here.}
\vspace{-.2in}
\end{table}

\begin{table}[t!]
  \vspace{-.0in}
\centering
  \caption{About Here.}
\vspace{-.2in}
\end{table}

The simulation results from Cases 2(a)-(d) are summarized in Table~\ref{tab:case2}.
A similar story line carries over into Case 2 where there are now $p=30$ (four informative) variables and $n=300$ observations.  Namely, DPM-vs is not significantly different from DPM-ind or Mclust-vs on any of the summary measures for Case 2(a), with the exception of computation time.  DPM-vs is the best method on all summary statistics (except {\em CompT}) by a sizeable margin on the remaining cases.  While Mclust is much faster than DPM-vs, the cases of the most interest in this paper are those with discrete variables, censoring and/or missing data (i.e., Cases 1(c), 1(d), 2(c), and 2(d)).  In these cases, the additional computation time of DPM-vs might seem inconsequential relative to the enormous gain in accuracy.  It is interesting that
%the computation time is twice as much for the DPM method, even though it doesn't have a variable selection piece in the MCMC.  This is because the split/merge proposals take much longer when there is one large cluster as the restricted Gibbs updates need to be done on the entire sample.  Another interesting observation is that
DPM-vs suffers far less from the missing values when moving from Case 2(c) to 2(d) than it did from Case 1(c) to 1(d).  This is likely due to the fact that there are a larger number of observations to offset the additional complexity of a larger $p$.  However, it is also likely that the additional (correlated) variables may help to reduce the posterior variance of the {\em imputed} values.

\vspace{-.2in}
\section{Application to Autism and Related Disorders}
\vspace{-.1in}
\label{sec:ASD}

%%%% From Scott
%
% Change section to ``Application to Autism and related disorders''
%
% Insert some text about who the PASD group is.
%
% These are the type of patients that a clinician will have to sift through to determine what type of disorder they have and how to best treat them.
%
% 

The cohort for this study consists of subjects falling in the criteria for ``potential ASD'' (PASD) on the basis of various combinations of developmental and psychiatric diagnoses obtained from comprehensive medical and educational records as described in \citet{Katusic16}.  The population of individuals with PASD is important because this group represents the pool of patients with developmental/behavioral symptoms from which clinicians have to determine who has ASD and/or other disorders.  Subjects 18 years of age or older were invited to participate in a face-to-face session to complete psychometrist-administered assessments of autism symptoms, cognition/intelligence, memory/learning, speech and language, adaptive functions, and maladaptive behavior.  In addition, guardians were asked to complete several self-reported, validated questionnaires.  The goal is to describe how the patients' test scores separate them in terms of clinical presentation and which test scores are the most useful for this purpose.  This falls in line with the new Research Domain Criteria (RDoC) philosophy that has gained traction in the field of mental health research.  RDoC is a new research framework for studying mental disorders.  It aims to integrate many levels of information (cognitive/self-report tests, imaging, genetics) to understand how all of these might be related to similar clinical presentations. 

A total of 87 test scores measuring cognitive and/or behavioral characteristics were considered from a broad list of commonly used tests for assessing such disorders.
%This list includes scores from the ADOS, Brief, OWL, SRS, WRAML, WJ, ABAS, WRAT4, WASI-II, Beery, CARS, ABC-C, SCQ, and Achenbach forms (
A complete list of the individual tests considered is provided in Web Appendix~\ref{sec:definitions}.
%However, many of these test scores measure similar things, and in fact some test scores are simply an aggregate of others.
Using expert judgment to include several commonly used aggregates in place of individual subtest scores, this list was reduced to 55 test score variables to be considered in the clustering procedure.  Five of the 55 variables have fewer than 15 possible values and are treated here as discrete, ordinal variables.  A majority (46) of the 55 variables also have a lower bound, which is attained by a significant portion of the individuals, and are treated as left censored.  Five of the variables have an upper bound that is attained by many of the individuals and are thus treated as right censored.  There are 486 observations (individuals) in the dataset, however, only 67 individuals have complete data, i.e., a complete case analysis would throw out 86\% of the observations.
%; it is therefore imperative to explicitly account for the missing data in this problem.

DPM-vs was applied to these data; four chains with random starting points were run in parallel for 85,000 iterations each, which took $\sim40$ hours on a 2.2GHz processor.  The first 10,000 iterations were discarded as burn-in.  More iterations were used here than in the simulation cases due to the fact that this analysis is slightly more complicated (e.g., more variables and observations) and it only needed to be performed once.  MCMC trace plots are provided in Web Appendix~\ref{sec:MCMCtrace}.  All chains converged to the same distribution (aside from relabeling) and were thus combined.% to generate the following summaries. 
Four of the tests (Beery standard, CompTsc\_ol, WJ\_Pass\_Comprehen, and Adaptive Composite) had a high ($>0.88$) posterior probability of being informative (Table~\ref{tab:inform_vars}).  There is also evidence that Ach\_abc\_Attention and Ach\_abc\_AnxDep are informative.  The posterior samples were split on which of these two should be included in the model (they were only informative together for 0.1\% of the MCMC samples).  The next highest posterior inclusion probability for any of the remaining variables was 0.17 and the sum of the inclusion probabilities for all remaining variables was only 0.28.  Thus, there is strong evidence to suggest that only five of the 55 variables are sufficient to inform the cluster membership.

\begin{table}[t!]
  \vspace{-.1in}
\centering
  \caption{About Here.}
\vspace{-.2in}
\end{table}

\begin{figure}[t!]
\vspace{-.0in}
\centering
\caption{About Here.}
\vspace{-.2in}
\end{figure}

A majority (54\%) of the posterior samples identified three components/clusters, with 0.12 and 0.25 posterior probability of two and four clusters, respectively.  The calculation of $\hphi$ also resulted in three components.  Figure~\ref{fig:scatter_ASD} displays the estimated cluster membership via pairwise scatterplots of the five most informative variables on a standardized scale.  Ach\_abc\_Attention has also been multiplied by minus one so that higher values imply better functioning for {\em all} tests.
%, (i.e., 0 means middle, 1 means high, -1 means low).
The corresponding mean vectors of the three main components are also provided in Table~\ref{tab:inform_vars}.  There are two groups that are very distinct (i.e., Clusters~1~and~2 are the ``high'' and ``low'' groups, respectively), but there is also a ``middle'' group (Cluster 3).
%that has some overlap with clusters 1 and 2.
Cluster 3 subjects generally have medium-to-high Adaptive\_Composite,  WJ\_Pass\_Comprehen, and  Ach\_abc\_Attention scores, but low-to-medium Beery\_standard and WJ\_Pass\_Comprehen.

Figure~\ref{fig:ASD_3D}(a) provides a 3D scatter plot on the three most informative variables, highlighting separation between Cluster 1 and Clusters 2 and 3. However, Clusters 2 and 3 are not well differentiated in this plot.  Figure~\ref{fig:ASD_3D}(b) shows a 3D scatter plot on the variables CompTsc\_ol, WJ\_Pass\_Comprehen, and Ach\_Attention, illustrating differentiation between Clusters 2 and 3.

The goal of this work is not necessarily to identify clusters that align with clinical diagnosis of ASD, i.e., it is not a classification problem.  The Research Domain Criteria (RDoC) philosophy is to get away from subjective based diagnosis of disease.  The hope is that these clusters provide groups of similar patients that may have similar underlying physiological causes and can be treated similarly (whether the clinical diagnosis was ASD or not).  That being said, the ``high'' cluster aligned with {\em no} clinical diagnosis of ASD for 92\% of its subjects, while the ``low'' cluster aligned with positive clinical ASD diagnosis for 50\% of its subjects.
As these clusters result from a bottom-up, data-driven method, they may prove useful to determine imaging biomarkers that correspond better with cluster assignment than a more subjective diagnosis provided by a physician.  This will be the subject of future work.
%The clustering results serve to generate hypotheses about what might show up in certain locations of the brain in a follow-up imaging study that may explain some of the differences between the clusters.  \cbs{John: could you please revise?}  This is a subject of further work.

\begin{figure}[t!]
\vspace{-.12in}
\centering
\caption{About Here.}
\vspace{-.23in}
\end{figure}

\vspace{-.24in}
\section{Conclusions \& Further Work}
\vspace{-.12in}
\label{sec:conclusions}

In this paper we developed a general approach to clustering via a Dirichlet process model that explicitly allows for discrete and censored variables via a latent variable approach, and missing data. This approach overcomes the assumption of (global) independence between informative and non-informative variables and the assumption of (local) independence of variables within cluster often assumed when clustering data of mixed type.  The MCMC computation proceeds via a split/merge algorithm by integrating out the component parameters.  This approach was shown to perform markedly better than other approaches on several simulated test cases.
%The additional generality also comes with a minimal cost (in terms of accuracy or computation for moderate $p$) when the actual data structure is simple (i.e., uninformative variables are independent of all other variables).
The approach was developed for moderate $p$ in the range of $\sim\!10 \!- \!300$.  The computation is $\cO(p^3)$, which makes it ill-suited for extremely large dimensions.  However, it may be possible to use a
%sparse matrix approach, i.e.,
graphical model \citep{Giudici1999,Wong2003} within the proposed framework to alleviate this burden for large $p$.

The approach was used to analyze test scores of individuals with potential ASD and identified three clusters.  Further, it was determined that only five of the 55 variables were informative to assess the cluster membership of an observation.  This could have a large impact for diagnosis of ASD as there are currently $\sim \! 100$  tests/subtest scores that could be used, and there is no universal standard.  Further, the clustering results have served to generate hypotheses about what might show up in brain imaging to explain some of the differences between potential ASD patients.  A follow-up study has been planned to investigate these possible connections.\\[-.43in]

\vspace{-.2in}
{\small
\begin{spacing}{1.69}
  %  \singlespacing
\bibliography{curt_ref.bib}
\bibliographystyle{unsrtnat}
\end{spacing}
\vspace{-.2in}
}

\clearpage
\pagestyle{empty}
\setcounter{figure}{0}
\setcounter{table}{0}

\begin{figure}
\vspace{-.1in}
\begin{center}
  \caption{3D scatter plot of three of the test score variables for potential ASD subjects.}
  \label{fig:scatter_intro}
\vspace{-0.05in}
\includegraphics[width=0.52\textwidth]{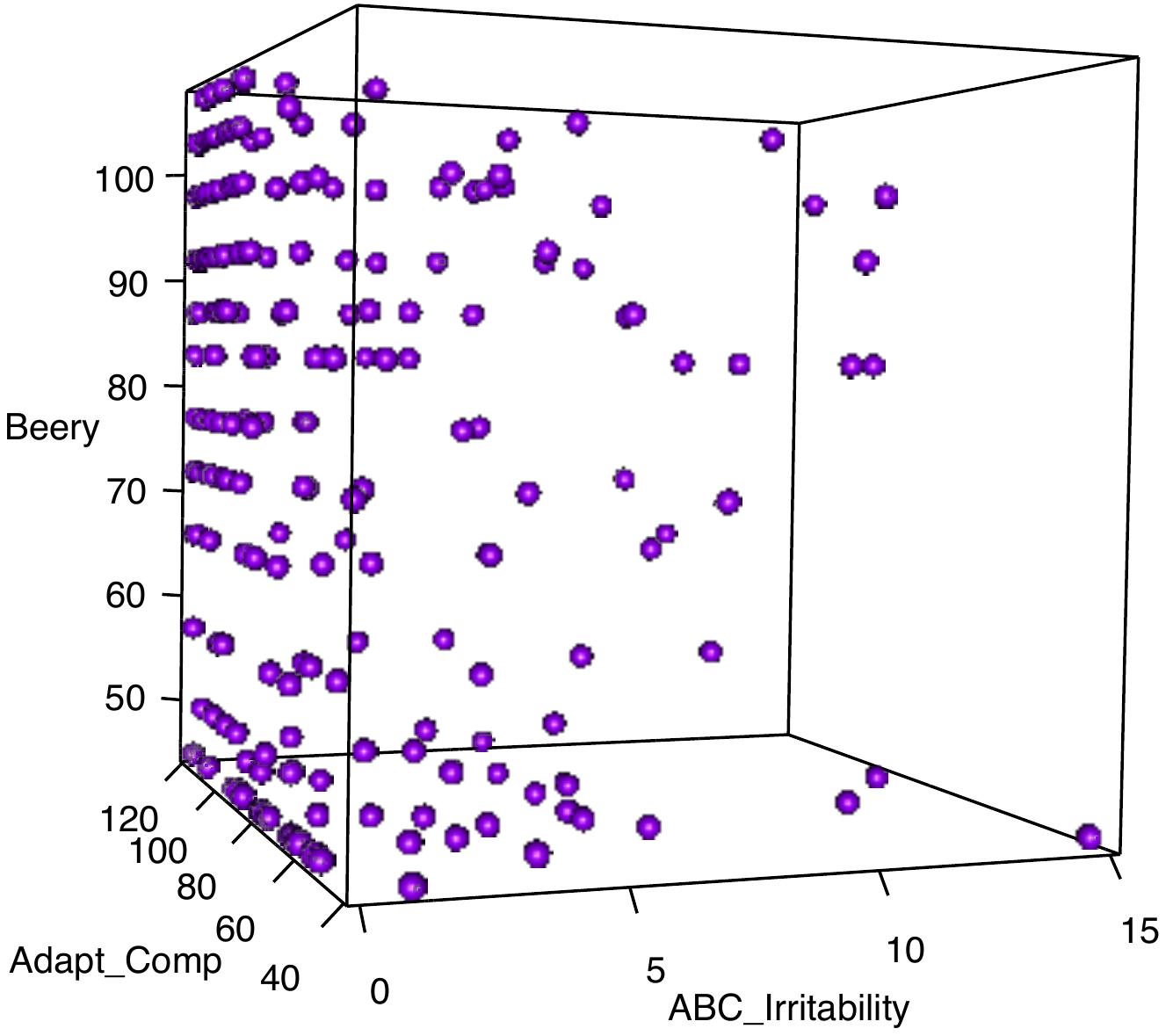}
\end{center}
\vspace{.0in}
\end{figure}

\vspace{-.08in}\begin{figure}[t!]
      \begin{widepage}
\vspace{-.05in}
\centering
\caption{Pairwise scatter plots of the first eight variables for simulation Case 2(c).}
\vspace{-.1in}
\includegraphics[width=.95\textwidth, height=.61\textheight]{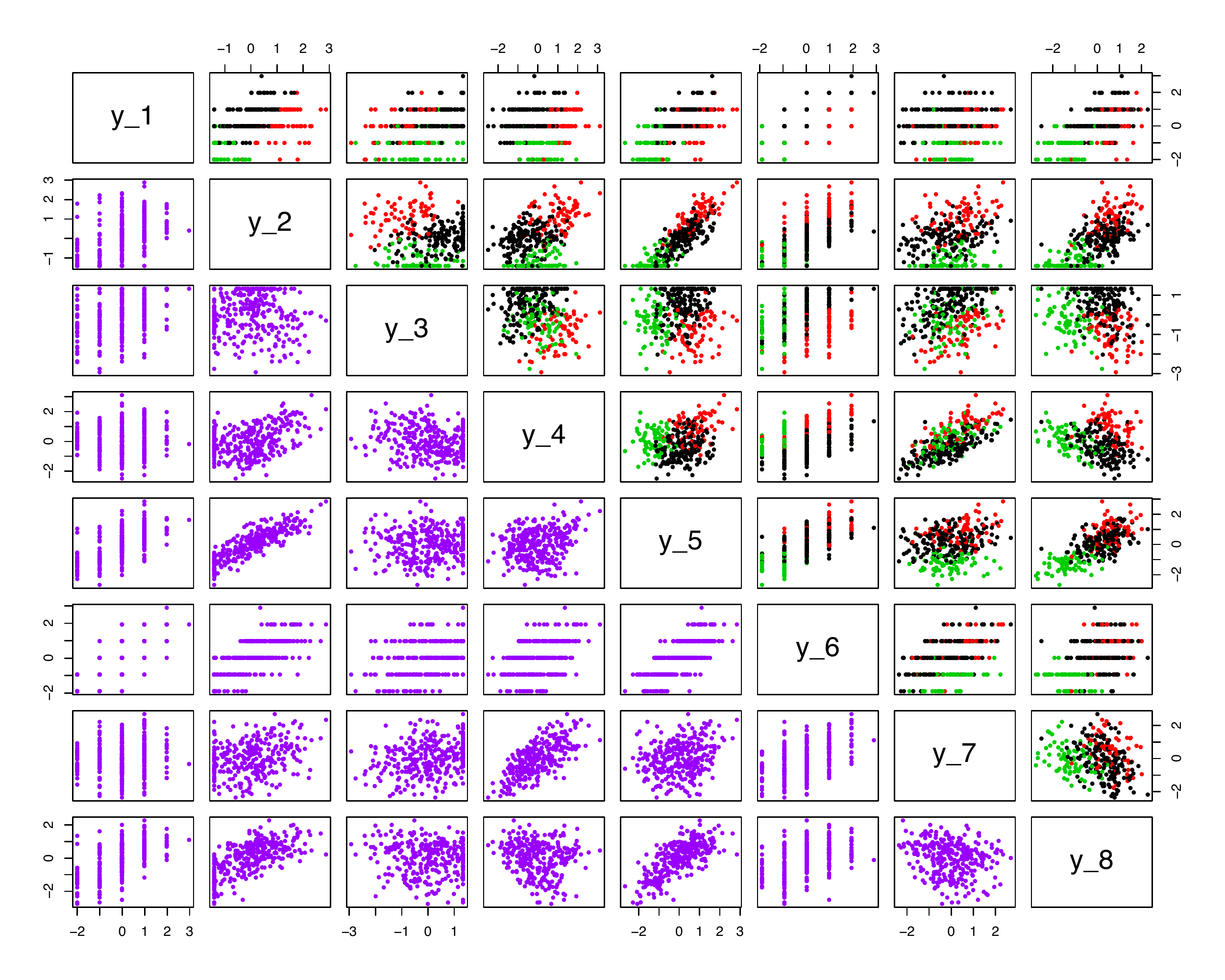}
\label{fig:sim_2c}
\vspace{-.6in}
      \end{widepage}
\end{figure}

\begin{figure}[t!]
\vspace{-.1in}
\centering
\caption{Pairwise scatter plots of the standardized version of the five most informative variables in Table~\ref{tab:inform_vars} with estimated cluster membership above the diagonal and the raw data below.}
\vspace{-.31in}
\includegraphics[width=.95\textwidth, height=.59\textheight]{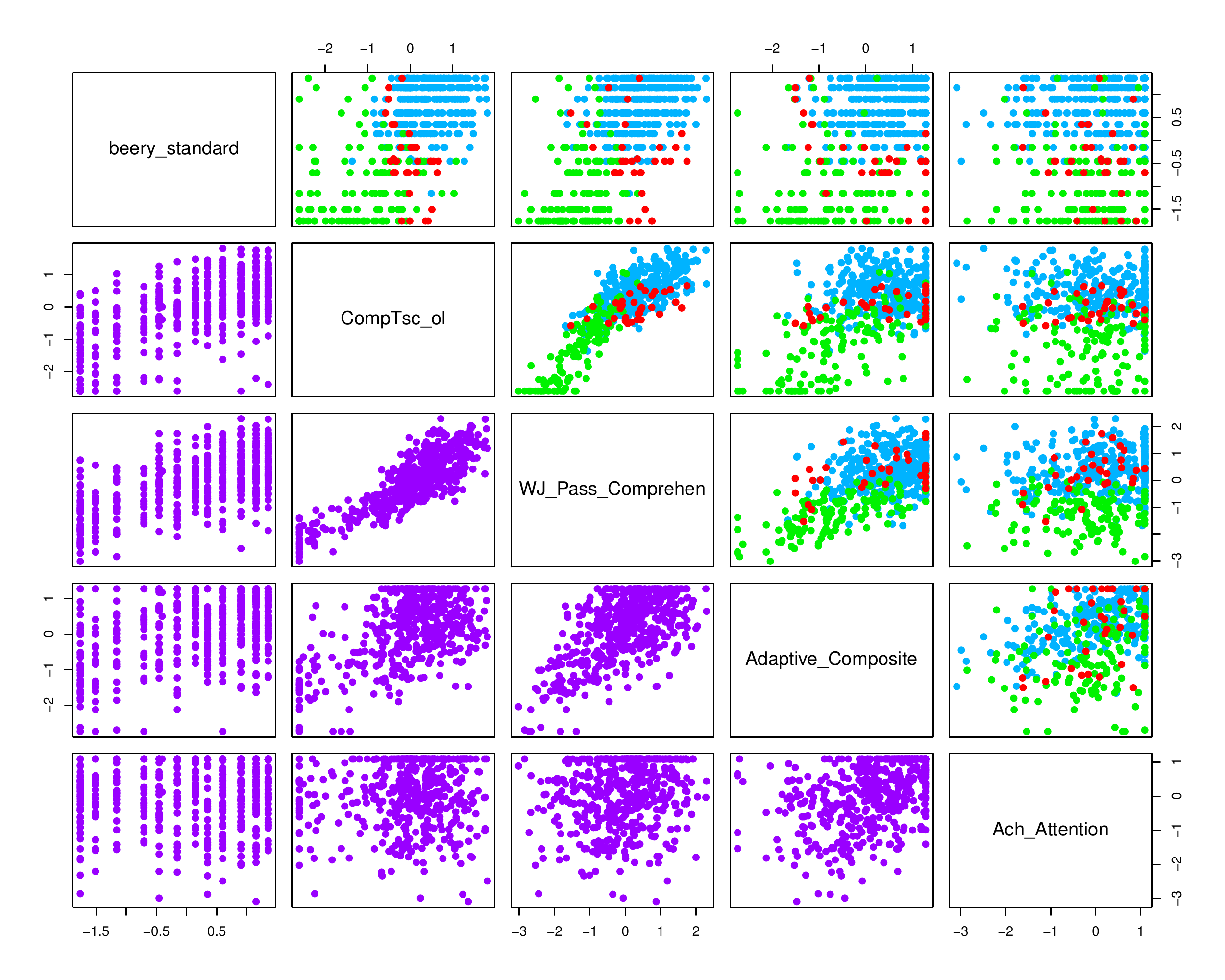}
\label{fig:scatter_ASD}
\vspace{-.18in}
\end{figure}

\begin{figure}[t!]
\vspace{-.0in}
\centering
\caption{Three dimensional scatter plots of the tests on standardized scale: (a) The most informative three variables with estimated cluster membership. (b)  Observations plotted on the variables CompTsc\_ol, WJ\_Pass\_Comprehen, and Ach\_abc\_Attention, to better illustrate the separation of clusters 2 and 3.
}
\label{fig:ASD_3D}
    \begin{subfigure}[b]{.489\textwidth}
      \centering
\vspace{-.22in}
      \caption{}
\vspace{-.16in}
      \includegraphics[width=.99\textwidth]{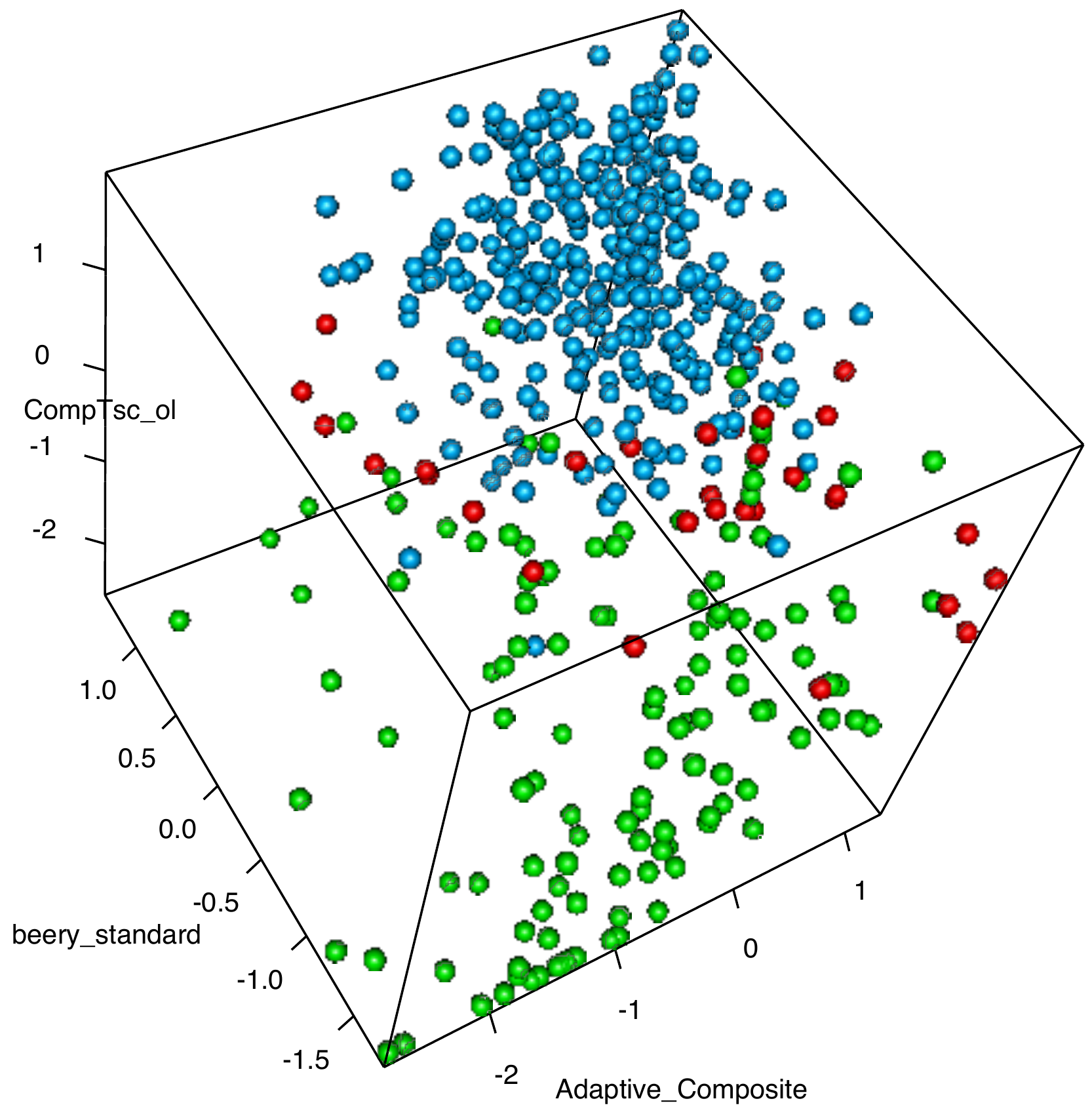}
    \end{subfigure}
    \begin{subfigure}[b]{.503\textwidth}
      \centering
\vspace{-.22in}
      \caption{}
\vspace{-.16in}
      \includegraphics[width=.99\textwidth]{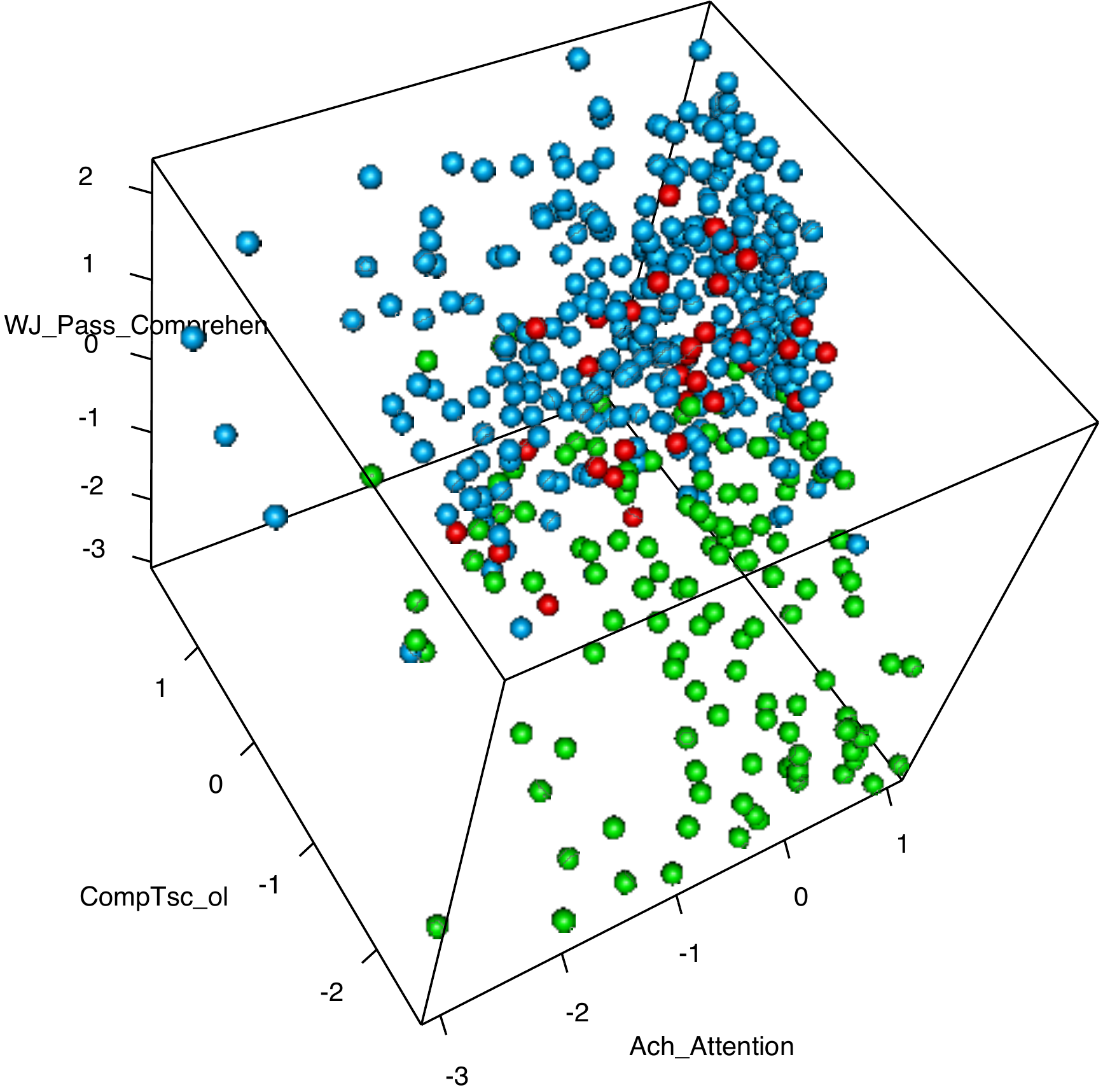}
    \end{subfigure}
\vspace{-.5in}\end{figure}

\renewcommand{\arraystretch}{1} 
\begin{table}[t!]
  \vspace{-.15in}
      \begin{widepage}
\centering
    {\footnotesize
  \caption{Simulation Case 1 Results.}
  \label{tab:case1}
  \vspace{-.05in}
  \input{table_1.tex}
    }
     \end{widepage}
\centering\footnotesize $^*\:$DPM-cont is identical to DPM-vs for cases 1(a) and 1(b) and is therefore not listed separately.
     \vspace{-.2in}   
\end{table}

  \begin{table}[t!]
  \vspace{.05in}
      \begin{widepage}
\centering
    {\footnotesize
  \caption{Simulation Case 2 Results.}
  \label{tab:case2}
  \vspace{-.05in}
  \input{table_2.tex}
}
      \end{widepage}
\centering\footnotesize $^*\:$DPM-cont is identical to DPM-vs for cases 1(a) and 1(b) and is therefore not listed separately.
%\centering\footnotesize $^*\:$DPM-cont is identical to DPM-vs for cases 2(a) and 2(b) and is therefore not listed separately.
     \vspace{-.2in}   
\end{table}

{\small
\renewcommand{\arraystretch}{.95} 
\begin{table}
  \vspace{-.07in}
  \centering
  \caption{Posterior inclusion probabilities and sample means for the six most informative tests.}
  \label{tab:inform_vars}
  \vspace{-.05in}
\footnotesize
  \begin{tabular}{|lcrrr|}
    \hline
    \multirow{2}{*}{\em Variable} & \multirow{2}{*}{$\Pr(\gamma_j=1)$} & \multicolumn{3}{c|}{\em Cluster Means} \\
    & & \multicolumn{1}{c}{1} & \multicolumn{1}{c}{2} & \multicolumn{1}{c|}{3} \\
    \hline
Beery\_standard & 1.000 & 0.77 & -1.02 & -0.44\\
CompTsc\_ol & 1.000 & 0.46 & -1.21 & -0.01\\
WJ\_Pass\_Comprehen & 0.944 & 0.38 & -1.26 & 0.30\\
Adaptive\_Composite & 0.889 & 0.44 & -0.68 & 0.16\\
ach\_abc\_Attention & 0.460 & -0.18 & 0.21 & 0.04\\
ach\_abc\_AnxDep & 0.427 & -0.04 & 0.06 & -0.01\\
\hline
  \end{tabular}
  \vspace{-.0in}
\end{table}
}

\clearpage
\pagestyle{plain}

\input{supp_mat_6.tex}

\end{document}

%% file: table_1.tex
\begin{tabular}{|lrrrrrrr|}
\hline
{\em Method} & \multicolumn{1}{c}{\em Acc} & \multicolumn{1}{c}{\em FI} & \multicolumn{1}{c}{\em ARI} & \multicolumn{1}{c}{\em M} & \multicolumn{1}{c}{$p_1$} & \multicolumn{1}{c}{\em PVC} & \multicolumn{1}{c|}{\em CompT} \\
\hline
\multicolumn{8}{|c|}{Case 1(a)} \\
\hline
\!DPM-vs$^*$ & {\bf 0.91}$\:\;\!\;\!\;\!\!$(0.11)\!\!\! & {\bf 0.86}$\:\;\!\;\!\;\!\!$(0.07)\!\!\! & {\bf 0.78}$\:\;\!\;\!\;\!\!$(0.16)\!\!\! & {\bf 2.9}$\;$(0.4)\!\!\! & {\bf 2.0}$\;$(0.3)\!\!\! & {\bf 0.99}$\:\;\!\;\!\;\!\!$(0.04)\!\!\! & 296$\:$(18)\!\!\\
\!DPM & 0.37$\:\;\!\;\!\;\!\!$(0.02)\!\!\! & 0.58$\:\;\!\;\!\;\!\!$(0.00)\!\!\! & 0.00$\:\;\!\;\!\;\!\!$(0.00)\!\!\! & 1.0$\;$(0.1)\!\!\! & 10.0$\;$(0.0)\!\!\! & 0.20$\:\;\!\;\!\;\!\!$(0.00)\!\!\! & 341$\:$(28)\!\!\\
\!DPM-ind & {\bf 0.90}$\:\;\!\;\!\;\!\!$(0.13)\!\!\! & {\bf 0.86}$\:\;\!\;\!\;\!\!$(0.08)\!\!\! & {\bf 0.76}$\:\;\!\;\!\;\!\!$(0.20)\!\!\! & {\bf 3.0}$\;$(0.6)\!\!\! & {\bf 1.9}$\;$(0.4)\!\!\! & {\bf 0.99}$\:\;\!\;\!\;\!\!$(0.04)\!\!\! & 295$\:$(23)\!\!\\
\!Mclust-vs & {\bf 0.91}$\:\;\!\;\!\;\!\!$(0.10)\!\!\! & {\bf 0.86}$\:\;\!\;\!\;\!\!$(0.07)\!\!\! & {\bf 0.78}$\:\;\!\;\!\;\!\!$(0.15)\!\!\! & {\bf 3.0}$\;$(0.4)\!\!\! & {\bf 2.0}$\;$(0.2)\!\!\! & {\bf 0.99}$\:\;\!\;\!\;\!\!$(0.06)\!\!\! & {\bf 1}$\:$(0)\!\!\\
\!VarSelLCM & 0.74$\:\;\!\;\!\;\!\!$(0.21)\!\!\! & 0.73$\:\;\!\;\!\;\!\!$(0.10)\!\!\! & 0.52$\:\;\!\;\!\;\!\!$(0.29)\!\!\! & {\bf 2.9}$\;$(1.2)\!\!\! & 3.5$\;$(3.2)\!\!\! & 0.84$\:\;\!\;\!\;\!\!$(0.32)\!\!\! & 6$\:$(2)\!\!\\
\hline
\multicolumn{8}{|c|}{Case 1(b)} \\
\hline
\!DPM-vs$^*$ & {\bf 0.89}$\:\;\!\;\!\;\!\!$(0.14)\!\!\! & {\bf 0.85}$\:\;\!\;\!\;\!\!$(0.08)\!\!\! & {\bf 0.76}$\:\;\!\;\!\;\!\!$(0.20)\!\!\! & {\bf 3.0}$\;$(0.7)\!\!\! & 1.9$\;$(0.4)\!\!\! & {\bf 0.98}$\:\;\!\;\!\;\!\!$(0.05)\!\!\! & 267$\:$(19)\!\!\\
\!DPM & 0.37$\:\;\!\;\!\;\!\!$(0.02)\!\!\! & 0.58$\:\;\!\;\!\;\!\!$(0.00)\!\!\! & 0.00$\:\;\!\;\!\;\!\!$(0.00)\!\!\! & 1.0$\;$(0.0)\!\!\! & 10.0$\;$(0.0)\!\!\! & 0.20$\:\;\!\;\!\;\!\!$(0.00)\!\!\! & 292$\:$(15)\!\!\\
\!DPM-ind & 0.37$\:\;\!\;\!\;\!\!$(0.02)\!\!\! & 0.58$\:\;\!\;\!\;\!\!$(0.00)\!\!\! & 0.00$\:\;\!\;\!\;\!\!$(0.00)\!\!\! & 1.0$\;$(0.0)\!\!\! & 8.5$\;$(0.8)\!\!\! & 0.35$\:\;\!\;\!\;\!\!$(0.08)\!\!\! & 243$\:$(18)\!\!\\
\!Mclust-vs & {\bf 0.87}$\:\;\!\;\!\;\!\!$(0.12)\!\!\! & 0.83$\:\;\!\;\!\;\!\!$(0.10)\!\!\! & {\bf 0.72}$\:\;\!\;\!\;\!\!$(0.19)\!\!\! & {\bf 2.8}$\;$(0.4)\!\!\! & {\bf 2.0}$\;$(0.1)\!\!\! & 0.94$\:\;\!\;\!\;\!\!$(0.12)\!\!\! & {\bf 1}$\:$(0)\!\!\\
\!VarSelLCM & 0.52$\:\;\!\;\!\;\!\!$(0.06)\!\!\! & 0.57$\:\;\!\;\!\;\!\!$(0.05)\!\!\! & 0.40$\:\;\!\;\!\;\!\!$(0.06)\!\!\! & 6.7$\;$(0.9)\!\!\! & 9.1$\;$(0.8)\!\!\! & 0.29$\:\;\!\;\!\;\!\!$(0.08)\!\!\! & 11$\:$(2)\!\!\\
\hline
\multicolumn{8}{|c|}{Case 1(c)} \\
\hline
\!DPM-vs & {\bf 0.85}$\:\;\!\;\!\;\!\!$(0.14)\!\!\! & {\bf 0.81}$\:\;\!\;\!\;\!\!$(0.08)\!\!\! & {\bf 0.68}$\:\;\!\;\!\;\!\!$(0.20)\!\!\! & {\bf 2.8}$\;$(0.6)\!\!\! & {\bf 1.9}$\;$(0.4)\!\!\! & {\bf 0.97}$\:\;\!\;\!\;\!\!$(0.07)\!\!\! & 254$\:$(18)\!\!\\
\!DPM-cont & 0.62$\:\;\!\;\!\;\!\!$(0.06)\!\!\! & 0.52$\:\;\!\;\!\;\!\!$(0.06)\!\!\! & 0.31$\:\;\!\;\!\;\!\!$(0.07)\!\!\! & 4.9$\;$(0.7)\!\!\! & {\bf 2.0}$\;$(0.5)\!\!\! & 0.89$\:\;\!\;\!\;\!\!$(0.11)\!\!\! & 296$\:$(25)\!\!\\
\!DPM & 0.37$\:\;\!\;\!\;\!\!$(0.02)\!\!\! & 0.58$\:\;\!\;\!\;\!\!$(0.00)\!\!\! & 0.00$\:\;\!\;\!\;\!\!$(0.00)\!\!\! & 1.0$\;$(0.0)\!\!\! & 10.0$\;$(0.0)\!\!\! & 0.20$\:\;\!\;\!\;\!\!$(0.00)\!\!\! & 285$\:$(19)\!\!\\
\!DPM-ind & 0.37$\:\;\!\;\!\;\!\!$(0.02)\!\!\! & 0.58$\:\;\!\;\!\;\!\!$(0.00)\!\!\! & 0.00$\:\;\!\;\!\;\!\!$(0.00)\!\!\! & 1.0$\;$(0.0)\!\!\! & 8.2$\;$(1.0)\!\!\! & 0.38$\:\;\!\;\!\;\!\!$(0.10)\!\!\! & 243$\:$(24)\!\!\\
\!Mclust-vs & 0.49$\:\;\!\;\!\;\!\!$(0.10)\!\!\! & 0.43$\:\;\!\;\!\;\!\!$(0.09)\!\!\! & 0.19$\:\;\!\;\!\;\!\!$(0.12)\!\!\! & 6.7$\;$(1.5)\!\!\! & 2.7$\;$(0.8)\!\!\! & 0.67$\:\;\!\;\!\;\!\!$(0.14)\!\!\! & {\bf 1}$\:$(0)\!\!\\
\!VarSelLCM & 0.48$\:\;\!\;\!\;\!\!$(0.08)\!\!\! & 0.51$\:\;\!\;\!\;\!\!$(0.06)\!\!\! & 0.34$\:\;\!\;\!\;\!\!$(0.07)\!\!\! & 6.8$\;$(1.1)\!\!\! & 8.4$\;$(0.8)\!\!\! & 0.36$\:\;\!\;\!\;\!\!$(0.08)\!\!\! & 10$\:$(1)\!\!\\
\hline
\multicolumn{8}{|c|}{Case 1(d)} \\
\hline
\!DPM-vs & {\bf 0.77}$\:\;\!\;\!\;\!\!$(0.19)\!\!\! & {\bf 0.76}$\:\;\!\;\!\;\!\!$(0.10)\!\!\! & {\bf 0.57}$\:\;\!\;\!\;\!\!$(0.25)\!\!\! & {\bf 2.6}$\;$(0.8)\!\!\! & {\bf 1.9}$\;$(0.6)\!\!\! & {\bf 0.95}$\:\;\!\;\!\;\!\!$(0.09)\!\!\! & 304$\:$(21)\!\!\\
\!DPM-cont & 0.62$\:\;\!\;\!\;\!\!$(0.04)\!\!\! & 0.52$\:\;\!\;\!\;\!\!$(0.05)\!\!\! & 0.31$\:\;\!\;\!\;\!\!$(0.06)\!\!\! & 4.8$\;$(0.8)\!\!\! & {\bf 1.9}$\;$(0.5)\!\!\! & 0.89$\:\;\!\;\!\;\!\!$(0.11)\!\!\! & 355$\:$(29)\!\!\\
\!DPM & 0.37$\:\;\!\;\!\;\!\!$(0.02)\!\!\! & 0.58$\:\;\!\;\!\;\!\!$(0.00)\!\!\! & 0.00$\:\;\!\;\!\;\!\!$(0.00)\!\!\! & 1.0$\;$(0.1)\!\!\! & 10.0$\;$(0.0)\!\!\! & 0.20$\:\;\!\;\!\;\!\!$(0.00)\!\!\! & 343$\:$(29)\!\!\\
\!DPM-ind & 0.37$\:\;\!\;\!\;\!\!$(0.02)\!\!\! & 0.58$\:\;\!\;\!\;\!\!$(0.00)\!\!\! & 0.00$\:\;\!\;\!\;\!\!$(0.00)\!\!\! & 1.0$\;$(0.0)\!\!\! & 7.6$\;$(1.3)\!\!\! & 0.44$\:\;\!\;\!\;\!\!$(0.13)\!\!\! & 277$\:$(40)\!\!\\
\!Mclust-vs & 0.50$\:\;\!\;\!\;\!\!$(0.10)\!\!\! & 0.43$\:\;\!\;\!\;\!\!$(0.08)\!\!\! & 0.20$\:\;\!\;\!\;\!\!$(0.11)\!\!\! & 7.0$\;$(1.1)\!\!\! & 3.1$\;$(0.9)\!\!\! & 0.64$\:\;\!\;\!\;\!\!$(0.14)\!\!\! & {\bf 1}$\:$(0)\!\!\\
\!VarSelLCM & 0.50$\:\;\!\;\!\;\!\!$(0.08)\!\!\! & 0.51$\:\;\!\;\!\;\!\!$(0.06)\!\!\! & 0.34$\:\;\!\;\!\;\!\!$(0.07)\!\!\! & 6.5$\;$(1.1)\!\!\! & 8.3$\;$(0.8)\!\!\! & 0.36$\:\;\!\;\!\;\!\!$(0.08)\!\!\! & 12$\:$(9)\!\!\\
\hline
\hline
\!True &   1.00$\:\:$\white{(0.00)$\!\!\!$}& 1.00$\:\:$\white{(0.00)$\!\!\!$}& 1.00$\:\:$\white{(0.00)$\!\!\!$}& 3.0$\:\:$\white{(0.0)$\!\!\!$}& 2.0$\:\:$\white{(0.0)$\!\!\!$}& 1.00$\:$\white{(0.00)$\!\!$}& \multicolumn{1}{c|}{$-$}\\
\hline
\end{tabular}

%% file: table_2.tex
\begin{tabular}{|lrrrrrrr|}
\hline
{\em Method} & \multicolumn{1}{c}{\em Acc} & \multicolumn{1}{c}{\em FI} & \multicolumn{1}{c}{\em ARI} & \multicolumn{1}{c}{\em M} & \multicolumn{1}{c}{$p_1$} & \multicolumn{1}{c}{\em PVC} & \multicolumn{1}{c|}{\em CompT} \\
\hline
\multicolumn{8}{|c|}{Case 2(a)} \\
\hline
\!DPM-vs$^*$ & {\bf 0.89}$\:\;\!\;\!\;\!\!$(0.11)\!\!\! & {\bf 0.85}$\:\;\!\;\!\;\!\!$(0.09)\!\!\! & {\bf 0.74}$\:\;\!\;\!\;\!\!$(0.20)\!\!\! & {\bf 3.0}$\;$(0.6)\!\!\! & {\bf 3.8}$\;$(0.7)\!\!\! & {\bf 0.99}$\:\;\!\;\!\;\!\!$(0.02)\!\!\! & 544$\:$(45)\!\!\\
\!DPM & 0.50$\:\;\!\;\!\;\!\!$(0.03)\!\!\! & 0.61$\:\;\!\;\!\;\!\!$(0.01)\!\!\! & 0.00$\:\;\!\;\!\;\!\!$(0.00)\!\!\! & 1.4$\;$(0.8)\!\!\! & 30.0$\;$(0.0)\!\!\! & 0.13$\:\;\!\;\!\;\!\!$(0.00)\!\!\! & 1158$\:$(120)\!\!\\
\!DPM-ind & {\bf 0.90}$\:\;\!\;\!\;\!\!$(0.10)\!\!\! & {\bf 0.85}$\:\;\!\;\!\;\!\!$(0.08)\!\!\! & {\bf 0.75}$\:\;\!\;\!\;\!\!$(0.17)\!\!\! & {\bf 3.1}$\;$(0.6)\!\!\! & {\bf 3.9}$\;$(0.4)\!\!\! & {\bf 1.00}$\:\;\!\;\!\;\!\!$(0.02)\!\!\! & 543$\:$(36)\!\!\\
\!Mclust-vs & {\bf 0.91}$\:\;\!\;\!\;\!\!$(0.12)\!\!\! & {\bf 0.88}$\:\;\!\;\!\;\!\!$(0.09)\!\!\! & {\bf 0.78}$\:\;\!\;\!\;\!\!$(0.23)\!\!\! & {\bf 2.9}$\;$(0.5)\!\!\! & {\bf 4.0}$\;$(0.8)\!\!\! & 0.98$\:\;\!\;\!\;\!\!$(0.07)\!\!\! & {\bf 8}$\:$(3)\!\!\\
\!VarSelLCM & 0.68$\:\;\!\;\!\;\!\!$(0.07)\!\!\! & 0.68$\:\;\!\;\!\;\!\!$(0.04)\!\!\! & 0.53$\:\;\!\;\!\;\!\!$(0.06)\!\!\! & 4.7$\;$(0.6)\!\!\! & {\bf 4.0}$\;$(0.0)\!\!\! & {\bf 1.00}$\:\;\!\;\!\;\!\!$(0.00)\!\!\! & 25$\:$(1)\!\!\\
\hline
\multicolumn{8}{|c|}{Case 2(b)} \\
\hline
\!DPM-vs$^*$ & {\bf 0.90}$\:\;\!\;\!\;\!\!$(0.10)\!\!\! & {\bf 0.85}$\:\;\!\;\!\;\!\!$(0.08)\!\!\! & {\bf 0.75}$\:\;\!\;\!\;\!\!$(0.17)\!\!\! & {\bf 3.1}$\;$(0.6)\!\!\! & {\bf 3.9}$\;$(0.5)\!\!\! & {\bf 0.99}$\:\;\!\;\!\;\!\!$(0.03)\!\!\! & 557$\:$(27)\!\!\\
\!DPM & 0.50$\:\;\!\;\!\;\!\!$(0.03)\!\!\! & 0.61$\:\;\!\;\!\;\!\!$(0.01)\!\!\! & 0.00$\:\;\!\;\!\;\!\!$(0.00)\!\!\! & 1.0$\;$(0.0)\!\!\! & 30.0$\;$(0.0)\!\!\! & 0.13$\:\;\!\;\!\;\!\!$(0.00)\!\!\! & 1009$\:$(67)\!\!\\
\!DPM-ind & 0.50$\:\;\!\;\!\;\!\!$(0.03)\!\!\! & 0.61$\:\;\!\;\!\;\!\!$(0.01)\!\!\! & 0.00$\:\;\!\;\!\;\!\!$(0.00)\!\!\! & 1.0$\;$(0.0)\!\!\! & 28.9$\;$(0.2)\!\!\! & 0.17$\:\;\!\;\!\;\!\!$(0.01)\!\!\! & 966$\:$(71)\!\!\\
\!Mclust-vs & 0.83$\:\;\!\;\!\;\!\!$(0.14)\!\!\! & 0.80$\:\;\!\;\!\;\!\!$(0.12)\!\!\! & 0.63$\:\;\!\;\!\;\!\!$(0.24)\!\!\! & 2.6$\;$(0.5)\!\!\! & 3.4$\;$(0.8)\!\!\! & 0.91$\:\;\!\;\!\;\!\!$(0.10)\!\!\! & {\bf 8}$\:$(3)\!\!\\
\!VarSelLCM & 0.43$\:\;\!\;\!\;\!\!$(0.04)\!\!\! & 0.46$\:\;\!\;\!\;\!\!$(0.04)\!\!\! & 0.26$\:\;\!\;\!\;\!\!$(0.05)\!\!\! & 7.3$\;$(0.6)\!\!\! & 27.3$\;$(1.4)\!\!\! & 0.22$\:\;\!\;\!\;\!\!$(0.05)\!\!\! & 53$\:$(3)\!\!\\
\hline
\multicolumn{8}{|c|}{Case 2(c)} \\
\hline
\!DPM-vs & {\bf 0.93}$\:\;\!\;\!\;\!\!$(0.03)\!\!\! & {\bf 0.89}$\:\;\!\;\!\;\!\!$(0.05)\!\!\! & {\bf 0.82}$\:\;\!\;\!\;\!\!$(0.07)\!\!\! & {\bf 3.3}$\;$(0.6)\!\!\! & {\bf 4.0}$\;$(0.2)\!\!\! & {\bf 1.00}$\:\;\!\;\!\;\!\!$(0.02)\!\!\! & 589$\:$(47)\!\!\\
\!DPM-cont & 0.62$\:\;\!\;\!\;\!\!$(0.16)\!\!\! & 0.57$\:\;\!\;\!\;\!\!$(0.14)\!\!\! & 0.34$\:\;\!\;\!\;\!\!$(0.20)\!\!\! & 4.4$\;$(1.2)\!\!\! & 3.4$\;$(1.4)\!\!\! & 0.87$\:\;\!\;\!\;\!\!$(0.05)\!\!\! & 597$\:$(49)\!\!\\
\!DPM & 0.50$\:\;\!\;\!\;\!\!$(0.03)\!\!\! & 0.61$\:\;\!\;\!\;\!\!$(0.01)\!\!\! & 0.00$\:\;\!\;\!\;\!\!$(0.00)\!\!\! & 1.0$\;$(0.0)\!\!\! & 30.0$\;$(0.0)\!\!\! & 0.13$\:\;\!\;\!\;\!\!$(0.00)\!\!\! & 1078$\:$(54)\!\!\\
\!DPM-ind & 0.50$\:\;\!\;\!\;\!\!$(0.03)\!\!\! & 0.61$\:\;\!\;\!\;\!\!$(0.01)\!\!\! & 0.00$\:\;\!\;\!\;\!\!$(0.00)\!\!\! & 1.0$\;$(0.0)\!\!\! & 28.8$\;$(0.5)\!\!\! & 0.17$\:\;\!\;\!\;\!\!$(0.02)\!\!\! & 1060$\:$(101)\!\!\\
\!Mclust-vs & 0.45$\:\;\!\;\!\;\!\!$(0.06)\!\!\! & 0.40$\:\;\!\;\!\;\!\!$(0.04)\!\!\! & 0.13$\:\;\!\;\!\;\!\!$(0.06)\!\!\! & 6.4$\;$(1.3)\!\!\! & 2.8$\;$(1.0)\!\!\! & 0.81$\:\;\!\;\!\;\!\!$(0.04)\!\!\! & {\bf 8}$\:$(24)\!\!\\
\!VarSelLCM & 0.42$\:\;\!\;\!\;\!\!$(0.05)\!\!\! & 0.43$\:\;\!\;\!\;\!\!$(0.05)\!\!\! & 0.21$\:\;\!\;\!\;\!\!$(0.06)\!\!\! & 7.3$\;$(0.8)\!\!\! & 26.6$\;$(1.5)\!\!\! & 0.24$\:\;\!\;\!\;\!\!$(0.05)\!\!\! & 59$\:$(3)\!\!\\
\hline
\multicolumn{8}{|c|}{Case 2(d)} \\
\hline
\!DPM-vs & {\bf 0.91}$\:\;\!\;\!\;\!\!$(0.08)\!\!\! & {\bf 0.86}$\:\;\!\;\!\;\!\!$(0.08)\!\!\! & {\bf 0.78}$\:\;\!\;\!\;\!\!$(0.14)\!\!\! & {\bf 3.2}$\;$(0.6)\!\!\! & {\bf 3.9}$\;$(0.4)\!\!\! & {\bf 0.99}$\:\;\!\;\!\;\!\!$(0.03)\!\!\! & 577$\:$(54)\!\!\\
\!DPM-cont & 0.61$\:\;\!\;\!\;\!\!$(0.16)\!\!\! & 0.55$\:\;\!\;\!\;\!\!$(0.14)\!\!\! & 0.32$\:\;\!\;\!\;\!\!$(0.19)\!\!\! & 4.3$\;$(1.1)\!\!\! & 3.2$\;$(1.2)\!\!\! & 0.87$\:\;\!\;\!\;\!\!$(0.05)\!\!\! & 581$\:$(41)\!\!\\
\!DPM & 0.50$\:\;\!\;\!\;\!\!$(0.03)\!\!\! & 0.61$\:\;\!\;\!\;\!\!$(0.01)\!\!\! & 0.00$\:\;\!\;\!\;\!\!$(0.00)\!\!\! & 1.3$\;$(0.5)\!\!\! & 30.0$\;$(0.0)\!\!\! & 0.13$\:\;\!\;\!\;\!\!$(0.00)\!\!\! & 1028$\:$(85)\!\!\\
\!DPM-ind & 0.50$\:\;\!\;\!\;\!\!$(0.03)\!\!\! & 0.61$\:\;\!\;\!\;\!\!$(0.01)\!\!\! & 0.00$\:\;\!\;\!\;\!\!$(0.00)\!\!\! & 1.0$\;$(0.0)\!\!\! & 28.2$\;$(1.1)\!\!\! & 0.19$\:\;\!\;\!\;\!\!$(0.04)\!\!\! & 958$\:$(81)\!\!\\
\!Mclust-vs & 0.46$\:\;\!\;\!\;\!\!$(0.06)\!\!\! & 0.41$\:\;\!\;\!\;\!\!$(0.04)\!\!\! & 0.14$\:\;\!\;\!\;\!\!$(0.05)\!\!\! & 6.7$\;$(1.4)\!\!\! & 3.0$\;$(1.2)\!\!\! & 0.81$\:\;\!\;\!\;\!\!$(0.04)\!\!\! & {\bf 6}$\:$(7)\!\!\\
\!VarSelLCM & 0.43$\:\;\!\;\!\;\!\!$(0.05)\!\!\! & 0.43$\:\;\!\;\!\;\!\!$(0.05)\!\!\! & 0.20$\:\;\!\;\!\;\!\!$(0.06)\!\!\! & 7.0$\;$(1.0)\!\!\! & 26.0$\;$(1.8)\!\!\! & 0.26$\:\;\!\;\!\;\!\!$(0.06)\!\!\! & 73$\:$(19)\!\!\\
\hline
\hline
\!True &  1.00$\:\:$\white{(0.00)$\!\!\!$}& 1.00$\:\:$\white{(0.00)$\!\!\!$}& 1.00$\:\:$\white{(0.00)$\!\!\!$}& 3.0$\:\:$\white{(0.0)$\!\!\!$}& 4.0$\:\:$\white{(0.0)$\!\!\!$}& 1.00$\:$\white{(0.00)$\!\!$}& \multicolumn{1}{c|}{$-$}\\
\hline
\end{tabular}

%% file: supp_mat_6.tex
\clearpage

%\setlength{\oddsidemargin}{-.0in} %.531in}
%\newgeometry{textwidth=6.5in,textheight=8.9in}

\begin{appendix}

\setcounter{equation}{0}
\setcounter{table}{0}
\setcounter{figure}{0}
\setcounter{page}{1}
\renewcommand{\theequation}{S.\arabic{equation}}
\renewcommand{\thetable}{S.\arabic{table}}
\renewcommand{\thefigure}{S.\arabic{figure}}

\thispagestyle{empty}

\vspace{-.2in}
\begin{Large}
\noindent
{\bf
Supplementary Material: ``Clustering and Variable Selection in the Presence of Mixed Variable Types and Missing Data''
}
\end{Large}
\vspace{-.2in}

\section{Cognitive/Behavioral Tests Descriptions}
\label{sec:definitions}

{\scriptsize
\singlespacing
\renewcommand{\arraystretch}{1.5} 
\vspace{-.15in}\begin{longtable}{|lll|}
    \caption{Test and self-report form descriptions for the 55 tests used in the analysis in the main paper.}\\
    \hline
$\!\!$Variable Name & Test/Form & Description \\
\hline
\hline && \\[-.1in]
$\!\!$General\_Adaptive\_Composite$\!\!\!\!\!\!$ & \parbox{.275\textwidth}{Adaptive Behavior Assessment System (ABAS-II) overall adaptive functioning composite score} & \parbox{.45\textwidth}{Includes all 9 skill areas in the 3 rows below plus Work (when applicable).}\\[0.1in]
\hline && \\[-.1in]
$\!\!$Conceptual\_Composite\_Score$\!\!\!\!\!\!$ & \parbox{.275\textwidth}{ABAS-II Conceptual Composite domain  } & \parbox{.45\textwidth}{Includes Communication, Functional Academics, and Self-Direction skill areas.  }\\[0.2in]
\hline && \\[-.1in]
$\!\!$Social\_Composite\_Score$\!\!\!\!\!\!$ & \parbox{.275\textwidth}{ABAS-II Social Composite domain  } & \parbox{.45\textwidth}{Includes Leisure and Social skill areas.  }\\[0.1in]
\hline && \\[-.1in]
$\!\!$Practical\_Composite\_Score$\!\!\!\!\!\!$ & \parbox{.275\textwidth}{ABAS-II Practical Composite domain  } & \parbox{.45\textwidth}{Includes Community Use, Home Living, Health and Safety, and Self-Care skill areas.  }\\[0.2in]
\hline && \\[-.1in]
$\!\!$ABC\_Irritability\_raw$\!\!\!\!\!\!$ & \parbox{.275\textwidth}{Aberrant Behavior Checklist, Irritability scale} & \parbox{.45\textwidth}{ABC is a maladaptive behavior rating scale.  Higher scores are worse on all of the ABC subscales.   Normed for individuals with significant developmental disabilities requiring special education, so does not capture mild issues in the general population.}\\[0.4in]
\hline && \\[-.1in]
$\!\!$ABC\_Lethargy\_raw$\!\!\!\!\!\!$ & \parbox{.275\textwidth}{ABC Lethargy/Social Withdrawal scale.  } & \parbox{.45\textwidth}{Items reflect underactivity or listlessness, social withdrawal (seeking isolation, unresponsiveness to social interactions, etc.).}\\[0.2in]
\hline && \\[-.1in]
$\!\!$ABC\_Stereotype\_raw$\!\!\!\!\!\!$ & \parbox{.275\textwidth}{ABC Stereotype scale.} & \parbox{.45\textwidth}{Stereotype scale measures compulsions and repetitive stereotyped behaviors.  }\\[0.2in]
\hline && \\[-.1in]
$\!\!$ABC\_Hyperactivity\_raw$\!\!\!\!\!\!$ & \parbox{.275\textwidth}{ABC Hyperactivity scale} & \parbox{.45\textwidth}{Includes ADHD symptoms (inattention, distractibility, impulsivity, hyperactivity) and also noncompliance/oppositional behavior.}\\[0.2in]
\hline && \\[-.1in]
$\!\!$ABC\_Inappropriate\_Speech\_raw$\!\!\!\!\!\!$ & \parbox{.275\textwidth}{ABC Inappropriate Speech scale} & \parbox{.45\textwidth}{Only 4 items, which capture the following aspects of speech:  excessive, repetitive (2 items), self-directed and loud. }\\[0.2in]
\hline && \\[-.1in]
$\!\!$COM\_RSI\_Total$\!\!\!\!\!\!$ & \parbox{.275\textwidth}{Autism Diagnostic Observation Schedule (ADOS), Communication + Reciprocal Social Interaction score} & \parbox{.45\textwidth}{The Communication + Reciprocal Social Interaction total score is used to determine ADOS-2 classification (autism, autism spectrum, or non-spectrum) based on cutoff scores.  This score corresponds with the DSM-5 Social Communication criteria (and does not include the restricted and repetitive behavior aspect).}\\[0.5in]
\hline && \\[-.1in]
$\!\!$SBRI\_Total$\!\!\!\!\!\!$ & \parbox{.275\textwidth}{Autism Diagnostic Observation Schedule (ADOS), Stereotyped Behaviors and Restricted Interests score} & \parbox{.45\textwidth}{This score includes unusual sensory interests/behaviors, stereotyped mannerisms, circumscribed or unusual interests, and compulsions/rituals.  The SBRI score is not used in the ADOS-2 diagnostic algorithm but informs determination of whether DSM-5 restricted interests/activities and repetitive behavior criteria are met.   Note: Sometimes these behaviors are not exhibited during testing yet are prominent at home and in the community - only observed behaviors can be scored.  So, it may underestimate RRB.  }\\[0.7in]
\hline && \\[-.1in]
$\!\!$Beery\_standard$\!\!\!\!\!\!$ & \parbox{.275\textwidth}{Beery-Buktenica Developmental Test of Visual-Motor Integration (Beery VMI)} & \parbox{.45\textwidth}{Assesses the extent to which individuals can integrate their visual and motor abilities (degree to which visual perception and finger-hand movements are well coordinated).  Important role in the development of handwriting and other skills.  }\\[0.4in]
\hline && \\[-.1in]
$\!\!$ Inhibit\_T$\!\!\!\!\!\!$ & \parbox{.275\textwidth}{BRIEF Inhibition} & \parbox{.45\textwidth}{Ability to inhibit impulsive responses (resist impulses, stop one's own behavior at the appropriate time).}\\[0.2in]
\hline && \\[-.1in]
$\!\!$Shift\_T$\!\!\!\!\!\!$ & \parbox{.275\textwidth}{BRIEF Task Shift} & \parbox{.45\textwidth}{Ability to adjust to changes in routine or task demands. Key aspects of shifting include the ability to make transitions, tolerate change, problem-solve flexibly, switch or alternate attention, and change focus from one mindset or topic to another.}\\[0.4in]
\hline && \\[-.1in]
$\!\!$Emotional\_Control\_T$\!\!\!\!\!\!$ & \parbox{.275\textwidth}{BRIEF Emotional Control} & \parbox{.45\textwidth}{Measures the impact of executive function problems on emotional expression and assesses a child's ability to modulate or control his or her emotional responses.}\\[0.3in]
\hline && \\[-.1in]
$\!\!$Self\_Monitor\_T$\!\!\!\!\!\!$ & \parbox{.275\textwidth}{BRIEF Self-monitoring} & \parbox{.45\textwidth}{Self-monitoring or interpersonal awareness (whether a child keeps track of the effect that his or her behavior has on others). }\\[0.2in]
\hline && \\[-.1in]
$\!\!$Initiate\_T$\!\!\!\!\!\!$ & \parbox{.275\textwidth}{BRIEF Task Initiation} & \parbox{.45\textwidth}{Ability to begin a task or activity and to independently generate ideas, responses, or problem-solving strategies. }\\[0.2in]
\hline && \\[-.1in]
$\!\!$Working\_Memory\_T$\!\!\!\!\!\!$ & \parbox{.275\textwidth}{BRIEF Working Memory} & \parbox{.45\textwidth}{Ability to hold information in mind for the purpose of completing a task, encoding information, or generating goals, plans, and sequential steps to achieving goals.}\\[0.3in]
\hline && \\[-.1in]
$\!\!$Plan\_Organize\_T$\!\!\!\!\!\!$ & \parbox{.275\textwidth}{BRIEF Task Organization} & \parbox{.45\textwidth}{Ability to manage current and future-oriented task demands (plan and organize problem solving approaches).}\\[0.2in]
\hline && \\[-.1in]
$\!\!$Task\_Monitor\_T$\!\!\!\!\!\!$ & \parbox{.275\textwidth}{BRIEF Task Monitoring} & \parbox{.45\textwidth}{Task-oriented monitoring or work-checking habits (whether a child assesses his or her own performance during or shortly after finishing a task to ensure accuracy or appropriate attainment of a goal).}\\[0.3in]
\hline && \\[-.1in]
$\!\!$Org\_of\_Materials\_T$\!\!\!\!\!\!$ & \parbox{.275\textwidth}{BRIEF Organization of Materials} & \parbox{.45\textwidth}{Ability to organize environment and materials - orderliness of work, play, and storage spaces (e.g., desks, lockers, backpacks, and bedrooms).}\\[0.2in]
\hline && \\[-.1in]
$\!\!$BRI\_T$\!\!\!\!\!\!$ & \parbox{.275\textwidth}{BRIEF Behavioral Regulation Index (BRI)} & \parbox{.45\textwidth}{Summary capturing ability to shift cognitive set and modulate emotions and behavior via appropriate inhibitory control.  Includes Inhibit, Shift, and Emotional Control subscales.  }\\[0.3in]
\hline && \\[-.1in]
$\!\!$MI\_T$\!\!\!\!\!\!$ & \parbox{.275\textwidth}{BRIEF Metacognition Index (MI)} & \parbox{.45\textwidth}{Summary capturing ability to initiate, plan, organize, self-monitor, and sustain working memory - relates directly to a child's ability to actively problem solve in a variety of contexts.  Includes Initiate, Working Memory, Plan/Organize, Organization of Materials, and Monitor subscales.  }\\[0.4in]
\hline && \\[-.1in]
$\!\!$GEC\_T$\!\!\!\!\!\!$ & \parbox{.275\textwidth}{BRIEF Global Executive Composite (GEC)} & \parbox{.45\textwidth}{Overall index of executive function; incorporates all of the BRIEF clinical scales. }\\[0.2in]
\hline && \\[-.1in]
$\!\!$Cars\_Total$\!\!\!\!\!\!$ & \parbox{.275\textwidth}{Childhood Autism Rating Scales (CARS2-ST and CARS2-HF)} & \parbox{.45\textwidth}{Structured interview and observation tool.  Scores are raw scores (T-scores and percentiles among population of individuals with ASD are available).  A measure of overall severity of ASD-related symptoms based on 15 items.   Ratings are based not only on frequency of the behavior in question, but also on its intensity, atypicality, and duration.}\\[0.5in]
\hline && \\[-.1in]
$\!\!$Tsc\_lc$\!\!\!\!\!\!$ & \parbox{.275\textwidth}{Oral and Written Language Scales (OWLS-II), Listening Comprehension (LC) subtest} & \parbox{.45\textwidth}{Measures oral language reception, or understanding of spoken language.  Examiner orally presents increasingly difficult words, phrases, and sentences; patient responds by pointing to or stating which of four picture choices is correct.  }\\[0.4in]
\hline && \\[-.1in]
$\!\!$Tsc\_oe$\!\!\!\!\!\!$ & \parbox{.275\textwidth}{Oral and Written Language Scales (OWLS-II), Oral Expression (OE) subtest} & \parbox{.45\textwidth}{Measures oral language expression, or use of spoken language.  Examiner presents a verbal prompt along with a picture and patient must respond orally to the prompt with increasingly difficult language.}\\[0.3in]
\hline && \\[-.1in]
$\!\!$CompTsc\_ol$\!\!\!\!\!\!$ & \parbox{.275\textwidth}{Oral and Written Language Scales (OWLS-II), Oral Language Composite} & \parbox{.45\textwidth}{Represents an overall level of oral language functioning.  Derived from the Listening Comprehension and Oral Expression scales.  }\\[0.2in]
\hline && \\[-.1in]
$\!\!$scq\_raw\_total$\!\!\!\!\!\!$ & \parbox{.275\textwidth}{Social Communication Questionnaire (Lifetime Version)} & \parbox{.45\textwidth}{40-item yes/no questionnaire; many items focus on the presence of symptoms during the period between the individual's 4th and 5th birthdays.  Scores are raw scores (no standardized scores available). Designed to assess for qualitative impairments in reciprocal social interaction and communication, as well as restricted, repetitive, and stereotyped behavior}\\[0.5in]
\hline && \\[-.1in]
$\!\!$T\_RRB$\!\!\!\!\!\!$ & \parbox{.275\textwidth}{Social Responsiveness Scale (SRS) Restricted Interests and Repetitive Behavior T-score} & \parbox{.45\textwidth}{Items assess restricted range of interests and activities, inflexibility, unusual sensory interests, perseveration on topics, atypicality (bizarre behavior, being regarded as odd by peers) as well as motor stereotypy. }\\[0.3in]
\hline && \\[-.1in]
$\!\!$T\_Score$\!\!\!\!\!\!$ & \parbox{.275\textwidth}{Social Responsiveness Scale  (SRS) Total T-score} & \parbox{.45\textwidth}{Reflects the sum of responses to all 65 SRS questions (including the SCI and RRB subscales).  Serves as an index of reciprocal social behavior across typical development, ASD, and other disorders.  A good single number rating of severity of ASD symptoms.  }\\[0.4in]
\hline && \\[-.1in]
$\!\!$T\_SCI$\!\!\!\!\!\!$ & \parbox{.275\textwidth}{Social Responsiveness Scale (SRS) Social Communication and Interaction (SCI) T-score} & \parbox{.45\textwidth}{Reflects 4 subscales: Social Awareness, Social Cognition, Social Communication, Social Motivation}\\[0.2in]
\hline && \\[-.1in]
$\!\!$wasi\_iq\_composite$\!\!\!\!\!\!$ & \parbox{.275\textwidth}{Wechsler Abbreviated Scale of Intelligence (WASI-II)} & \parbox{.45\textwidth}{IQ composite score based on Vocabulary, Similarities, Block Design, Matrix Reasoning subtests. }\\[0.2in]
\hline && \\[-.1in]
$\!\!$WJ\_Basic\_Read\_Skills\_z\_Score$\!\!\!\!\!\!$ & \parbox{.275\textwidth}{Woodcock-Johnson Test of Achievement, Basic Reading cluster.} & \parbox{.45\textwidth}{Measures sight vocabulary and the ability to apply phonic and structural analysis skills.  Combination of Letter-Word Identification and Word Attack.  }\\[0.3in]
\hline && \\[-.1in]
$\!\!$WJ\_Pass\_Comprehen\_z\_Score$\!\!\!\!\!\!$ & \parbox{.275\textwidth}{Woodcock-Johnson Test of Achievement, } & \parbox{.45\textwidth}{Reading comprehension.  Measures understanding of written text. The majority of items require a student to supply a missing word to sentences and then paragraphs of increasing complexity.}\\[0.3in]
\hline && \\[-.1in]
$\!\!$WJ\_Word\_Attack\_z\_Score$\!\!\!\!\!\!$ & \parbox{.275\textwidth}{Woodcock-Johnson Test of Achievement, Word Attack subtest} & \parbox{.45\textwidth}{Measures ability to apply phonic/decoding skills to unfamiliar words.   The majority of items require students to pronounce nonsense words of increasing complexity. }\\[0.3in]
\hline && \\[-.1in]
$\!\!$wraml\_Verbal\_Memory\_Index\_Sum$\!\!\!\!\!\!$ & \parbox{.275\textwidth}{Wide Range Assessment of Memory and Learning (WRAML-2)} & \parbox{.45\textwidth}{Measures ability to learn and recall both meaningful verbal information and relatively rote verbal information.  Derived from the sum of the Story Memory and Verbal Learning subtests.}\\[0.3in]
\hline && \\[-.1in]
$\!\!$wrat\_spelling\_standard$\!\!\!\!\!\!$ & \parbox{.275\textwidth}{Wide Range Achievement Test (WRAT4), Spelling subtest} & \parbox{.45\textwidth}{Measures ability to identify sounds and transfer them into written form from dictated words.  Standard spelling test - word is stated, used in a sentence, and repeated and patient writes it.}\\[0.3in]
\hline && \\[-.1in]
$\!\!$wrat\_math\_standard$\!\!\!\!\!\!$ & \parbox{.275\textwidth}{Wide Range Achievement Test (WRAT4), Math Computation subtest} & \parbox{.45\textwidth}{Measures ability to count, identify numbers, solve simple oral math problems, and calculate written math problems.  Problems are presented in a range of domains, including arithmetic, algebra, geometry, and advanced operations.  }\\[0.4in]
\hline && \\[-.1in]
$\!\!$ach\_abc\_AnxDep$\!\!\!\!\!\!$ & \parbox{.275\textwidth}{Achenbach Assessment of Empirically Based Assessment (ASEBA) - Adult Behavior Checklist (ABCL) or Adult Self-Report (ASR) - Anxious/Depressed scale } & \parbox{.45\textwidth}{Measures behaviors such as nervousness, worrying, fearfulness, loneliness, sadness, feeling worthless, feeling too guilty, feeling persecuted, lacking self-confidence.}\\[0.3in]
\hline && \\[-.1in]
$\!\!$ach\_abc\_Withdrawn$\!\!\!\!\!\!$ & \parbox{.275\textwidth}{ASEBA - Adult Behavior Checklist (ABCL) or Adult Self-Report (ASR) - Withdrawn scale} & \parbox{.45\textwidth}{Measures behaviors such as poor relationships, not getting along with others, preferring to be alone, anhedonia, being secretive.  }\\[0.2in]
\hline && \\[-.1in]
$\!\!$ach\_abc\_Somatic$\!\!\!\!\!\!$ & \parbox{.275\textwidth}{ASEBA - Adult Behavior Checklist (ABCL) or Adult Self-Report (ASR) - Somatic Complaints scale} & \parbox{.45\textwidth}{Measures complaints of discomfort or illness.  }\\[0.1in]
\hline && \\[-.1in]
$\!\!$ach\_abc\_Thought$\!\!\!\!\!\!$ & \parbox{.275\textwidth}{ASEBA - Adult Behavior Checklist (ABCL) or Adult Self-Report (ASR) - Thought Problems scale} & \parbox{.45\textwidth}{Measures symptoms such as hallucinations, obsessions, compulsions, strange thoughts and behaviors, self-harm, and suicide attempts.}\\[0.2in]
\hline && \\[-.1in]
$\!\!$ach\_abc\_Attention$\!\!\!\!\!\!$ & \parbox{.275\textwidth}{ASEBA - Adult Behavior Checklist (ABCL) or Adult Self-Report (ASR) - Attention Problems scale} & \parbox{.45\textwidth}{Measures attention problems, forgetfulness, daydreaming, failing to finish things, avoiding work, disorganization, lateness, difficulty planning and prioritizing.  }\\[0.3in]
\hline && \\[-.1in]
$\!\!$ach\_abc\_Agressive$\!\!\!\!\!\!$ & \parbox{.275\textwidth}{ASEBA - Adult Behavior Checklist (ABCL) or Adult Self-Report (ASR) - Aggressive Behavior scale} & \parbox{.45\textwidth}{Measures behaviors such as meanness, arguing, threatening, blaming others, fighting, temper outbursts, screaming, sulking.}\\[0.2in]
\hline && \\[-.1in]
$\!\!$ach\_abc\_RuleBreak$\!\!\!\!\!\!$ & \parbox{.275\textwidth}{ASEBA - Adult Behavior Checklist (ABCL) or Adult Self-Report (ASR) - Rule-Breaking Behavior scale} & \parbox{.45\textwidth}{Measures behaviors such as irresponsibility, substance abuse, lacking feelings of guilt, lying or cheating, stealing, difficulty keeping a job.  }\\[0.2in]
\hline && \\[-.1in]
$\!\!$ach\_abc\_Intrusive$\!\!\!\!\!\!$ & \parbox{.275\textwidth}{ASEBA - Adult Behavior Checklist (ABCL) or Adult Self-Report (ASR) - Intrusive scale} & \parbox{.45\textwidth}{Measures behaviors such as bragging, showing off, attention-seeking, being boisterous, teasing.}\\[0.2in]
\hline
  \end{longtable}
}

%\clearpage

\renewcommand{\arraystretch}{.7} 

\vspace{.0in}\section{Marginalized Likelihood}
\vspace{.0in}
\label{sec:like_deriv}

The derivation of Result~3 in the main paper is provided below.  Let the component parameters be denoted as $\theta = \{\bmu_{11}, \dots, \bmu_{M1}, \bSigma_{111}, \dots, \bSigma_{M11}, \bb_2, \bQ_{21}, \bQ_{22} \} $.  We wish to obtain a closed form result for,
\vspace{-.15in}\bdm
f(\bZ \mid \bgamma, \bphi) = \int f(\bZ \mid \bgamma, \bphi, \theta) f(\theta \mid \bgamma) d\theta.
\vspace{-.05in}\edm
However, after a little bit of algebra we have,
\vspace{-.15in}\begin{eqnarray}
f(\bZ \mid \bgamma, \bphi, \theta) &=& 
\prod_{m=1}^M \prod_{\{i:\phi_i=m\}} \frac{1}{(2\pi)^{p/2}} \left| \bSigma_m \right|^{-1/2} \exp\left\{  -\frac{1}{2} (\bz_i - \bmu_m)' \bSigma_m^{-1}(\bz_i - \bmu_m)  \right\} \nonumber \\
&=& \left[ \prod_{m=1}^M A_m \right] B, \nonumber \\[-.38in] \nonumber
\end{eqnarray}
where $M=\max_i \{\phi_i\}$, and 
\vspace{-.1in}\bdm
\small
A_m = (2\pi)^{-\frac{n_mp_1}{2}} \left| \bSigma_{m11} \right|^{-\frac{n_m}{2}} \exp\left\{  \!-\frac{1}{2} \sum_{i:\phi_i=m}(\bz^{(1)}_{i} \!-\! \bmu_{m1})' \bSigma_{m11}^{-1}(\bz^{(1)}_{i} \!-\! \bmu_{m1})  \right\} \mbox{, and}\;\;\;\;\;\;\;\;\;\;\;\;\;\;\;\;\;\;\;\;\;\;\;\;\;\;\;
\edm
\vspace{-.2in}\bdm
\small
\!B = (2\pi)^{-\frac{np_2}{2}} \!\left| \bQ_{22} \right|^{\frac{n}{2}}  \!\exp\!\left\{ \! -\frac{1}{2} \sum_{i=1}^n\!\left[ {\bz^{(2)}_{i}}'\bQ_{22}\bz^{(2)}_{i} \!-\! 2{\bz^{(2)}_{i}}'(\bb_2 \!-\! \bQ_{21}\bz^{(1)}_{i}) \!+\!  (\bb_2 - \bQ_{21}\bz^{(1)}_{i})'\bQ_{22} (\bb_2 \!-\! \bQ_{21}\bz^{(1)}_{i})\right] \!\right\}\!. \nonumber
\edm
Combining this with the prior independence of $(\bmu_{m1}, \bSigma_{m11})$, $m=1,\dots...$ and $(\bb_{2}, \bQ_{21}, \bQ_{22})$ we have,
\vspace{-.1in}\bdm
\int f(\bZ \mid \bgamma, \bphi, \theta) f(\theta \mid \bgamma) d\theta\;\;\;\;\;\;\;\;\;\;\;\;\;\;\;\;\;\;\;\;\;\;\;\;\;\;\;\;\;\;\;\;\;\;\;\;\;\;\;\;\;\;\;\;\;\;\;\;\;\;\;\;\;\;\;\;\;\;\;\;\;\;\;\;\;\;\;\;\;\;\;\;\;\;\;\;\;\;\;\;\;\;\;\;\;\;\;\;\;\;\;\;\;\;\;\;\;\;\;\;\;\;\;\;\;\;\;\;\;\;\;\;
\edm
\vspace{-.35in}\beq
\;\;\;\;\;\;\;\;= \left[\prod_{m=1}^M \int A_m f(\bmu_{m1}, \bSigma_{m11}) d(\bmu_{m1}, \bSigma_{m11}) \right]\!
\int \!B f(\bb_{2}, \bQ_{21}, \bQ_{22}) d(\bb_{2}, \bQ_{21}, \bQ_{22}).\!
\label{eq:marg_like_1}
\vspace{.0in}\eeq
Now
\bdm
f(\bmu_{m1}, \bSigma_{m11}) = f(\bmu_{m1} \mid \bSigma_{m11}) f(\bSigma_{m11}),
\edm
with
\vspace{-.2in}\begin{eqnarray}
f(\bmu_{m1} \mid \bSigma_{m11}) &\!=\! & (2\pi)^{-\frac{p_1}{2}} \left| \frac{1}{\lambda}\bSigma_{m11} \right|^{-\frac{1}{2}} \exp \left\{ -\frac{\lambda}{2} \bmu_{m1}'\bSigma_{m11}^{-1}\bmu_{m1}\right\} \nonumber \\
f(\bSigma_{m11}) &\!=\! &
\frac{\left| \bPsi_{11} \right|^{-\frac{\eta-p_2}{2}} \left| \bSigma_{m11} \right|^{-\frac{\eta-p_2+p_1+1}{2}} }
     {2^\frac{(\eta-p_2)p_1}{2} \Gamma_{p_1}(\frac{\eta-p_2}{2})}
\exp \left\{ -\frac{1}{2} \mbox{tr}\left(\bPsi_{11} \bSigma_{m11}^{-1}\right) \right\}.  \nonumber
\end{eqnarray}
After some tedious algebra,
\vspace{-.2in}\bdm
A_m f(\bmu_{m1}, \bSigma_{m11}) = A_m^{(1)} A_m^{(2)} A_m^{(3)},
\vspace{-.3in}\edm
with
\vspace{-.15in}\begin{eqnarray}
A_m^{(1)} &\!\! = \!\!& (2\pi)^{-\frac{n_mp_1}{2}} \left( \frac{\lambda}{n_m+\lambda}\right)^{\frac{p_1}{2}}
\frac{\left| \bPsi_{11} \right|^{\frac{\eta-p_2}{2}}  \Gamma_{p_1}(\frac{n_m+\eta-p_2}{2})}
     {\left| \bV_{m11} \right|^{\frac{n_m+\eta-p_2}{2}}  \Gamma_{p_1}(\frac{\eta-p_2}{2})}, \nonumber \\
     A_m^{(2)} & \!\!= \!\!&  (2\pi)^{-\frac{p_1}{2}}  \!\left| \frac{1}{n_m\!+\!\lambda}\bSigma_{m11} \right|^{-\frac{1}{2}}\!\!
     \exp \left\{-\frac{n_m\!+\!\lambda}{2} \left(\bmu_{m1}\!-\!\frac{n_m}{n_m\!+\!\lambda}\bar{\bz}_{m1} \right)' \!\!\bSigma_{m11}^{-1}\!\!\left(\bmu_{m1}\!-\!\frac{n_m}{n_m\!+\!\lambda}\bar{\bz}_{m1} \right) \right\}\nonumber \\
A_m^{(3)} & \!\!= \!\!& \frac{\left| \bV_{m11} \right|^{\frac{n_m+\eta-p_2}{2}}  \left| \bSigma_{m11} \right|^{-\frac{n_m+\eta-p_2+p_1+1}{2}}}
{2^{\frac{(n_m+\eta-p_2)p_1}{2}} \Gamma_{p_1}(\frac{n_m+\eta-p_2}{2})} \exp\left\{ -\frac{1}{2} \mbox{tr}\left( \bV_{m11} \bSigma_{m11}^{-1}\right) \right\}, \nonumber
     \nonumber
\end{eqnarray}
where $\bV_{m11}$ and $\bar{\bz}_{m1}$ are as defined in Result~3 of the main paper.
As a function of $\bmu_{m1}$, we recognize $A_m^{(2)}$ to be the multivariate normal density with mean $\frac{n_m}{n_m\!+\!\lambda}\bar{\bz}_{m1}$ and covariance $\frac{1}{n_m+\lambda}\bSigma_{m11}$.  Also, as a function of $\bSigma_{m11}$ we recognize $A_m^{(3)}$ to be the density of an inverse-Wishart distribution with parameters $\eta^* = n_m+\eta-p_2$ and $\bPsi^* = \bV_{m11}$.  Thus,
\beq
\mbox{\white .}\!\!\!\!\!\!\!\!\!\!\!\!\int \!\!A_m f(\bmu_{m1}, \bSigma_{m11}) d(\bmu_{m1}, \bSigma_{m11}) = (\;\!\!2\pi\;\!\!)^{-\frac{n_m p_1}{2}} 
%\left[\!
 \left( \!\frac{\lambda}{n_m\!+\!\lambda}\!\right)^{\!\!\!\frac{p_1}{2}}\!\!
\frac{\left| \bPsi_{11} \right|^{\frac{\eta-p_2}{2}}  \Gamma_{\!p_1\!}(\frac{n_m\!+\eta-p_2}{2})}
     {\!\left| \bV_{\!m11} \right|^{\frac{n_m+\eta-p_2}{2}}  \Gamma_{\!p_1\!}(\frac{\eta-p_2}{2})}.
%     \right]\!.
\label{eq:marg_like_2}\!\!
\eeq

Now the prior distribution corresponding to the second term in (\ref{eq:marg_like_1}) is 
\vspace{-.1in}\bdm
f(\bb_{2}, \bQ_{21}, \bQ_{22}) = f(\bb_{2} \mid \bQ_{22})f(\bQ_{21} \mid \bQ_{22})f(\bQ_{22}),
\vspace{-.15in}\edm
where
{\small
\vspace{-.1in}  \begin{eqnarray}
  f(\bb_{2} \mid \bQ_{22}) &\!\!\!\!=\!\!\!\!& (2\pi)^{-\frac{p_2}{2}} \left| \frac{1}{\lambda}\bQ_{22} \right|^{-\frac{1}{2}} \exp \left\{ -\frac{\lambda}{2} \bb_2'\bQ_{2}^{-1}\bb_{2}\right\}  \nonumber \\
\mbox{\white{.}}\!\!\!\!\!\!\!f(\bQ_{21} \!\mid\! \bQ_{22}) &\!\!\!\!=\!\!\!\!& (2\pi)^{-\frac{p_1p_2}{2}} \!\left| \bPsi_{\!11} \right|^{\frac{p_2}{2}} \!\left| \bQ_{22} \right|^{-\frac{p_1}{2}}\! \exp \!\left\{ \!-\frac{1}{2} \mbox{tr}\!\left[ \bPsi_{\!11} \!\left( \bQ_{21} \!+\! \bQ_{22}\bPsi_{\!21} \bPsi_{11}^{-1} \right)'\!\! \bQ_{22}^{-1}\!\! \left( \bQ_{21} \!+\! \bQ_{22}\bPsi_{21} \bPsi_{11}^{-1} \right) \right]\!\right\}  \nonumber \\
f(\bQ_{22}) &\!\!\!\!=\!\!\!\!&
\frac{\left| \bPsi_{22 \mid 1} \right|^{\frac{\eta}{2}} \left| \bQ_{22} \right|^{\frac{\eta+p_2+1}{2}} }
     {2^\frac{\eta p_2}{2} \Gamma_{p_2}(\frac{\eta}{2})}
\exp \left\{ -\frac{1}{2} \mbox{tr}\left(\bPsi_{22 \mid 1} \bQ_{22}\right) \right\},  \nonumber \\[-.37in] \nonumber
  \end{eqnarray}
  }
where $\bPsi_{22 \mid 1} = \bPsi_{22} - \bPsi_{21}\bPsi_{11}\bPsi_{12}$.  After some more tedious algebra,
\vspace{-.1in}\bdm
B f(\bb_{2}, \bQ_{21}, \bQ_{22}) = B^{(1)} B^{(2)} B^{(3)} B^{(4)},
\edm
with
\begin{eqnarray}
B^{(1)} &\!\!\! \!\!=\!\!\!\!& (2\pi)^{-\frac{np_2}{2}} \left( \frac{\lambda}{n+\lambda}\right)^{\frac{p_2}{2}}
\frac{\left| \bPsi_{11} \right|^{\frac{\eta-p_2}{2}}  \left| \bPsi_{22 \mid 1} \right|^{\frac{\eta}{2}}  \Gamma_{p_2}(\frac{n+\eta}{2})}
     {\left| \bV_{\!11} \right|^{\frac{p_2}{2}}  \left| \bV_{\!22 \mid 1} \right|^{\frac{n+\eta}{2}}  \Gamma_{p_2}(\frac{\eta}{2})}, \nonumber \\
B^{(2)} & \!\!\!\!\!=\!\! \!\!&  (2\pi)^{-\frac{p_2}{2}}  \!\left| \frac{1}{n\!+\!\lambda}\bQ_{22} \right|^{-\frac{1}{2}}\!\!
\exp \left\{-\frac{1}{2} \left[\bb_{2}'\bQ_{22}^*\bb_2 - 2\bb_2'\bb^* + {\bb^*}'{\bQ^*}^{-1}\bb^* \right] \right\}  \nonumber \\
B^{(3)} & \!\!\!\!=\! \!\!& (2\pi)^{-\frac{p_1p_2}{2}}\! \left| \bV_{\!\!11} \right|^{-\frac{p_2}{2}} \!\left| \bQ_{22} \right|^{-\frac{p_1}{2}} \!\exp \!\left\{ \!-\frac{1}{2} \mbox{tr}\!\left[ \bV_{\!\!11} \!\left( \bQ_{21} \!\!+\! \bQ_{22}\bV_{\!\!21} \bV_{\!11}^{-1} \right)' \!\!\bQ_{22}^{-1} \!\!\left( \bQ_{21} \!\!+\! \bQ_{22}\bV_{\!\!21} \bV_{\!11}^{-1} \right)\right]\!\right\} 
\nonumber \\
B^{(4)} & \!\!\!\!\!=\!\! \!\!& \frac{\left| \bV_{\!\!22 \mid 1} \right|^{-\frac{n+\eta}{2}}  \left| \bQ_{22} \right|^{-\frac{n+\eta+p_2+1}{2}}}
{2^{\frac{(n+\eta)p_2}{2}} \Gamma_{p_2}(\frac{n+\eta}{2})} \exp\left\{ -\frac{1}{2} \mbox{tr}\left( \bV_{\!\!22\mid 1} \bQ_{22}\right) \right\}, \nonumber
%     \nonumber
\end{eqnarray}
where $\bb^* = n(\bar{\by}_2 + \bQ_{22}^{-1} \bQ_{21} \bar{y}_1)$ and $\bQ^* =(n\!+\!\lambda)\bQ_{22}^{-1}$, and $\bV_{11}$, $\bV_{22 \mid 1}$ are as defined in Result~3 of the main paper.

As a function of $\bb_2$, we recognize $B^{(2)}$ to be the multivariate normal density in canonical form with precision $\bQ^*$ and mean ${\bQ^*}^{-1}\bb^*$.  Also, as a function of $\bQ_{21}$ we recognize $B^{(3)}$ to be the density of a $\cM\cN(\bM^*, \bU^*, \bV^*)$ with parameters $\bM^* = -\bQ_{22}\bV_{\!\!21} \bV_{\!11}^{-1}$, $\bU^*=\bQ_{22}$, and $\bV^* = \bV_{\!\!11}^{-1}$.  Finally, we can recognize $B^{(4)}$ to be the density of $\bQ_{22}$: a Wishart distribution with parameters $\eta^* = n+\eta$ and $\bPsi^* = \bV_{\!22 \mid 1}$.  Thus,
\beq
\!\!\!\int \!B f(\bb_{2}, \bQ_{21}, \bQ_{22}) d(\bb_{2}, \bQ_{21}, \bQ_{22})= (2\pi)^{-\frac{np_2}{2}}\! \left( \!\frac{\lambda}{n+\lambda}\!\right)^{\!\!\frac{p_2}{2}}
\frac{\left| \bPsi_{\!11} \right|^{\!\!\frac{\eta-p_2}{2}}  \!\left| \bPsi_{\!22 \mid 1} \right|^{\frac{\eta}{2}}  \Gamma_{\!p_2}(\frac{n+\eta}{2})}
     {\left| \bV_{\!11} \right|^{\frac{p_2}{2}}  \left| \bV_{\!22 \mid 1} \right|^{\frac{n+\eta}{2}}  \Gamma_{\!p_2}(\frac{\eta}{2})}.\!\!
\label{eq:marg_like_3}
\eeq
Combining (\ref{eq:marg_like_1}), (\ref{eq:marg_like_2}), and (\ref{eq:marg_like_3}) gives the desired result.

\section{MCMC Algorithm}
\vspace{-.0in}
\label{sec:MCMC}

This section describes the MCMC sampling scheme for the full model described in Section~\ref{sec:model_descr} of the main paper.  Since the component parameters are integrated out, the entire collection of parameters to be sampled in the MCMC is
\beq
\Theta = \left\{
\bgamma,
\bphi,
\lambda,
\eta,
\bPsi,
\alpha,
\tilde{\bZ}
\right\},
\label{eq:param_set}
\eeq
where $\tilde{\bZ}$ contains any latent element of $\bZ$ (i.e., are either missing data, or correspond to a discrete variable or censored observation).

The MCMC algorithm proceeds by performing Metropolis Hastings (MH) updates for each of the elements listed in $\Theta$ in a Gibbs fashion.  The only update that depends on the raw observed data $\bY=[\by_1',\dots,\by_n']'$ is the update of $\tilde{\bZ}$.  All other parameters, conditional on $\bY$ and $\bZ$, only depend on $\bZ$.  Therefore, $\bY$ does not appear in the notation of any of the updates below, except that for $\tilde{\bZ}$.  The $\bgamma$ vector is updated with add/delete/swap proposals.  The $\tilde{\bZ}$ and $\bPsi$ are high dimensional and thus some creativity is needed to ensure good proposals.  To accomplish this we first sample some component parameters from their conjugate distribution given the other parameters (for a fixed $\gamma$ this is not difficult), and then use these component parameters to obtain good proposals for $\tilde{\bZ}$ and $\bPsi$, respectively.  As mentioned in the main paper the $\phi_i$ can be updated individually with simple Gibbs sampling.  However, this approach has known mixing issues and, thus, a modified split-merge algorithm \citep{Jain04} will be described below.  The remaining updates for $\lambda, \eta, \alpha$ are more straight-forward random walk MH updates.\\[-.1in]

\noindent
{\underline{MH update for $\bgamma$}}

\noindent
The $\bgamma$ vector is updated with MH by proposing an add, delete, or swap move. That is, the proposal $\bgamma^*$ is generated as follows.
\begin{itemize}
\item[(i)] Set the proposal $\bgamma^*=\bgamma$
\item[(ii)] Randomly choose an integer $j^*$ from $1,\dots,p$.
\item[(iii)] Flip the value of $\gamma_{j^*}$, i.e., $\gamma_{j^*}^*=1-\gamma_{j^*}$.
\item[(iv)] If the set $\{j:\gamma_j \neq \gamma_{j^*}\}$ is not empty, draw a Bernoulli $B^*$ with probability $\pi$.
\item[(v)] If $B^*=1$ randomly choose another $j^{**}$ from the set $\{j:\gamma_j \neq \gamma_{j^*}\}$ and also set $\gamma^*_{j^{**}}=1-\gamma_{j^{**}}$, i.e., a swap proposal.  If $B^*=0$, leave $\bgamma^*$ as a single variable add/delete proposal.
\end{itemize}
Let $d(\bgamma^* \mid \bgamma)$ represent the density of this proposal.  The MH ratio is then
\bdm
MH = \frac{f(\bZ \mid \bgamma^*, \bphi, \lambda, \eta, \bPsi, \alpha) f(\bgamma^*) d(\bgamma \mid \bgamma^*)}
{f(\bZ \mid \bgamma, \bphi, \lambda, \eta, \bPsi, \alpha) f(\bgamma) d(\bgamma^* \mid \bgamma)},
\edm
where $f(\bZ \mid \bgamma, \bphi, \lambda, \eta, \bPsi, \alpha)$ is the marginal likelihood provided in Result~3 and $f(\bgamma)$ is the prior distribution for $\bgamma$, i.e., independent Bernoulli$(\rho)$.  In the results of the main paper, $\rho$ was set to 0.5.
\\[-.1in]

\noindent
{\underline{MH update for $\alpha$}}

\noindent
The update for $\alpha$ was conducted via a MH random walk proposal on log scale. Draw a proposal $\mbox{log}(\alpha^*) = \mbox{log}(\alpha)+\epsilon$ for a deviate $\epsilon {\sim} N(0,s^2)$.
The tuning parameter was set to $s=1$ to achieve an acceptance rate $\approx$ 40\%, and resulted in good mixing.  Let the density of the proposal, given the current value of $\alpha$ be denoted $d(\alpha^* \mid \alpha)$.  The only portion of the posterior that differs between the current value and the proposal is in the term
\bdm
f(\bphi \mid \alpha) = \prod_{i=2}^n \frac{n_{i,\phi_i} I_{\{n_{i,\phi_i} >0\}} + \alpha I_{\{n_{i,\phi_i} =0\}}}{i-1+\alpha} \propto \frac{\alpha^M \Gamma(\alpha)}{\Gamma(\alpha+n)}.
\edm
The MH ratio is then
\bdm
MH = \frac{f(\bphi \mid \alpha^*) f(\alpha^*) d(\alpha \mid \alpha^*)}
{f(\bphi \mid \alpha) f(\alpha) d(\alpha^* \mid \alpha)},
\edm
where $d(\alpha)$ is the density for a Gamma$(A_\alpha, B_\alpha)$ random variable.
\\[-.1in]

\noindent
{\underline{MH update for $\lambda$}}

\noindent
The update for $\lambda$ was conducted via a MH random walk proposal on log scale. Draw a proposal $\mbox{log}(\lambda^*) = \mbox{log}(\lambda)+\epsilon$ for a deviate $\epsilon {\sim} N(0,s^2)$.
The tuning parameter was set to $s=0.5$ to achieve an acceptance rate $\approx$ 40\%.  Let the density of the proposal, given the current value of $\lambda$ be denoted $d(\lambda^* \mid \lambda)$. The MH ratio is then
\bdm
MH = \frac{f(\bZ \mid \bgamma, \bphi, \lambda^*, \eta, \bPsi, \alpha) f(\lambda^*) d(\lambda \mid \lambda^*)}
{f(\bZ \mid \bgamma, \bphi, \lambda, \eta, \bPsi, \alpha) f(\lambda) d(\lambda^* \mid \lambda)},
\edm
where $f(\lambda)$ is the density for a Gamma$(A_\lambda, B_\lambda)$ random variable.
\\[-.1in]

\noindent
{\underline{MH update for $\eta$}}

\noindent
The update for $\eta$ is entirely analogous to that for $\lambda$.  A tuning parameter of $s=1$ was used for $\eta$ updates to encourage $\approx$ 40\% acceptance. 
\\[-.1in]

\noindent
{\underline{MH update for $\bPsi$}}

\noindent
The prior distribution is $\bPsi \sim \cW(\bP,N)$.  If the component parameters $\theta = \{\bmu_{11}, \dots, \bmu_{M1}$, $\bSigma_{111}$, $\dots$, $\bSigma_{M11}$, $\bb_2$, $\bQ_{21}$,$\bQ_{22} \}$, were given then $\bPsi$ would have a conjugate update of the form,
\begin{eqnarray}
\bPsi_{22 \mid 1} \mid \theta &\sim& \cW(\bQ_{22}+\bP_{22}, N+\eta), \nonumber \\
\bPsi_{11} \mid \theta &\sim& \cW(\bP^*,M\eta+N+p_2), \label{eq:Psi_conj} \\
\bPsi_{21} \mid \theta, \bPsi_{11} &\sim& \cM\cN(-(\bP_{22}+\bQ_{22})(\bP_{21}+\bQ_{21})\bPsi_{11}\;,\; (\bP_{22}+\bQ_{22})^{-1}, \bPsi_{11}), \nonumber
\end{eqnarray}
where
\bdm
\!\bP^* \!= \!\left[\bP_{11}^{-1}\!\! +\! (\bP_{\!21} \!+\! \bQ_{\!21})'(\bP_{\!22}\!+\!\bQ_{\!22})^{-1}(\bP_{\!21} \!+\! \bQ_{\!21}) \!+\! \bP_{\!21}'\bP_{\!22}^{-1}\bP_{\!21} \!+\! \bQ_{\!21}'\bQ_{\!22}^{-1}\bQ_{\!21} \!+\! \sum_{m=1}^M \bSigma_{m11}^{-1} \!\right]^{-1}\!\!,
\edm
and $\bPsi_{22 \mid 1}$ is independent of $\bPsi_{11}, \bPsi_{21}$ given $\theta$.

We do not sample $\theta$, so $\bPsi$ does not have such a conjugate update in the MCMC routine.  However, we can generate a very good proposal $\Psi^*$ in the following manner.  Conditional on the current values of $\bgamma, \bphi, \lambda, \eta, \alpha, \tilde{\bZ}$, and $\bPsi$, one could draw component parameters $\theta = \{\bmu_{11}, \dots, \bmu_{M1}, \bSigma_{111}, \dots, \bSigma_{M11}, \bb_2, \bQ_{21}, \bQ_{22} \}$, from their conjugate distribution provided in Section~\ref{sec:like_deriv}.  Conditional on the value of $\theta$ we could then draw from the distribution of $\bPsi \mid \theta$ provided above.  However, we do not want to have the current $\Psi$ value involved in the update as this complicates the proposal density calculation.  A simple fix is to draw component parameters $\theta^*$ conditional on the current values of $\bgamma, \bphi, \lambda, \eta, \alpha, \tilde{\bZ}$, but with $\bPsi$ fixed at some value $\tilde{\bPsi}$, independent of the current (or previous) values in the chain.  This way, the proposal density for $\bPsi^*$ is selected at random from a set of possible proposal distributions.  The proposal density $d(\bPsi^* \mid \bPsi)=d(\bPsi^*)$ is then conditional on $\theta^*$ and is simply the product of the densities in (\ref{eq:Psi_conj}).  This is allowable under the same principle used by \cite{Jain04} for the split-merge algorithm.  Thus the MH ratio is then
\bdm
MH = \frac{f(\bZ \mid \bgamma, \bphi, \lambda, \eta, \bPsi^*, \alpha) f(\bPsi^*) d(\bPsi)}
{f(\bZ \mid \bgamma, \bphi, \lambda, \eta, \bPsi, \alpha) f(\bPsi) d(\bPsi^*)},
\edm
where $f(\bPsi)$ is the density for a $\cW(P, N)$ random variable. Note that this update would be made much easier if $\theta$ were just sampled in the MCMC as well.  However, this makes it very difficult to update $\bgamma$ since the dimension of $\theta$ will be changing with $\bgamma$.  Reversible jump (RJ) MCMC could be used to overcome this issue by updating $\bgamma, \theta$ jointly, but this comes with its own challenges.  For a given $\bgamma$ and $\bphi$, however, drawing a $\theta$ to determine the proposal distribution as above poses no issues.
\\[-.1in]

\noindent
{\underline{MH update for $\tilde{\bZ}$}}

\noindent
The same logic used in the update of $\bPsi$ is used here as well.  Conditional on the component parameters $\theta$, the elements of $\tilde{\bZ}$ have simple Gibbs updates.  Namely, for a given observation $i$ with missing values, the update for the elements of $\bz_i$ that correspond to missing data elements in $\by_i$ would be to draw from the normal distribution specified by $\phi_i$ and $\theta$, conditional on the observed variables in $\bz_i$.  If a value $y_{ij}$ is discrete or censored, then the update for $z_{ij}$ would be to draw from the normal distribution specified by $\phi_i$ and $\theta$, conditional on the other variables in $\bz_i$ {\bf and the conditional limits imposed by $y_{ij}$}.  Thus, a very similar trick as above is used.  First divide $\tilde{\bZ}$ up into $K$ partitions and denote them $\tilde{\bZ}_1,\dots,\tilde{\bZ}_K$.  We draw a separate $\theta_k^*$ for each partition by conditioning on the current values of $\bgamma, \bphi, \lambda, \eta, \alpha, \bPsi$, and all $\tilde{\bZ}$ values {\bf except} $\tilde{\bZ}_k$.  Update each of the elements of $\tilde{\bZ}_k$ conditional on $\theta_k^*$ as described above to produce a proposal $\tilde{\bZ}_k^*$.  Denote the density of this proposal as $d(\tilde{\bZ}_k^* \mid \tilde{\bZ}_k)= d(\tilde{\bZ}_k^*)$. The MH ratio is then,
\bdm
MH = \frac{f(\bY \mid \bZ^*) f(\bZ^* \mid \bgamma, \bphi, \lambda, \eta, \bPsi, \alpha) d(\tilde{\bZ}_k)}
{f(\bY \mid \bZ) f(\bZ \mid \bgamma, \bphi, \lambda, \eta, \bPsi, \alpha, \bY) d(\tilde{\bZ}_k^*)}.
\edm
The likelihood of the raw observed data conditional on the latent variables $f(\bY \mid \bZ)$ is either 1 or 0, depending on whether or not the $z_{ij}$ corresponding to discrete or censored $y_{ij}$ is consistent with the conditional limits imposed by $y_{ij}$, or not.  With the proposal strategy discussed above this will always be the case, but it is still denoted in the $MH$ above for completeness.  
\\[-.1in]

\noindent
{\underline{MH update for $\bphi$}}

\noindent
This is the most complex of the parameter updates as it uses a less standard split-merge MH approach \citep{Jain04} as this improves mixing dramatically over one-at-a-time Gibbs updates for the $\bphi_i$.  However, the split-merge update does make use of the individual Gibbs update for the $\phi_i$.  This is provided as
\vspace{-.1in}\bdm
  \!\!\!\!\!\!\!\!\!\!\!\!\!\!\!\!\!\!\!\!\!\!\!\!\!\!\!\Pr (\phi_i \!=\! m \!\mid \!\mbox{\em rest})  \; \propto \; f(\bZ \mid \bgamma, \bphi, \lambda, \eta, \bPsi, \alpha) f(\phi_i \mid \bphi_{-i})\;\;\;\;\;\;\;\;\;\;\;\;\;\;\;\;\;\;\;\;\;\;\;\;\;\;\;\;\;\;\;\;\;\;\;\;\;\;\;
  \edm
\vspace{-.4in}\beq
  \;\;\;\;\;\;\;\;\;\;\propto \: \left\{
  \begin{array}{ll}
    \!\frac{\mbox{\small $\nmmi\!-\!1$}}{\mbox{\small $n\!-\!1\!+\!\alpha$}}\!\!\left(\!\frac{\mbox{\small $\nmmi\!+\!\lambda$}}{\mbox{\small $n_m\!+\!\lambda$}} \!\right)^{\!\!\!\frac{p_1}{2}} \!\!\!\frac{\mbox{$\left| \bPsi_{11} \right|^{\frac{\nmmi+\eta-p_2}{2}}\Gamma_{\!p_1}\!\!\left(\frac{n_m+\eta-p_2}{2}\right)$}}{\mbox{$\left| \bV_{m11} \right|^{\frac{n_m+\eta-p_2}{2}}\Gamma_{\!p_1}\!\!\left(\frac{n_{m(-i)+\eta-p_2}}{2}\right)$}} &
\begin{array}{l}
  \mbox{\!if $m=\phi_l$ for some} \\
  \mbox{\!$\phi_l \in \bphi_{-i}$},
\end{array}   \\
    \!\frac{\mbox{\small $\alpha$}}{\mbox{\small $n\!-\!1\!+\!\alpha$}}\!\left(\!\frac{\mbox{\small $\lambda$}}{\mbox{\small $\lambda\!+\!1$}} \!\right)^{\!\!\frac{p_1}{2}} \!\!\!\frac{\mbox{$\left| \bPsi_{11} \right|^{\frac{\eta-p_2}{2}}\Gamma_{\!p_1}\!\!\left(\frac{\eta+1-p_2}{2}\right)$}}{\mbox{$\left| \bV_{m11} \right|^{\frac{1+\eta-p_2}{2}}\Gamma_{\!p_1}\!\!\left(\frac{\eta-p_2}{2}\right)$}} & \!\mbox{for $m\!=\!M\!+\!1$}, \\    
    \!0 & \!\mbox{otherwise},
  \end{array}
  \right.\!\!\!\!\!\!\!\!\!\!
  \label{eq:phi_gibbs}
\eeq
where $\bphi_{-i}$ is the $\bphi$ vector with out the $i\tth$ element and $\bphi_{-i}$ has been relabeled if necessary so that it has at least one $\phi_l = m$ for $m=1,\dots,M$.  The split-merge MH update then works as follows.
\begin{itemize}
\item[1.] Set $\bphi^*=\bphi$. Select two points $i$ and $i'$ at random.  Let $\cC = \{l : \phi_l = \phi_{i} \mbox{ or } \phi_l = \phi_{i'}$\}.
\item[2.]
  \begin{itemize}
  \item[(a)] If $\phi_i = \phi_{i'}$, then propose a split move to divide $\cC$ into two groups in $\bphi^*$.\\[-.05in]
    \begin{itemize}
%    \item[(i)] Set $\phi_i^{\launch} = \phi_i$ and $\phi_{i'}^{\launch} = M+1$.\\[.1in]
    \item[(i)] \white{.}
      \vspace{-.64in}\bdm
      \!\!\mbox{For $l \in \cC$, set } \phi_l^{\launch} = \left\{
      \begin{array}{ll}
        \phi_i & \mbox{if }\|\bz^{(1)}_l - \bz^{(1)}_{i}\| \leq \|\bz^{(1)}_{l} - \bz^{(1)}_{i'}\|,\;\;\;\;\;\;\;\;\;\;\;\;\\[.05in]
        M+1 & \mbox{otherwise}
      \end{array}
      \right.
      \edm
    \item[(ii)] Conduct a Gibbs update sweep (\ref{eq:phi_gibbs}) to all $\phi_l : l \in \cC$, restricted to $\phi_l = \phi_i$ or $\phi_l=M+1$.
    \item[(iii)] Repeat step (iii) for a total of $L$ passes through $\cC$.  This determines $\bphi^\launch$ and the randomly chosen proposal distribution to be used next in step 2(a)(iv).
    \item[(iv)] Set $\bphi^* = \bphi^\launch$ and conduct one further restricted Gibbs sweep to the $\phi^\launch_l : l \in \cC$.  The proposal density $d(\bphi^* \mid \bphi)$ is the product of the restricted Gibbs sampling probabilities in this final sweep, whereas $d(\bphi \mid \bphi^*) = 1$.
    \end{itemize}
  \item[(b)] If $\phi_i \neq \phi_{i'}$, then propose a merge move to combine the observations in $\cC$ into one group $\bphi^*$.
    \begin{itemize}
    \item[(i)] Set $\phi^*_l = \phi_i$ for all $l \in \cC$.  The proposal density is $d(\bphi^* \mid \bphi)=1$.
    \item[(ii)] Conduct steps 2(a)(i) - 2(a)(iv) in order to evaluate the reverse proposal density $d(\bphi \mid \bphi^*)$.
    \item[(iii)] The reverse proposal density $d(\bphi \mid \bphi^*)$ is the product of restricted Gibbs sampling probabilities for moving from $\bphi^\launch$ to $\bphi$.
    \end{itemize}
  \end{itemize}
\item[3.] The MH ratio is then,
    \begin{eqnarray}
MH & \!\!=\!\! & \frac{f(\bZ \mid \bgamma, \bphi^*, \lambda, \eta, \bPsi, \alpha) f(\bphi^*) d(\bphi \mid \bphi^*)}
{f(\bZ \mid \bgamma, \bphi, \lambda, \eta, \bPsi, \alpha) f(\bphi) d(\bphi^* \mid \bphi)}\nonumber \\
&\!\!=\!\! & \frac{\!\prod_{m \in \{\phi^*_l:l\in \cC\}} 
\left[    \alpha(n^*_m-1)!
 \left( \!\frac{\lambda}{n^*_m\!+\!\lambda}\!\right)^{\!\!\frac{p_1}{2}}
\!\!\frac{\left| \bsPsi_{11} \right|^{\frac{\eta-p_2}{2}}  \Gamma_{\!p_1\!}(\frac{n_m\!+\!\eta-p_2}{2})}
     {\left| \bsV^*_{\!m11} \right|^{\frac{n^*_m+\eta-p_2}{2}}  \Gamma_{\!p_1\!}(\frac{\eta-p_2}{2})}
     \right]d(\bphi \!\mid\! \bphi^*)}
{\!\prod_{m \in \{\phi_l:l\in \cC\}} 
\left[  \alpha (n_m-1)!
 \left( \!\frac{\lambda}{n_m\!+\!\lambda}\!\right)^{\!\!\frac{p_1}{2}}
\!\!\frac{\left| \bsPsi_{11} \right|^{\frac{\eta-p_2}{2}}  \Gamma_{\!p_1\!}(\frac{n_m\!+\!\eta-p_2}{2})}
     {\left| \bsV_{\!m11} \right|^{\frac{n_m+\eta-p_2}{2}}  \Gamma_{\!p_1\!}(\frac{\eta-p_2}{2})}
     \right]d(\bphi^* \!\mid\! \bphi)},  \nonumber
\end{eqnarray}
where $n_m$ is the number of $\phi_l=m$ and $n_m^*$ is the number of $\phi^*_l=m$.  Draw a Uniform(0,1) and accept or reject in the usual manner on the basis of the MH ratio.

\item[4.] Perform one final (unrestricted) Gibbs update over {\em all} observations, i.e., for each $\phi_l$, $l=1,\dots,n$.  As discussed in \cite{Jain04}, alternating between split-merge and Gibbs updates produces an ergodic Markov chain.
\end{itemize}
\white{.}\\[-.1in]

\noindent
    {\underline{Joint MH update for $\gamma$ and $\bphi$}}
    
\noindent
The MCMC routine then consists of applying each of the above updates in turn to complete a single MCMC iteration, with the exception that the $\bgamma$ update be applied $L_g$ times each iteration.
Also, as discussed in the main paper, to improve mixing we recommend using the following joint update for $\bgamma$ and $\bphi$ in place of the individual updates for $\bgamma$ and $\bphi$ every other iteration (or simply in addition to them every iteration).  This is fairly straight-forward as the proposals are generated by simply generating a $\bgamma^*$ as above, then a $\bphi^*$ conditional on $\bgamma^*$ as above.  The MH ratio for such an update is then,
\bdm
MH = \frac{f(\bZ \mid \bgamma^*, \bphi^*, \lambda, \eta, \bPsi, \alpha) f(\bgamma^*) f(\bphi^*) d(\bgamma \mid \bgamma^*)d(\bphi \mid \bphi^*, \bgamma)}
{f(\bZ \mid \bgamma, \bphi, \lambda, \eta, \bPsi, \alpha) f(\bgamma)  f(\bphi) d(\bgamma^* \mid \bgamma)d(\bphi^* \mid \bphi, \bgamma^*)},
\edm
By the same rational as that used in \cite{Jain04}, both the individual updates and the joint update leave the possibility for the states to remain unchanged, therefore applying each of these transitions in turn will produce an ergodic Markov chain.

\vspace{-.2in}
\section{MCMC Trace Plots}
\label{sec:MCMCtrace}

In order to get a big picture view of the mixing of the MCMC algorithm, the MCMC trace plots for the number of informative variables $p_1$ and the number of clusters $M$ are provided below in Figure~\ref{fig:S1} for two separate MCMC chains (in blue and red respectively) of 75,000 iterations each.  The discrete nature of the variables makes it slightly more difficult to assess steady state than for continuous variables.  However, this assessment can be conducted via the following questions.  For parameters that spend a lot of time at multiple values, are they changing values frequently (i.e., good mixing)? Also, are they spending the same amount of relative time in a particular value through the life of the chain (i.e., a good indication that steady state has been reached). When looking at these plots, the answer to both questions appears to “yes”.

\begin{figure}[h!]
\caption{MCMC Trace plots for $\gamma_j$}
\label{fig:S2}
\includegraphics[width=.99\textwidth]{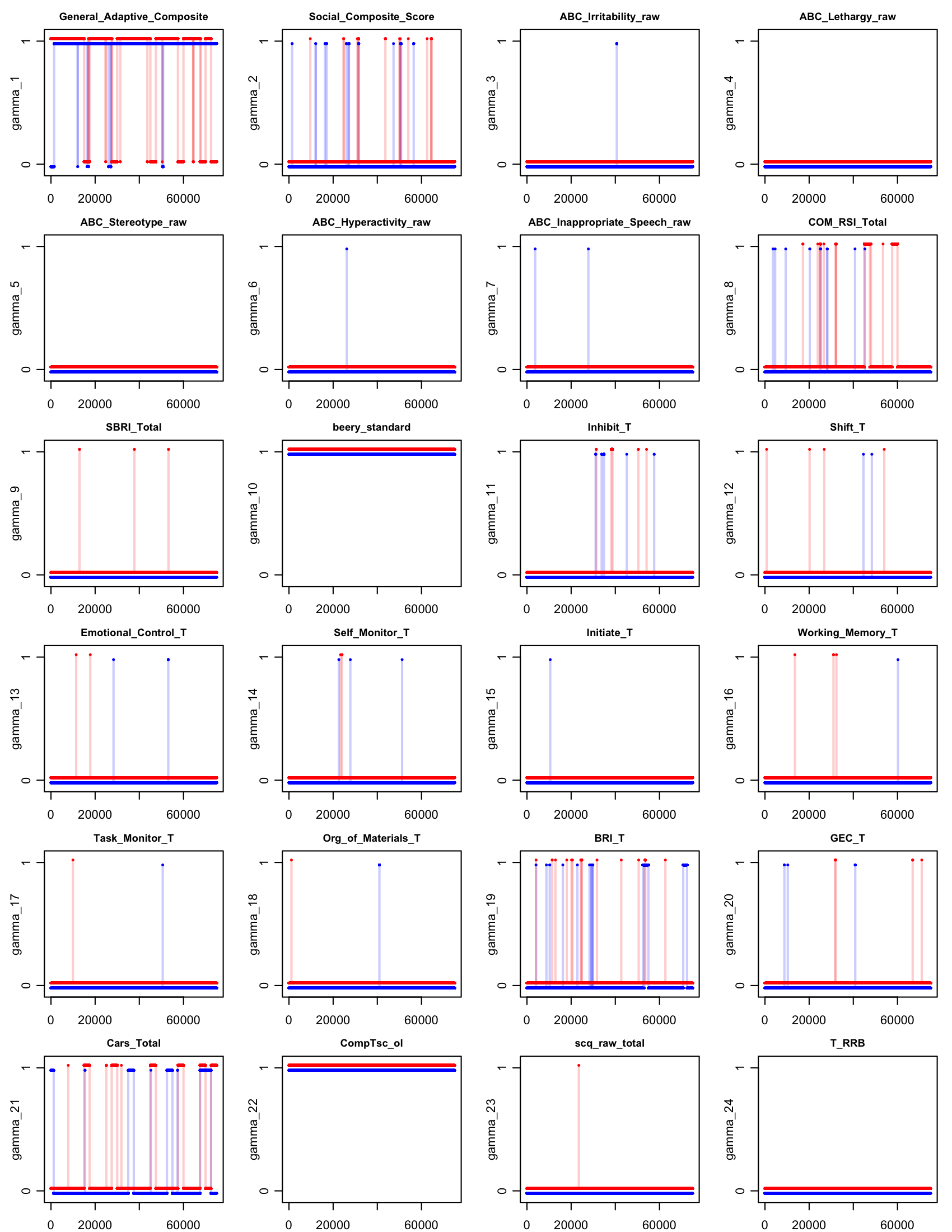}
\end{figure}

It is apparent the the mixing is slower for the variable selection than for the cluster membership, however, over 75,000 iterations, if a chain spends a non-negligible number of iterations at a value for $p_1$ (i.e., 4, 5, or 6), then their are dozens of switches to that model size that occur all throughout the life of the chains.  A more granular view of the mixing is also provided by the MCMC trace plots for the individual $\gamma_j$ (for $j=1,\dots,46$) in Figure~\ref{fig:S2} and each of the $\phi_i$ (for $i=1,\dots,96$; the exhaustive list of all 487 subjects trace plots all looked similar to these) in Figure~\ref{fig:S3}.  Once again results are provided for two chains.  Mixing is slow for $\gamma_j$, thus the need for so many MCMC iterations.  Over 75,000 iterations, if a chain spends a non-negligible number of iterations at either 0 or 1 for a given variable, then their are generally many switches that occur all throughout the life of the chains.  Mixing for the $\phi_i$ on the other hand is quite good in comparison; all observations that spend a non-negligible number of iterations with more than one cluster switch back and forth between cluster memberships quite regularly.  While there were a maximum of 12 clusters observed over all MCMC iterations from both chains, labels 7 - 12 only accounted for a total of 0.0036 of the posterior probability, so only labels 1-6 are displayed for clearer presentation.   The $\phi_i$ displayed in these plots are {\bf not} the raw $\phi_i$ that are subject to label switching.  Rather, they have been relabeled for clearer interpretation according to the information theoretic approach discussed in Section~\ref{sec:inference} so that, for instance, a label of 1 can be interpreted as ``belonging to the same cluster 1'' regardless of the MCMC iteration number.

%\begin{comment}

\vspace{-.0in}\begin{figure}[h!]
\begin{center}
\caption{MCMC Trace plots for $p_1$ and $m$}
\label{fig:S1}
\vspace{-.07in}
\includegraphics[width=.49\textwidth]{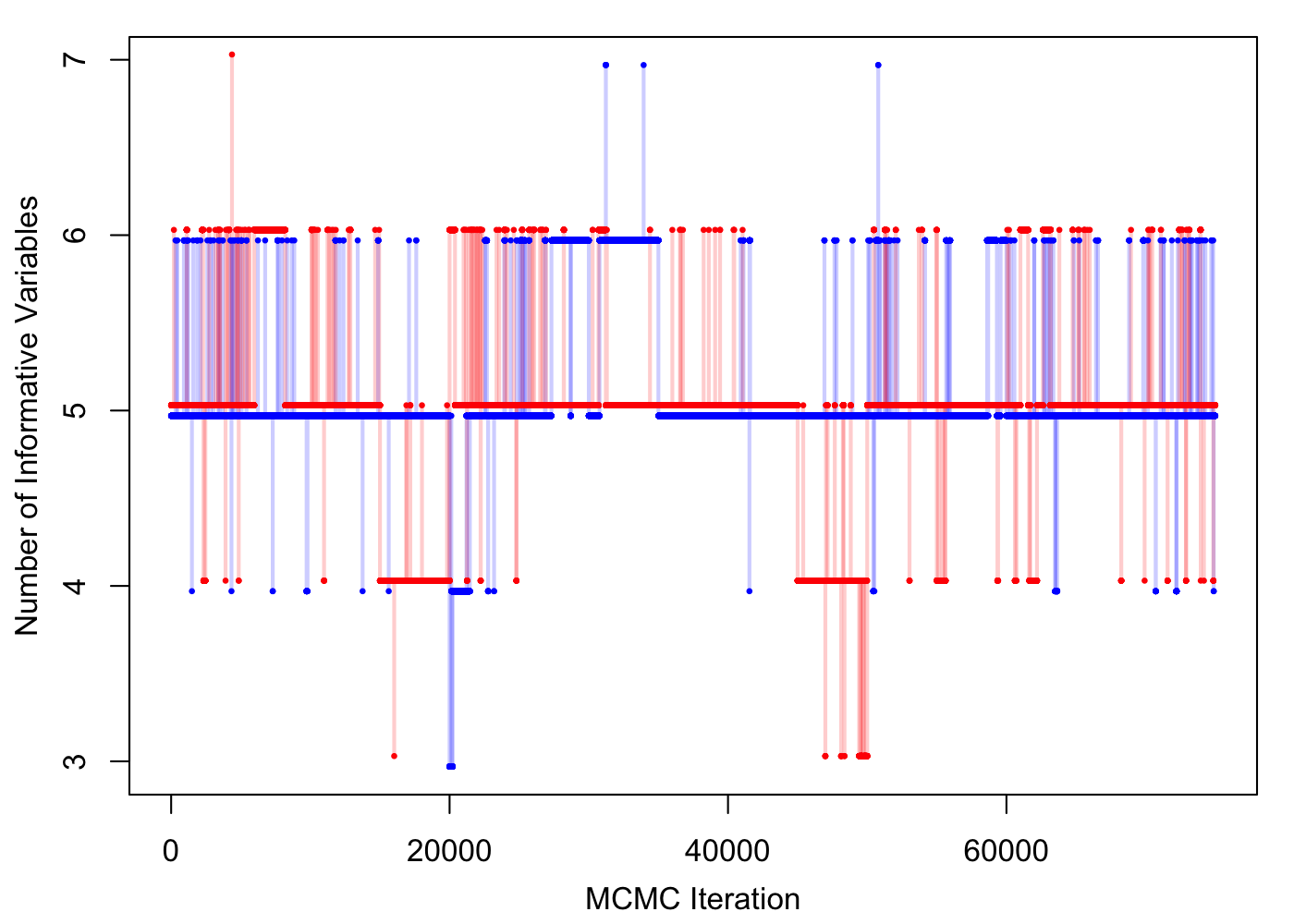}
\includegraphics[width=.49\textwidth]{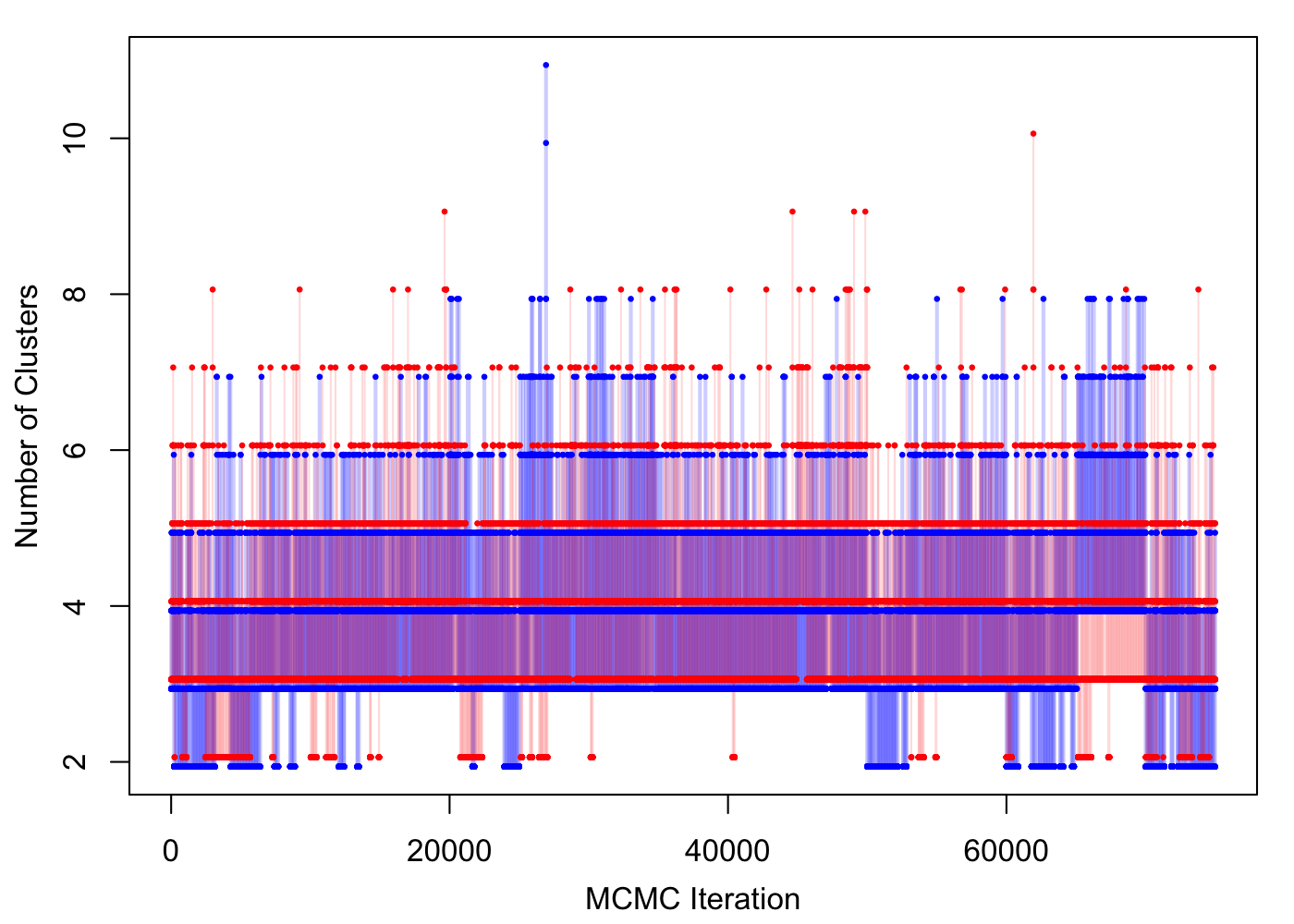}
\end{center}
\vspace{-.5in}
\end{figure}

\setcounter{figure}{1}
\begin{figure}[h!]
\caption{MCMC Trace plots for $\gamma_j$}
\includegraphics[width=.99\textwidth]{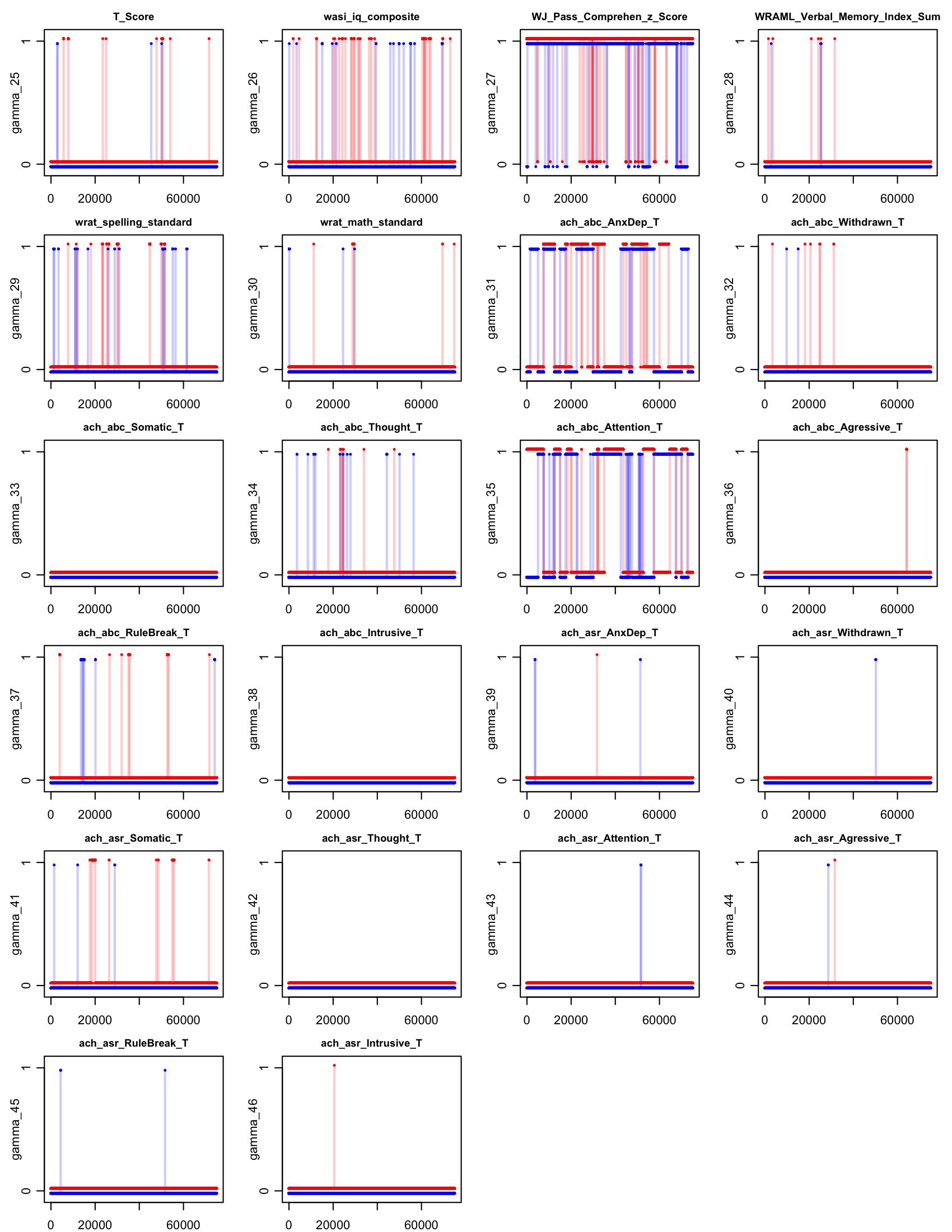}
\end{figure}

\begin{figure}[h!]
\caption{MCMC Trace plots for $\phi_i$}
\label{fig:S3}
\includegraphics[width=.99\textwidth]{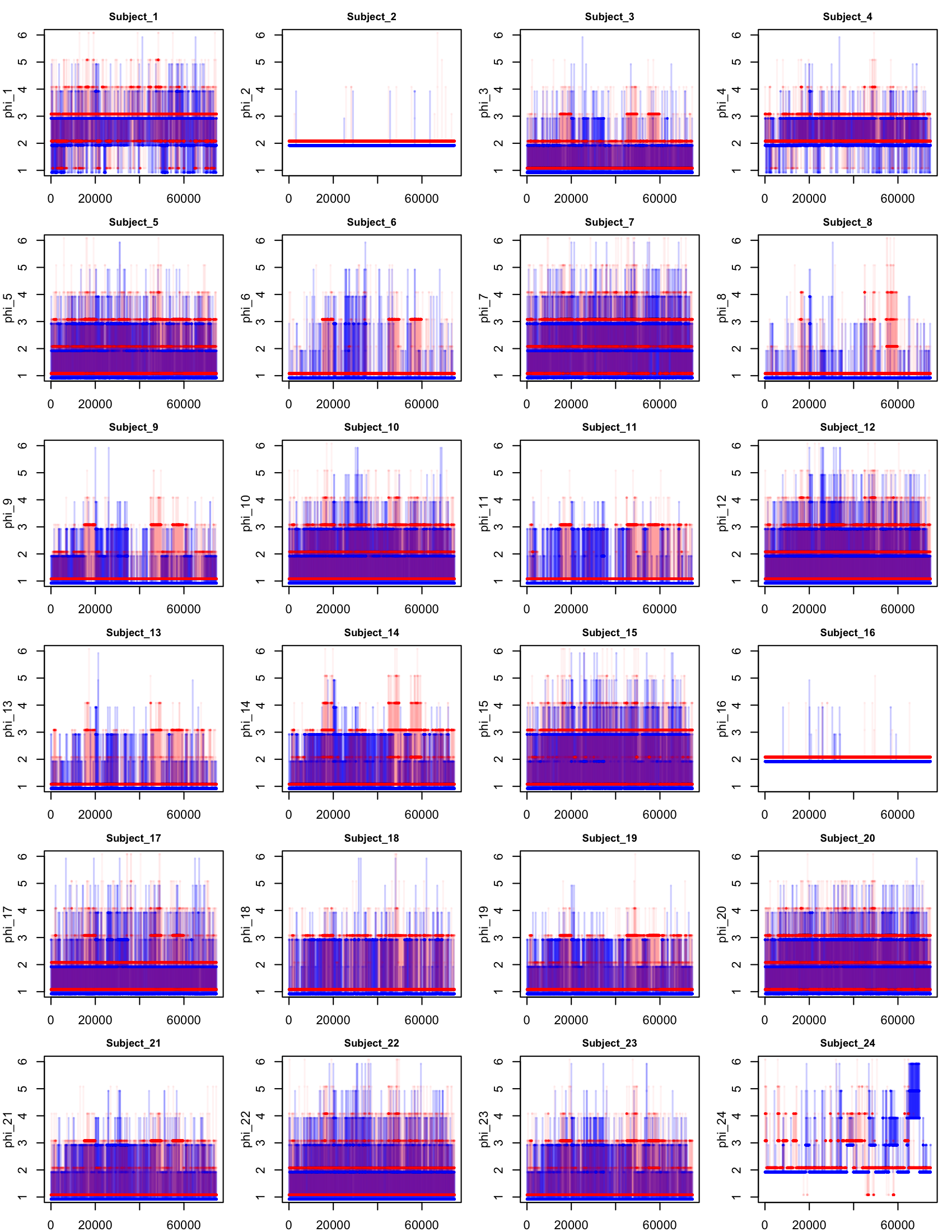}
\end{figure}

\setcounter{figure}{1}
\begin{figure}[h!]
\caption{MCMC Trace plots for $\phi_i$}
\includegraphics[width=.99\textwidth]{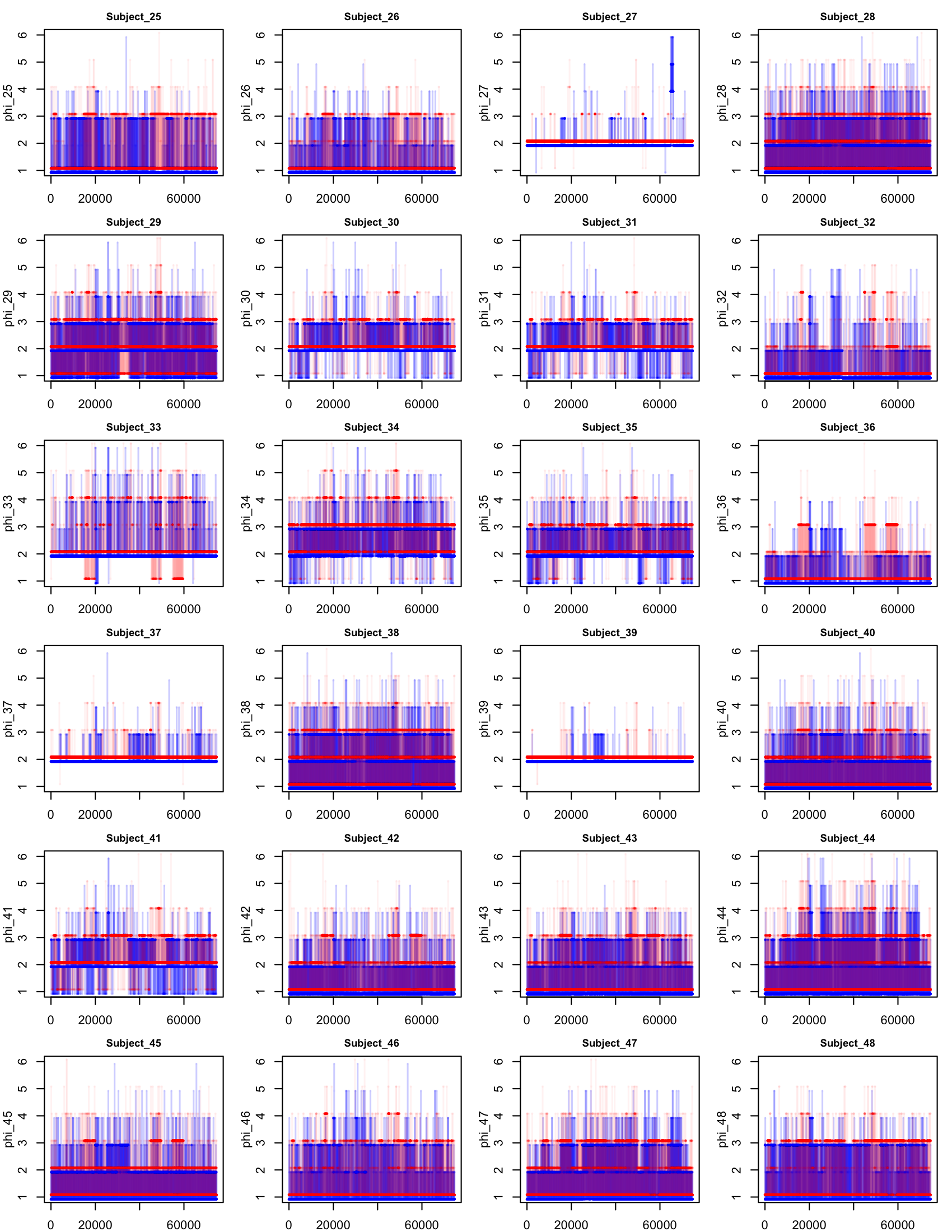}
\end{figure}
\setcounter{figure}{1}
\begin{figure}[h!]
\caption{MCMC Trace plots for $\phi_i$}
\includegraphics[width=.99\textwidth]{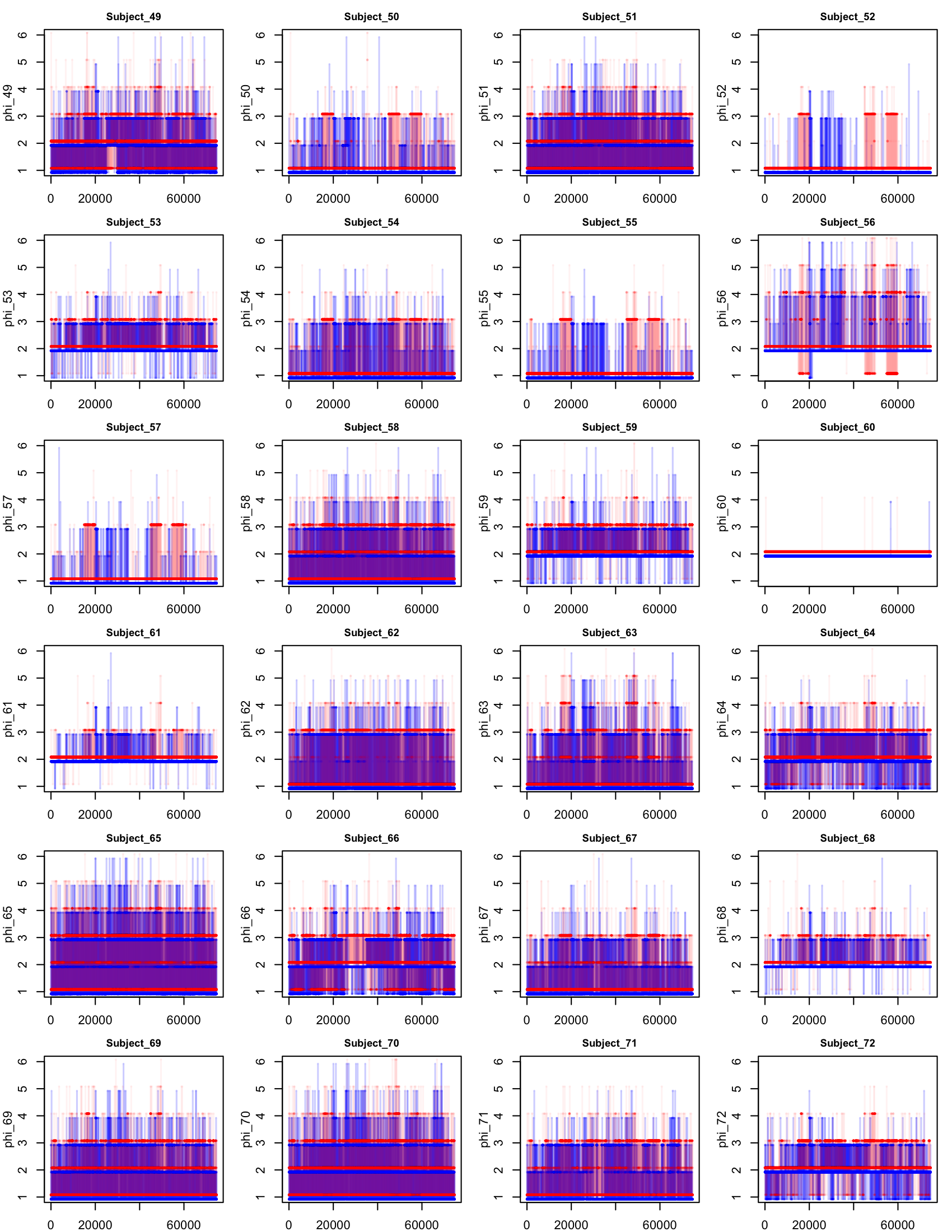}
\end{figure}
\setcounter{figure}{1}
\begin{figure}[h!]
\caption{MCMC Trace plots for $\phi_i$}
\includegraphics[width=.99\textwidth]{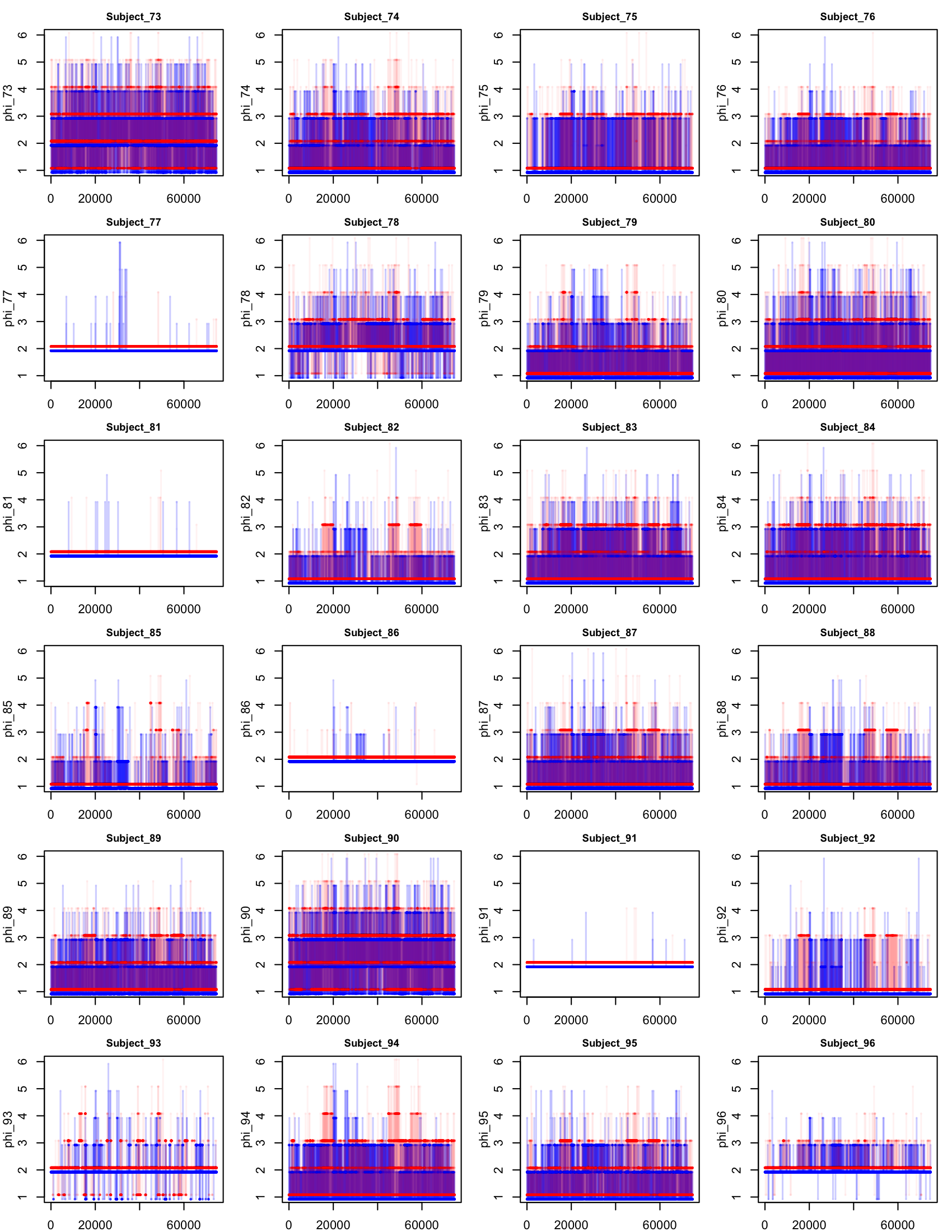}
\end{figure}

%\end{comment}

\end{appendix}